\documentclass[journal]{ieeeaccess}
\usepackage[noadjust]{cite}
\usepackage{amsmath,amsthm,graphicx,cite,booktabs,amssymb,gensymb, color, xcolor, fancyhdr,float,perpage, colortbl}
\usepackage[ruled,vlined]{algorithm2e}
\usepackage{algorithmic,array,enumerate,rotating}
\usepackage{subfigure,epsfig,dblfloatfix}
\usepackage{multirow}
\usepackage[affil-it]{authblk}
\usepackage{hyperref}
\hypersetup{
	colorlinks   = true,
	linkcolor    = blue,
	citecolor    = green
}

\makeatletter

\newtheorem{rem}{Remark}

\newcommand{\thickhline}{%
    \noalign {\ifnum 0=`}\fi \hrule height 1pt
    \futurelet \reserved@a \@xhline }

\makeatother
\def\bsx{{\boldsymbol{x}}}
\def\bsy{{\boldsymbol{y}}}
\def\bsz{{\boldsymbol{z}}}
\def\bsu{{\boldsymbol{u}}}
\def\bsv{{\boldsymbol{v}}}
\def\bsr{{\boldsymbol{r}}}
\def\bsH{{\boldsymbol{H}}}

\def\bsA{{\boldsymbol{A}}}
\def\bsB{{\boldsymbol{B}}}
\def\bsc{{\boldsymbol{c}}}

\def\bsI{{\boldsymbol{I}}}
\def\bs0{{\mathbf{0}}}
\def\bpsi{{\boldsymbol{\psi}}}
\def\balpha{{\boldsymbol{\alpha}}}

\begin{document}

\history{Date of submission 20 September 2021.}
\doi{xxxx/xxxx.xxxx.DOI}

\title{Plug-and-Play Quantum Adaptive Denoiser for Deconvolving Poisson Noisy Images}

\author{\uppercase{Sayantan Dutta} \href{https://orcid.org/0000-0002-8187-4193}{\includegraphics[scale=0.07]{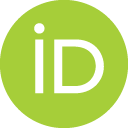}} \authorrefmark{1,2}, \IEEEmembership{Student Member, IEEE}, \uppercase{Adrian Basarab} \href{https://orcid.org/0000-0002-5642-7244}{\includegraphics[scale=0.07]{ORCID128x128.png}} \authorrefmark{1}, \IEEEmembership{Senior Member, IEEE},
\uppercase{Bertrand Georgeot} \href{https://orcid.org/0000-0002-5886-9118}{\includegraphics[scale=0.07]{ORCID128x128.png}} \authorrefmark{2}, 
\uppercase{and Denis Kouam\'e}\authorrefmark{1}, \IEEEmembership{Senior Member, IEEE}}

\address[1]{Institut de Recherche en Informatique de Toulouse, UMR CNRS 5505, Universit\'e de Toulouse, Toulouse, France}

\address[2]{Laboratoire de Physique Th\'eorique, Universit\'e de Toulouse, CNRS, UPS, Toulouse, France}

\tfootnote{This work was supported in part by Centre National de la Recherche Scientifique (CNRS) through the 80 prime program.}

\markboth
{Dutta \headeretal: Plug-and-Play Quantum Adaptive Denoiser for Deconvolving Poisson Noisy Images}
{Dutta \headeretal: Plug-and-Play Quantum Adaptive Denoiser for Deconvolving Poisson Noisy Images}

\corresp{Corresponding author: Sayantan Dutta (e-mail: sayantan.dutta@irit.fr).}


%
%

\begin{abstract}

A new Plug-and-Play (PnP) alternating direction of multipliers (ADMM) scheme is proposed in this paper, by embedding a recently introduced adaptive denoiser using the Schroedinger equation’s solutions of quantum physics. The potential of the proposed model is studied for Poisson image deconvolution, which is a common problem occurring in number of imaging applications, such as limited photon acquisition or X-ray computed tomography. Numerical results show the efficiency and good adaptability of the proposed scheme compared to recent state-of-the-art techniques, for both high and low signal-to-noise ratio scenarios. This performance gain regardless of the amount of noise affecting the observations is explained by the flexibility of the embedded quantum denoiser constructed without anticipating any prior statistics about the noise, which is one of the main advantages of this method. The main novelty of this work resided in the integration of a modified quantum denoiser into the PnP-ADMM framework and the numerical proof of convergence of the resulting algorithm.

\end{abstract}




\maketitle

\section{Introduction}
\label{sec:intro}

\PARstart{R}{estoration} of a distorted image is one of the most fundamental tasks in inverse problems related to imaging applications such as denoising, deblurring, super-resolution, compression or compressed sensing. In number of applications such as limited photon acquisition, X-ray computed tomography, positron emission tomography, etc., the noise degrading the acquired data follows a Poisson distribution. These Poissonian models have been extensively studied in the fields of astronomical \cite{starck2002deconvolution, starck2003astronomical, starck2007astronomical}, photographic \cite{foi2005spatially, guo2019toward} or biomedical \cite{fessler1995penalized, nowak2000statistical, dey2006richardson, willett2004fast, sarder2006deconvolution, de2011alternating} imaging. The inversion process is expressed as the estimation of a clean image $\bsx \in \mathbb{R}^n$ from observed degraded image $\bsy \in \mathbb{R}^m$. The estimation of the underlying hidden image from this distorted observation is often formulated as the optimization of a cost function implementing the idea of the maximum a posteriori (MAP) estimator \cite{poor2013introduction}, i.e., the maximization of the posterior probability, defined as
\begin{equation}
\hat{\bsx} = \underset{\bsx} {\textrm{arg max}}~~ P(\bsx|\bsy),
\label{eq:PostProb}
\end{equation}
where $P(\bsx|\bsy)$ is the posterior probability density function that defines $\bsx$ for a given measurement $\bsy$ and $\hat{\bsx}$ represents the estimation of the unobserved image $\bsx$. Taking $- log(\cdot)$ element wise and applying the Bayes' theorem, the maximization problem above becomes
\begin{align}
\hat{\bsx} 
& = \underset{\bsx} {\textrm{arg min}} \big( - log \left(  P(\bsy|\bsx) \right) - log \left(  P(\bsx) \right) + log \left(  P(\bsy) \right) \big).
\label{eq:logPostProb}
\end{align}
$f(\bsx) = - log \left(  P(\bsy|\bsx) \right) $ is the negative log-likelihood function whose expression depends on the observation (degradation) model, and $g(\bsx) = - log \left(  P(\bsx) \right) $ is the \textit{a priori} log-distribution of $\bsx$, that only depends on some prior knowledge on the image to estimate and is also called regularization function. Note that $ P(\bsy) $ does not depend on $\bsx$ and is usually ignored in the estimation of $\hat{\bsx}$. With these notations, the optimization problem to solve can be expressed as
\begin{equation}
\hat{\bsx} = \underset{\bsx} {\textrm{arg min}}~~ \big( f(\bsx) + g(\bsx) \big).
\label{eq:xhat}
\end{equation}

Using a suitable choice of the regularization function, based for example on the \textit{a priori} statistics of the image to estimate, proximal operator- \cite{bauschke2011convex} based iterative schemes have been extensively studied to solve \eqref{eq:xhat} \cite{beck2009fast, yang2009efficient, eckstein1992douglas, setzer2010deblurring, boyd2011distributed, afonso2010fast, yang2010fast, chan2013constrained, almeida2013deconvolving, tao2009alternating, zhang2020signal}. In particular, the alternating direction method of multipliers (ADMM) \cite{boyd2011distributed, afonso2010fast, yang2010fast, chan2013constrained, almeida2013deconvolving, tao2009alternating, zhang2020signal} has been largely used, by redefining the optimization problem \eqref{eq:xhat} into a constrained optimization framework. During the last decade, a new approach was proposed in the literature, enabling the use of state-of-the-art denoisers instead of the proximal operator, known as the plug-and-play (PnP) scheme \cite{venkatakrishnan2013plug}. PnP paves the way of using a wide range of state-of-the-art denoisers such as patch-based dictionary learning methods \cite{elad2006image}, block-matching 3D filtering (BM3D) \cite{dabov2007image}, non-local means (NLM) \cite{buades2005non}, high-order variational models \cite{lu2016implementation}, etc. The interest of PnP schemes in image restoration have been shown by number of studies, e.g., \cite{azzari2016variance, azzari2017variance, rond2016poisson, kwan2017resolution, wang2017parameter, unni2018linearized, brifman2016turning, teodoro2016image, sreehari2016plug, chan2017plug, chan2019performance, ryu2019plug, zhang2020plug, xu2020provable, cohen2020regularization, teodoro2017scene}. Interestingly, these PnP-ADMM methods do not require any prior information about the hidden image $\bsx$, as a consequence of the intrinsic association between the regularizer and the external denoiser.

More recently, alternative learning-based approaches were developed in the literature using deep learning (convolutional neural network (CNN)) techniques for tackling inverse problems \cite{heide2014flexisp, borgerding2017amp, lucas2018using, liang2020deep, yanna2020deep}. During the past few years the implementation of these Deep-CNN networks has been introduced for image denoising \cite{zhang2017beyond, zhang2017learning} and further extended to the PnP schemes \cite{chen2021deep}. These Deep-CNN networks give several advantages such as reconstruction accuracy and convergence speed \cite{giryes2018tradeoffs}. However, more often they suffer from some drawbacks. First, such denoisers should be trained using the noise variance in each iteration. Hence, during the iterative process of the PnP framework, the noise variance is usually unknown since it varies at each iteration, and leads to a divergence of the algorithm for a pre-trained Deep-CNN architecture \cite{sommerhoff2019energy}. Second, the training procedure is very costly since Deep-CNN denoisers require expensive retraining whenever the noise level or noise type change. Also, each iteration involves a Deep-CNN denoising process, so using a large neural network and/or too many iterative operations leads to a time consuming task. Third, the theoretical aspects of Deep-CNN denoiser-based PnP models are still not clear.

This work focuses on PnP-ADMM algorithms applied to Poisson deconvolution problems, i.e., recover an image from a blurred observation contaminated by Poisson noise. Since the state-of-the-art denoisers (\textit{e.g.}, BM3D \cite{dabov2007image}) used within PnP schemes were primarily designed for additive Gaussian noise, they consequently exhibit inconsistency with a non-Gaussian model.
Furthermore, decoupling the restoration and denoising steps within PnP frameworks alternatively converts the noise distribution affecting the observed distorted image into a possibly different noise model, and in particularly into a non-Gaussian noise.
To mitigate this limitation, a variance stabilizing transformation (VST) \cite{anscombe1948transformation, dupe2009proximal, makitalo2010optimal, makitalo2011closed}, known as the Anscombe transformation, was embedded in several PnP-ADMM algorithms to adapt them to a data-dependent model. Indeed, VST was designed to remodel approximately a random data-dependent noise into an additive Gaussian noise, before processing through a Gaussian denoiser.
Although these refined VST-based PnP schemes exhibit very good performance for low-intensity noise \cite{azzari2016variance, azzari2017variance, rond2016poisson} and outperform existing state-of-the-art prior based models, they are less accurate while dealing with high-intensity noise  (\textit{i.e.,} low SNR) \cite{salmon2014poisson}. Furthermore, the nonuniform nature of the convolution operator under a VST leads to fundamental flaws in the deconvolution algorithms \cite{ azzari2017variance, rond2016poisson, deledalle2012compare}.

In this paper, we address these shortcomings by embedding into a PnP-ADMM scheme a new adaptive denoiser \cite{dutta2021quantum, smith2018adaptive} designed by borrowing tools from quantum mechanics. The adaptive nature of this denoiser makes it highly efficient at selectively eliminating noise from higher intensity pixels, without relying on any statistical assumption about the noise \cite{dutta2021poisson}. Its efficiency regardless of the assumption of Gaussian noise represents the main motivation of its interest in Poisson deconvolution PnP-ADMM algorithms, discarding the necessity of a VST. To summarize, the main novelty of the paper is the use of quantum mechanical concepts in the field of image restoration. The primary contributions are the quantum denoiser, its integration into a PnP-ADMM scheme, and the experimental proof of convergence of the final algorithm.

The remainder of the paper is organized as follows. After a brief discussion on PnP-ADMM algorithms in Section~\ref{sec:background}, the construction of the proposed method referred to as QAB-PnP is illustrated for Poisson inverse problems in Section~\ref{sec:probcons}. Section~\ref{sec:simu} regroups the numerical experiments and Section~\ref{sec:conclusion} draws the conclusions and the perspectives.

\section{Background}
\label{sec:background}

\subsection{Alternating direction method of multipliers}
\label{sec:ADMM}


ADMM is an iterative convex optimization algorithm, resulting from the fusion of the dual decomposition  method with the method of multipliers \cite{dantzig1960decomposition, dantzig2016linear, bnnobrs1962partitioning, hestenes1969multiplier, powell1969method, gabay1976dual}. Several developments have been proposed during the last few decades, resulting into a rapidly growing literature \cite{eckstein1992douglas, boyd2011distributed, yang2010fast, afonso2010fast, setzer2010deblurring}.
ADMM algorithm is able to solve constrained optimization problems of the form
\begin{equation}
\begin{array}{l}
\underset{\bsx,\bsz} {\textrm{minimize}}~~~~  f(\bsx) + g(\bsz)   \\
\textrm{subject to}~~ \bsA \bsx + \bsB \bsz = \bsc,
\label{eq:ADMMgeneral}
\end{array}
\end{equation}
where $f$ and $g$ are assumed to be closed convex functions of variables $\bsx \in \mathbb{R}^n$ and $\bsz \in \mathbb{R}^m$, with $\bsA \in \mathbb{R}^{p \times n}$, $\bsB \in \mathbb{R}^{p \times m}$ and $\bsc \in \mathbb{R}^p$. 
The associated augmented Lagrangian function is defined as 
\begin{align}
\mathcal{L}_\lambda (\bsx,\bsz,\bsv) = f(\bsx) + g(\bsz) & + \dfrac{\lambda}{2} \left\| \bsA \bsx + \bsB \bsz  - \bsc + \dfrac{\bsv}{\lambda} \right\|_2^2  \nonumber \\
& - \dfrac{1}{2\lambda} \left\| \bsv \right\|_2^2,
\label{eq:auglag_0}
\end{align}
\noindent where $\bsv \in \mathbb{R}^p$ is the Lagrangian multiplier, and $\lambda \in \mathbb{R}^+$ is the penalty parameter of the augmented Lagrangian. An equivalent expression of the augmented Lagrangian $\mathcal{L}_\lambda (\bsx,\bsz,\bsv)$ can be obtained by scaling the Lagrangian multiplier $\bsu = (1/\lambda)\bsv$, as follows:
\begin{align}
\mathcal{L}_\lambda (\bsx,\bsz,\bsv) & =  f(\bsx) + g(\bsz) + (\lambda/2) \left\| \bsA \bsx + \bsB \bsz - \bsc + \bsu \right\|_2^2  \nonumber \\
& ~~~~~~~~~~~~~~~~~~ - \textrm{constant}_\bsv  \nonumber \\
& \overset{\textrm{def}}{=} \mathcal{L}_\lambda (\bsx,\bsz,\bsu)
\label{eq:auglag_1}
\end{align}

ADMM algorithm decouples the augmented Lagrangian into three iterative steps as follows:
\begin{align}
& \bsx^{k+1} = \underset{\bsx} {\textrm{arg min}}~~ \mathcal{L}_\lambda (\bsx,\bsz^k,\bsu^k) 
 \label{eq:step1} \\
& \bsz^{k+1} = \underset{\bsz} {\textrm{arg min}}~~ \mathcal{L}_\lambda (\bsx^{k+1},\bsz,\bsu^k) 
 \label{eq:step2} \\
& \bsu^{k+1} = \bsu^k + \bsA \bsx^{k+1} + \bsB \bsz^{k+1} - \bsc.
\label{eq:step3}
\end{align}

The convergence of this iterative scheme has been widely discussed in the literature of convex programming and within various statistical problems \cite{gabay1983chapter, he2015non, fadili2009monotone}. ADMM technique has a broad spectrum of applications in the context of signal and image restoration applications \cite{figueiredo2010restoration, steidl2010removing, chan2011augmented, morin2012alternating, hourani2018restoration}.

\subsection{ADMM application to image restoration}
\label{sec:ADMMimrest}

Let us consider the following general image restoration problem, characterized by the forward model
\begin{equation}
\bsy = \boldsymbol{O} \bsx,
\label{eq:imres}
\end{equation}
where $\bsy$ is the observed image related to the underlying image $\bsx$ through the degradation operator $\boldsymbol{O}$. ADMM can be used to estimate the MAP solution of such an image restoration task by reformulating it as \eqref{eq:ADMMgeneral} using the following parameterization: $\bsz = \bsx$, thus $\bsA = -\bsB =  \bsI_{n \times n}$, $\bsc = \bs0_n$, where $\bsI_{n \times n}$ is the identity matrix of size $n \times n$ and $\bs0_n$ is a zero vector of size $n$. The associated augmented Lagrangian is given by
\begin{equation}
\mathcal{L}_\lambda (\bsx,\bsz,\bsu) = f(\bsx) + g(\bsz) + \dfrac{\lambda}{2} \left\| \bsx - \bsz + \bsu \right\|_2^2,
\label{eq:auglag}
\end{equation}
where $f(\bsx) = - log \left(  P(\bsy|\bsx) \right) $ is the data fidelity term depending on $\boldsymbol{O}$ and $g(\bsz)$ the regularization function. To accelerate the convergence, the penalty parameter $\lambda$ is usually increased at each iteration, by multiplication by a factor of $\gamma > 1$ \cite{chan2017plug}, instead of using a fixed value. At each iteration, ADMM performs the following steps:
\begin{align}
& \bsx^{k+1} = \underset{\bsx} {\textrm{arg min}}~~ f(\bsx) + \dfrac{\lambda^k}{2} \left\| \bsx - \bsz^k + \bsu^k \right\|_2^2
 \label{eq:admm1} \\
& \bsz^{k+1} = \underset{\bsz} {\textrm{arg min}}~~ g(\bsz) + \dfrac{\lambda^k}{2} \left\| \bsx^{k+1} - \bsz + \bsu^k \right\|_2^2
 \label{eq:admm2} \\
& \bsu^{k+1} = \bsu^k + \bsx^{k+1} - \bsz^{k+1}
\label{eq:admm3} \\
& \lambda^{k+1} =\gamma \lambda^k
\label{eq:admm4}
\end{align}

\begin{figure*}[h!]
\centering
\includegraphics[width=1\textwidth]{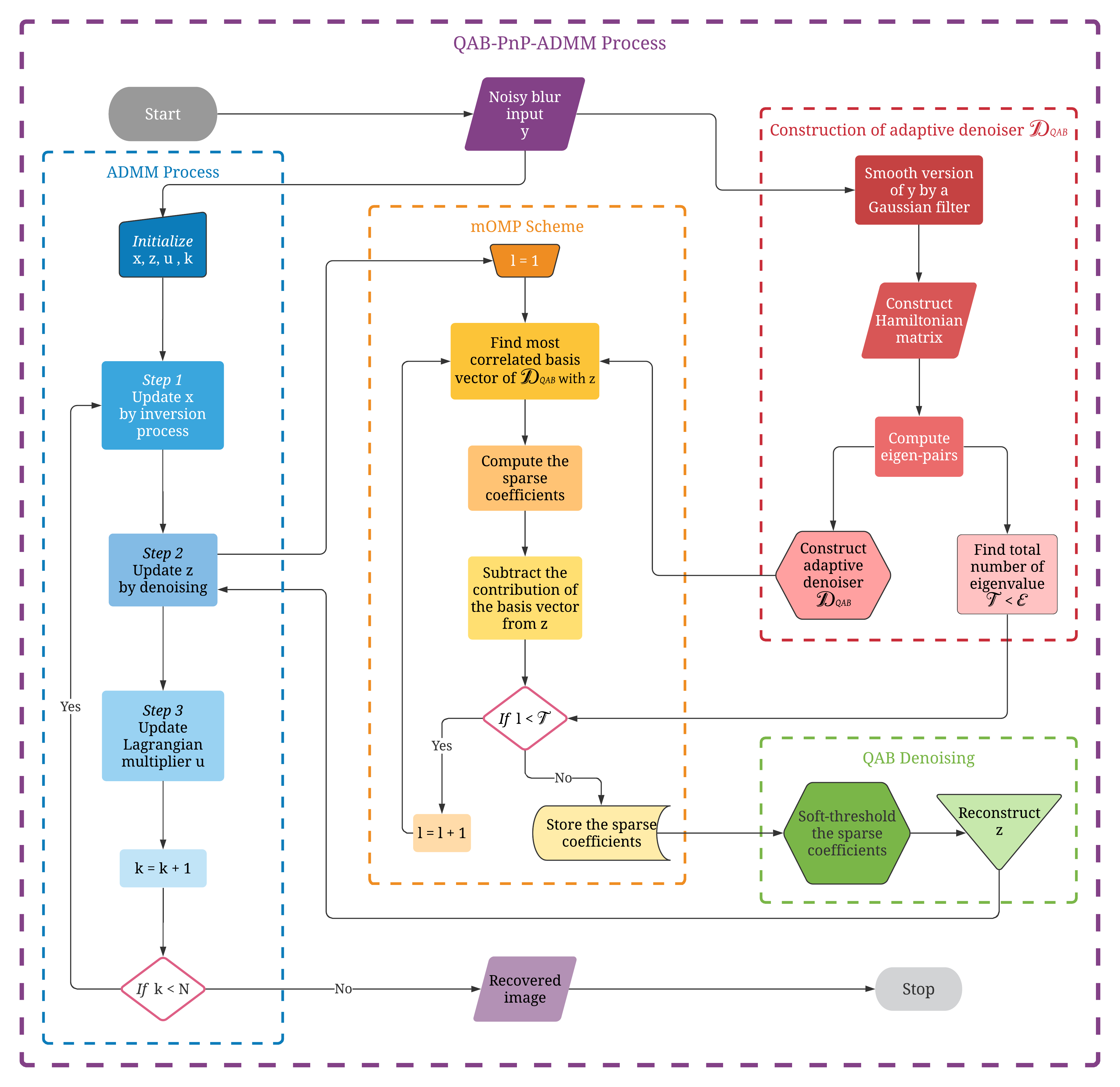}

\caption{Flowchart of the proposed QAB-PnP algorithm.}
\label{fig:flowc}
\end{figure*}

\subsection{Plug-and-Play (PnP) framework}
\label{sec:pnp}

Since its initial development, the PnP scheme \cite{venkatakrishnan2013plug} is largely accepted for signal and image restoration problems due to its extremely promising performance \cite{azzari2016variance, azzari2017variance, rond2016poisson, kwan2017resolution, wang2017parameter, unni2018linearized, brifman2016turning, teodoro2016image, sreehari2016plug, chan2017plug, chan2019performance, ryu2019plug, zhang2020plug, xu2020provable, cohen2020regularization, teodoro2017scene}. The primary goal of PnP is to consider a state-of-the-art denoiser as the prior of a constrained optimization process. Interestingly, no prior knowledge is required about the image to estimate to derive the regularization function $g$, since $g$ is intrinsically defined through the external denoiser used.

The efficiency of ADMM algorithm mainly reposes on its ability of decoupling the optimization processes over each variable, as shown in the previous section. ADMM steps performed at each iteration, \eqref{eq:admm1}, \eqref{eq:admm2} and \eqref{eq:admm3}, can be interpreted as follows. \eqref{eq:admm1} is originally an inversion step to get the best possible primary image satisfying the data through the data fidelity function $f(\bsx)$, while the third step \eqref{eq:admm3} updates the Lagrangian multiplier. The second step \eqref{eq:admm2} can be rewritten as
\begin{equation}
\bsz^{k+1} = \underset{\bsz} {\textrm{arg min}}~~ g(\bsz) + \dfrac{\lambda^k}{2} \left\| \bsz - (\bsx^{k+1} + \bsu^k) \right\|_2^2.
\label{eq:admm22}
\end{equation}

The expression on the right hand side of \eqref{eq:admm22} fundamentally intends to find the solution that optimizes the compromise between the difference between $\bsz$ and $(\bsx^{k+1} + \bsu^k)$ and the regularization function $g(\bsz)$. Thus, it can be associated to a denoising problem designed to denoise $(\bsx^{k+1} + \bsu^k)$. Therefore it is possible to rewrite this step as
\begin{equation}
\bsz^{k+1} =  \mathcal{D} \Big( \bsx^{k+1} + \bsu^k \Big),
\label{eq:admm222}
\end{equation}
where $\mathcal{D}(\cdot)$ is a denoising operator. Hence it is feasible to implement a state-of-the-art denoiser to handle the denoising operation as proposed in \cite{venkatakrishnan2013plug}. The most interesting feature representing the key benefit of this approach is that this PnP model does not require the prior term $g(\bsz)$ explicitly, rather it is indirectly related to the choice of the denoiser $\mathcal{D}(\cdot)$ (see, e.g., \cite{buades2005non, dabov2007image, chatterjee2011patch, knaus2014progressive}). 

Despite its interest shown in number of imaging applications, PnP-ADMM still presents important theoretical challenges while dealing with Poisson deconvolution. Indeed, most advanced denoisers available in the literature generally consider additive Gaussian models and cannot be implemented directly for other noise removal processes which do not follow Gaussian statistics. Furthermore, despite observing an image degraded by a specific noise model (\textit{e.g.,} Poisson in our case), the image $( \bsx^{k+1} + \bsu^k )$ to be denoised at each iteration in \eqref{eq:admm222} does not necessarily follow the same noise distribution. Therefore, handling an inverse problem using the PnP-ADMM algorithm requires to transform the unknown noise distribution of the noisy image $( \bsx^{k+1} + \bsu^k )$ into an additive Gaussian distribution before implementing a Gaussian denoiser. In this context VST-like \cite{anscombe1948transformation} transformations propose an efficient way of estimating approximately a Gaussian distribution from other types of data-dependent models. The convolution product is however not invariant under this VST and consequently leads to theoretical flaws. Therefore, a versatile denoiser adapted to different noise models, without \textit{a priori} hypothesis about the noise statistics, is desirable to be efficient regardless of the prior noise distribution in this PnP framework.

In this work, our primary focus will be on the formulation of a PnP-ADMM model using an adaptive denoiser, constructed from the principles of quantum mechanics \cite{dutta2021quantum, smith2018adaptive}, and its implementation into Poisson deconvolution processes. This quantum adaptive basis (QAB)-based denoiser does not require any explicit noise model. Therefore, while included in an PnP-ADMM scheme, it does not need the use of a VST before denoising and mitigates this  theoretical limitation.

\subsection{Convergence of PnP-ADMM algorithms}
\label{sec:convPnPADMM}

One major challenge of PnP-ADMM algorithms is to prove their convergence, due to the implicit relation between the regularization function $g(\bsz)$ and the denoising operator $\mathcal{D}(\cdot)$. Note that the convergence of conventional ADMM has been largely discussed in the literature, primarily in \cite{gabay1983chapter} and \cite{eckstein1992douglas} and more recently in \cite{boyd2011distributed} based on the proximal operator \cite{moreau1965proximite} or in \cite{hong2016convergence}. The proof of global convergence of PnP-ADMM algorithm \cite{sreehari2016plug} has been shown in the case of non-expansive denoisers belonging to the family of symmetric smoothing filters \cite{milanfar2013symmetrizing, kheradmand2013general, chan2015understanding, teodoro2018convergent}. Yet these conditions are too restrictive for generalisation to all the denoisers. To overcome this issue, a series of works has been published during the last few years showing the fixed point convergence of PnP-ADMM algorithms for bounded denoisers not necessarily symmetric and non-expansive \cite{chan2017plug, chan2019performance, cohen2020regularization, zhang2020plug, ryu2019plug, xu2020provable, teodoro2017scene}, but we stress that all these algorithms were constructed for Gaussian noise.

\section{Proposed PnP-ADMM algorithm}
\label{sec:probcons}

\subsection{Poissonian deconvolution model}
\label{sec:poissdecon}

Let us denote by $\bsx \in \mathbb{R}^{n^2}$ the image to be recovered from the observation $\bsy \in \mathbb{R}^{n^2}$, a degraded version by a point spread function (PSF) and Poisson process denoted by $\mathcal{P}(\cdot)$. Without loss of generality, we consider herein square images of size $n \times n$, written as vectors in lexicographical order. The resulting image formation model is 
\begin{equation}
\bsy = \mathcal{P}( \bsH \bsx ),
\label{eq:noisemodel}
\end{equation}
where $ \bsH \in \mathbb{R}^{n^2 \times n^2} $ is a block circulant with circulant blocks (BCCB) matrix acounting for 2D circulant convolution with the PSF. The pixels of the observed blurry and noisy image $\bsy$ are denoted by $\bsy[i], i=1,2,\cdots,n^2$, and are contemplated as the independent realizations of a Poisson process with parameter $(\bsH \bsx)[i] \geq 0$ given by
\begin{eqnarray}
\underset{ \mbox{for} \; i = 1,2,\cdots, n^2} { P\big(\bsy[i] \big| \bsx[i] \big) } = \left \{
   \begin{array}{c c l}
   		\dfrac{e^{-(\bsH \bsx)[i]}  {(\bsH \bsx)[i]}^{\bsy[i]}  }{\bsy[i]!}  &  \mbox{if} \; \bsy[i] \geq 0,\\
  	    				 0  & \mbox{elsewhere},
   \end{array}
   \right.
\label{eq:Poissnoisedist}
\end{eqnarray}
where $(\cdot)[i]$ represents the $i$-th component of a vectorized image. The restoration of $\bsx$ from the noisy-blurred observation $\bsy$ is the primary objective of Poisson deconvolution methods.

One standard way to estimate $\bsx$ from the observation model \eqref{eq:noisemodel} is to use the MAP estimator in \eqref{eq:PostProb}. The Poisson noise probability density function is defined as
\begin{equation}
P(\bsy|\bsx) = \displaystyle\prod_i \dfrac{e^{-(\bsH \bsx)[i]}  {(\bsH \bsx)[i]}^{\bsy[i]}  }{\bsy[i]!}.
\end{equation}
Thus, the log-likelihood term, \textit{i.e.,} the data fidelity term $f(\bsx)$ used within the MAP estimator, is given by
\begin{align}
f(\bsx) & = - log\left(  P(\bsy|\bsx) \right) \nonumber \\
& = - \displaystyle\sum_i  log\left(  \dfrac{e^{-(\bsH \bsx)[i]}  {(\bsH \bsx)[i]}^{\bsy[i]}  }{\bsy[i]!}  \right)  \nonumber \\
& = - \bsy^T log(\bsH \bsx) + {\boldsymbol{1}}^T \bsH \bsx + \text{constant},
\label{eq:datafidelity}
\end{align}
where ${\boldsymbol{1}}$ is a vector of length $n^2$ with all elements equal to $1$. As explained previously, the function $g(\bsx)$ in \eqref{eq:xhat}, a prior of $\bsx$, depends on some prior knowledge on the image to estimate. In a PnP framework, this prior is intrinsically defined through the external denoiser, removing the fact of defining the prior term $g(\bsx)$ explicitly.
Hence, using the data fidelity term $f(\bsx)$ in \eqref{eq:datafidelity}, the PnP-ADMM steps depicted in \eqref{eq:admm1}, \eqref{eq:admm3}, \eqref{eq:admm4} and \eqref{eq:admm222} become:
\begin{align}
&\bsx^{k+1}  = \underset{\bsx} {\textrm{arg min}}\Big(- \bsy^T log(\bsH \bsx) + {\boldsymbol{1}}^T \bsH \bsx \nonumber \\
& ~~~~~~~~~~~~~~~~~~~~~~~~~~~ + \dfrac{\lambda^k}{2} \left\| \bsx - \bsz^k + \bsu^k  \right\| _2^2\Big)
\label{eq:admmPnP1}\\
& \bsz^{k+1} =  \mathcal{D} \Big(\bsx^{k+1} + \bsu^k \Big)
\label{eq:admmPnP2} \\
& \bsu^{k+1} = \bsu^k + \bsx^{k+1} - \bsz^{k+1}
\label{eq:admmPnP3} \\
& \lambda^{k+1} =\gamma \lambda^k,
\label{eq:admmPnP4}
\end{align}
where $\mathcal{D}(\cdot)$ is the denoising operator considered within the PnP-ADMM algorithm. In this work, following \cite{ruder2016overview}, a gradient descent algorithm is used to solve the minimization problem \eqref{eq:admmPnP1}, that requires the use of the gradient of the augmented Lagrangian $ \mathcal{L}_\lambda $ given by
\begin{equation}
\nabla_\bsx \mathcal{L}_\lambda = - \bsH^T \big(\bsy/(\bsH \bsx)\big) + \bsH^T {\boldsymbol{1}} + \lambda^k ( \bsx - \bsz^k + \bsu^k ),
\label{eq:admmPnP1Grad}
\end{equation}
where $ \nabla_\bsx $ represents the derivative with respect to $\bsx$ and $\bsy/(\bsH \bsx)$ stands for element-wise division.

The following subsection describes the Poisson denoiser inspired from quantum mechanics used within the proposed PnP-ADMM algorithm for Poisson image deconvolution, to solve the step in \eqref{eq:admmPnP2}.

\begin{figure*}[h!]
\centering
\includegraphics[width=1\textwidth]{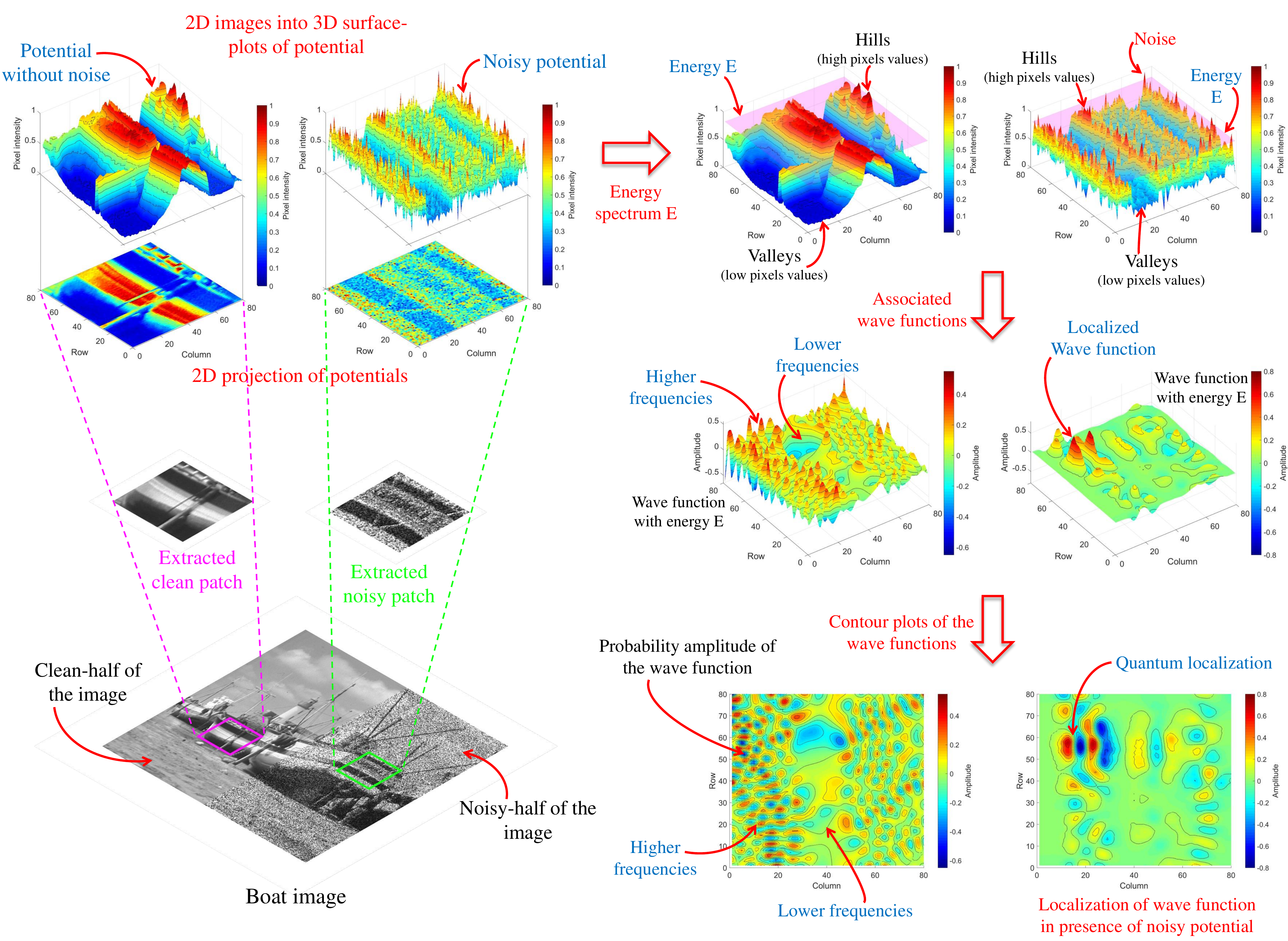}

\caption{Relationship between the clean and noisy images under the quantum mechanical framework and their effects on the wave functions: example on Boat image.}
\label{fig:relation}
\end{figure*}

\subsection{Quantum adaptative basis (QAB) denoiser}
\label{sec:QAB}

In the last decade, several works have been conducted to use quantum mechanical principles in signal \cite{eldar2002quantum} and image processing applications. More precisely, the interest in image segmentation \cite{gabbouj2013, youssry2015quantum, youssry2016quantum, youssry2019continuous}, restoration \cite{yuan2013quantum, LalegKirati15} and denoising \cite{LalegKirati16, smith2018adaptive, dutta2021quantum} have been studied in the literature.

The denoiser embedded in the proposed method is based on the construction of an adaptive basis inspired from quantum mechanics, as originally proposed in \cite{dutta2021quantum, smith2018adaptive}.
An illustration of the adaptive basis construction is given in Fig.~\ref{fig:relation}. It  displays the relationship between a clean and a noisy image in the quantum mechanical framework. The basic idea is to use the image as a potential of a quantum system, where the height of the potential is determined by the pixel intensity. For illustration purpose, we considered the Boat image with half of it contaminated by Gaussian noise. Two patches, one clean and one noisy, are extracted from the image and plotted as 3D surfaces, which will ultimately act as the potentials of the system. In this system, the wave function governs the probability of presence of a quantum particle with energy $E$ at some position on the surface. For a clean image, the wave function uses a broad range of frequencies to probe the surface. In presence of random noise, the wave function collapses and becomes localized at some particular position on the surface, as highlighted in Fig.~\ref{fig:relation}. The salient feature of the adaptive basis is the fact that the pixel intensity is directly linked to the local frequency of the wave. The localization property in the presence of noise is actually a hindrance, cured by performing a pre-smoothing of the noisy potential in order to create an adaptive basis extended over the whole image. For more details on the construction of the basis, we refer the reader to \cite{dutta2021quantum}. For self-consistency, we recall hereafter the main steps of the QAB (quantum adaptive basis) technique.

\subsubsection{Background on the adaptive QAB transform}
\label{sec:adapttransf}

In the non-relativistic quantum mechanics, the time-independent Schroedinger equation yields an equation for the stationary wave solution $\bpsi(a)$, given by
\begin{equation}
 - \frac{\hbar ^2}{2m} \nabla ^2 \psi = - V(a)  \bpsi + E \bpsi,
\label{eq:schrodinger}
\end{equation}
where $\hbar$ is the Planck constant and $\bpsi(a)$ characterizes the energy state $E$ of the particle with mass $m$ in a potential $V$. The probability amplitude of the particle is given by $|\bpsi (a) |^2$, normalized under $\int |\bpsi(a)|^2 d a =1$. The wave function $\bpsi(a)$ is an element of the Hilbert space of square-integrable functions. It is possible to rewrite the equation \eqref{eq:schrodinger} as
\begin{equation}
H_{\mathcal{QAB}} \bpsi =  E \bpsi,
\label{eq:hamiltonian}
\end{equation}
where $H_{\mathcal{QAB}} = - \frac{\hbar ^2}{2m} \nabla ^2 +V$ is the Hamiltonian operator. One can conclude from \eqref{eq:hamiltonian} that the solution $\bpsi(a)$ of the equation \eqref{eq:schrodinger} represents an eigen-state of the system described by the Hamiltonian operator. These eigen-states of \eqref{eq:hamiltonian} are oscillatory functions and primarily have two properties: i) the oscillation frequency increases with energy and ii) for the same eigen-function, the local frequency depends on the local value of the potential, and this dependence is regulated by the value of $\hbar ^2/2m$ which acts as a hyperparameter herein.

In the perspective of designing an adaptive transformation for image processing, one may consider the image pixels' values as the potential $V$ in the Schroedinger equation \eqref{eq:schrodinger} for a discretized space. The stationary solutions of \eqref{eq:schrodinger} can be obtained by computing the eigen-pairs of the discretized Hamiltonian operator defined as:
\begin{eqnarray}
H_{\mathcal{QAB}}[i,j]= \left \{
   \begin{array}{c c l}
      \bsx[i]+ 4 \dfrac{\hbar ^2}{2m} & & \mbox{for} \; i=j,\\
       -\dfrac{\hbar ^2}{2m} & & \mbox{for} \; i = j \pm 1,\\
        -\dfrac{\hbar ^2}{2m} & & \mbox{for} \; i = j \pm n,\\
      0 & & \mbox{otherwise},
   \end{array}
   \right.
\label{eq:H}
\end{eqnarray}
where $\bsx \in \mathbb{R}^{n^2}$ is an image (\textit{i.e.,} $V = \bsx$), vectorized in lexicographical order and $H_{\mathcal{QAB}}[i,j]$ represents the $(i,j)$-th element of the operator $H_{\mathcal{QAB}} \in \mathbb{R}^{ n^2 \times n^2 }$. Note that zero padding is used to handle the boundary conditions. As a consequence some violations of the rule \eqref{eq:H} can be observed. More precisely, $H_{\mathcal{QAB}}[i,j] = \bsx[i]+ 2 \frac{\hbar ^2}{2m}$ for $i=j$ and $i \in \{1,n,n^2-n+1,n^2\}$, $H_{\mathcal{QAB}}[i,j] = \bsx[i]+ 3 \frac{\hbar ^2}{2m}$ for $i=j$ and $i \in \{ 2,3,...,n-1,n^2-n+2,n^2-n+3,...,n^2-1\}$, $H_{\mathcal{QAB}}[i,j] = \bsx[i]+ 3 \frac{\hbar ^2}{2m}$ for $i=j$ and $i ~\mod~ n\in \{0,1\}$, except for $i \in \{ 1,2,...,n,n^2-n+1,n^2-n+2,...,n^2\}$ in order to respect the boundary conditions, and $H_{\mathcal{QAB}}[i,i+1]=H_{\mathcal{QAB}}[i+1,i]=0$ for any $i$ multiple of $n$ apart from $n^2$. More details about the construction of the Hamiltonian operator associated to an image can be found in \cite{dutta2021quantum}.

The corresponding eigenbasis of the Hamiltonian operator \eqref{eq:H} represents the adaptive transform. In the seminal works \cite{dutta2021quantum, smith2018adaptive}, it was shown that this adaptive basis gives an efficient way of image denoising, especially in the presence of Gaussian, Poisson or speckle noise.
In this work, this adaptive basis, referred to as quantum adaptive basis (QAB), is used to construct the denoiser $\mathcal{D_{QAB}}(\cdot)$ embedded in the proposed PnP-ADMM scheme.

These basis vectors belong to the family of oscillating functions along with the Fourier and wavelet bases, but with a local frequency depending on the local value of $\sqrt{2m(E-V)}/\hbar$. Due to its dependence on the difference between the energy $E$ and potential $V$, in the same basis vector the lower values of the potential are associated with oscillations of higher frequency. Thus, the property of these adaptive basis vectors able to describe different image pixels' values using different frequency levels, makes it fundamentally distinct from the Fourier and wavelet bases. From the above discussion it is understandable that the local frequency depends on the value of $\hbar ^2/2m$, which is a hyperparameter. Apart from that, the level of noise also has an impact on the basis vectors. Indeed, the presence of random noise in the system leads to a subtle quantum phenomenon \cite{anderson1958absence} which makes these vectors localize exponentially at different positions of the potential in the system. To mitigate this phenomenon which degrades the denoising, it is important to low-pass the corrupted image using, for example, a Gaussian filter with suitable standard deviation $\sigma_{\mathcal{QAB}}$, before the computation of the QAB from the Hamiltonian operator \eqref{eq:H}. The reader may refer to \cite{dutta2021quantum} for an in-depth discussion about the QAB vector localization in the presence of noise.

The QAB explained above is used to denoise an image, as suggested in \cite{dutta2021quantum}, as follows: project the noisy image onto the QAB to identify the valuable information and the noise, followed by a soft-thresholding of the projection coefficients, before taking the inverse projection of the modified coefficients to recover the noise-free image. The denoised image is retrieved as following:
\begin{eqnarray}
\hat{\bsx}= \sum _{i = 1} ^{n^2} \tau_{i}\balpha_{i}\bpsi_{i},
\label{eq:recons}
\end{eqnarray}
with 
\begin{eqnarray}
\tau_{i}= \left \{
   \begin{array}{c c l}
      1 &  & \mbox{for} \; i \leq s ,\\
      1 - \dfrac{i-s}{\rho} & & \mbox{for} \; i > s \;  \mbox{and for}\; 1 - \dfrac{i - s}{\rho} > 0 ,\\
      0 & & \mbox{otherwise},
   \end{array}
   \right .
\label{eq:thr}
\end{eqnarray}
where $\balpha_{i}$ are the coefficients representing the image $\bsx$ in QAB, whose basis vectors are $\bpsi_{i}$. $s$ and $\rho$ are two thresholding hyperparameters. The denoising process thus corresponds to expanding the signal in the adaptative basis and thresholding the coefficients according to an energy criterion (see \cite{dutta2021quantum} for a detailed discussion of this procedure).

\subsection{QAB-PnP algorithm}
\label{sec:algo}

This section illustrates, in the context of Poisson image deconvolution, the proposed PnP-ADMM algorithm, denoted as QAB-PnP, incorporating the QAB denoiser introduced in the previous section. In this particular context, various state-of-the-art denoisers have been introduced in the literature, such as Gaussian denoisers (\textit{e.g.,} BM3D \cite{dabov2007image}, etc) fused with VST-like transforms or not. Using QAB  $\mathcal{D_{QAB}}$ instead of a classical denoiser is the main contribution of this work. It consists in including a modified version of the QAB denoiser into the deconvolution PnP-ADMM method from Section~\ref{sec:poissdecon}, more precisely to solve \eqref{eq:admmPnP2}.

The denoising process integrated in the proposed QAB-PnP algorithm requires the computation of the coefficients $\balpha_i$, obtained by projecting the noisy image onto the QAB. This is a time consuming task for a large image and affects the computational load of the deconvolution algorithm given that the denoising process is performed at each iteration. However, one may note that most of the $\balpha_i$ are not used for reconstructing the denoised image given that they are discarded by the threshohlding operation. To increase the computational efficiency of the proposed algorithm, only the coefficients which contribute the most in the restoration process are computed. To this end, let us focus on $\mathcal{T}$ basis vectors $\balpha_i$ from $\mathcal{D_{QAB}}$, corresponding to an energy level below $\mathcal{E}$, assuming that higher energy levels naturally correspond to higher frequencies, where $\mathcal{E}$ is considered as a free hyperparameter. The corresponding $\mathcal{T}$ coefficients will be the most significant for the reconstruction of the clean image, and can be computed using the orthogonal matching pursuit (OMP) algorithm \cite{sahoo2015signal, davenport2010analysis, tropp2007signal, needell2009cosamp}.

    
\begin{algorithm}[h!]
\label{Algo:OMP}
\BlankLine
\KwIn{ $\bsv$ , $\mathcal{T}$ , $\mathcal{D_{QAB}}$}
\BlankLine\BlankLine

{\textbf{Initialization:}  $\bsr^0 = \bsv$ , $\Lambda^0 =  \emptyset$ , $\Phi^0$ is an empty matrix}\\


 \For{ $l$ from $0$ to $\mathcal{T} - 1$}{
 
 {$l = l + 1 $}\\
 
 {$\lambda^l = \underset{j = 1,2,...,\mathcal{T}} {\textrm{arg max}}~~ | \langle \bsr^{l-1},\bpsi_j \rangle |$, for $ \bpsi_j \in \mathcal{D_{QAB}}$ ~~~~(Break ties deterministically)}\\
 
 {$\Lambda^l = \Lambda^{l-1} \bigcup {\lambda^l}$ }\\
 
 {$\Phi^l = [\Phi^{l-1} ~~~~ \psi_{\lambda^l}]$}\\
 
 {$ a^l = \underset{a} {\textrm{arg min}}~~ \left\| \bsv - \Phi^l a \right\|^2_2$}\\

 {$ \bsr^l = \bsv - \Phi^l a^l$}\\
 
 }
\BlankLine\BlankLine
\KwOut{$\hat{\balpha}$, which has nonzero elements only at $\Lambda^l$, \textit{i.e.,} $\hat{\balpha}_{\Lambda^l} = a^l$}

\caption{Modified Orthogonal Matching Pursuit algorithm.}

\DecMargin{1em}
\end{algorithm}

\begin{algorithm}[h!]
\label{Algo:QAB}
\BlankLine
\KwIn{ $\bsz$ , $\mathcal{D_{QAB}}$, $\mathcal{T}$, $s$ , $\rho$}
\BlankLine\BlankLine

 {Compute the sparse coefficients $\hat{\balpha}_i$ with sparsity $\mathcal{T}$ by using the measurement data $\bsz$ and the operator $\mathcal{D_{QAB}}$ following the modified orthogonal matching pursuit method as illustrated in Algorithm \ref{Algo:OMP}.}\\
 
 {Threshold the coefficients  $\hat{\balpha}_{i}$.}\\
 
 {Compute $\hat{\bsz}$ following \eqref{eq:thr} and \eqref{eq:recons}.}\\
 
\BlankLine\BlankLine
\KwOut{$\hat{\bsz}$}

\caption{QAB denoising algorithm.}

\DecMargin{1em}
\end{algorithm}

\begin{algorithm}[h!]
\label{Algo:QABPnPADMM}
\BlankLine
\KwIn{ $\bsy$ , $\mathcal{E}$ , $\lambda_0$ , $\gamma$, $\dfrac{\hbar ^2}{2m}$ , $\sigma_{\mathcal{QAB}}$ , $N$}
\BlankLine\BlankLine

{\textbf{Initialization:}  $\bsx^0$ , $\bsz^0$ , $\bsu^0$}\\

{Compute a smooth version of $\bsy$ by low-pass Gaussian filter with standard deviation $\sigma_{\mathcal{QAB}}$}\\
 
{Form the Hamiltonian matrix $H_{\mathcal{QAB}}$ based on the smoothed version of $\bsy$ using \eqref{eq:H}}\\
 
 {Calculate the eigen-pairs of $H_{\mathcal{QAB}}$}\\
 
 {Construct $\mathcal{D_{QAB}}$ using the eigenvectors $\bpsi_{i}$ of $H_{\mathcal{QAB}}$}\\
 
 {Find the total number of eigenvalue $\mathcal{T}$, less than the energy level $\mathcal{E}$ }\\

 \Begin{\textbf{ADMM process:}

 \For{ $k$ from $0$ to $N-1$}{
 
 {\textit{Step 1:}}\\
 {$ \bsx^{k+1} = \underset{\bsx} {\textrm{arg min}}~~ - \bsy^T log(\bsH \bsx) + {\boldsymbol{1}}^T \bsH \bsx + \dfrac{\lambda^k}{2} \left\| \bsx - \bsz^k + \bsu^k \right\|_2^2 $ }\\
 
 {\textit{Step 2:}}\\
 {$ \bsz^{k+1} = \mathcal{D_{QAB}}( \bsx^{k+1} + \bsu^{k} )$, following QAB denoising Algorithm \ref{Algo:QAB}}\\
 
 {\textit{Step 3:}}\\
 {$ \bsu^{k+1} = \bsu^{k} + \bsx^{k+1} - \bsz^{k+1}$}\\

 {$ \lambda^{k+1} = \gamma \lambda^{k}$}\\
 
 }
\BlankLine\BlankLine
\KwOut{$\hat{\bsx} = \bsx^N$}
}

\caption{Poisson deconvolution using QAB-PnP algorithm.}

\DecMargin{1em}
\end{algorithm}

The OMP algorithm was fundamentally designed to obtain a sparse approximation $\hat{\balpha}_i$ with sparsity $\mathcal{T}$ of the corresponding coefficients $\balpha_i$ while projecting the noisy image, say $\bsv \in \mathbb{R}^{n^2}$ onto the denoising basis $\mathcal{D_{QAB}}$. Therefore the primary goal of OMP is to recover coefficients $\hat{\balpha}_i$ with $\mathcal{T}$ non-zero elements, such that $\bsv \simeq \mathcal{D_{QAB}} \hat{\balpha}_i$. To get the best possible approximation, it is important to identify the columns $ \bpsi_i \in \mathcal{D_{QAB}}$ which contribute in the reconstruction of $\bsv$. The basic idea is to choose the column of $\mathcal{D_{QAB}}$ which is mostly correlated with $\bsv$, followed by subtracting its contribution and repeat the step on the residual. After $\mathcal{T}$ iterations one can have the desired set of basis vectors and projection coefficients. Within the adaptive basis $\mathcal{D_{QAB}}$, the basis eigenvectors are organized in ascending order, the first $\mathcal{T}$ basis vectors with energy less than $\mathcal{E}$ being the most correlated with $\bsv$. Therefore, the OMP algorithm is modified herein so that it estimates only the projection coefficients onto the subspace formed by these $\mathcal{T}$ basis vectors. This modified OMP algorithm is detailed in Algorithm~\ref{Algo:OMP}.

The sparse coefficients $\hat{\balpha}_i$ estimated by Algorithm~\ref{Algo:OMP} are further used by the denoising method detailed in Algorithm~\ref{Algo:QAB}, integrated in the proposed QAB-PnP deconvolution method in Fig.~\ref{fig:flowc} and Algorithm ~\ref{Algo:QABPnPADMM} \footnote{The Matlab code of the proposed Plug-and-Play-ADMM algorithm using the quantum-adaptive-basis denoiser is [Online]. Available: \href{https://github.com/SayantanDutta95/QAB-PnP-ADMM-Deconvolution.git}{https://github.com/SayantanDutta95/QAB-PnP-ADMM-Deconvolution.git}}.

\begin{figure}[h!]
\centering
\subfigure[Lena]{\includegraphics[width=0.15\textwidth]{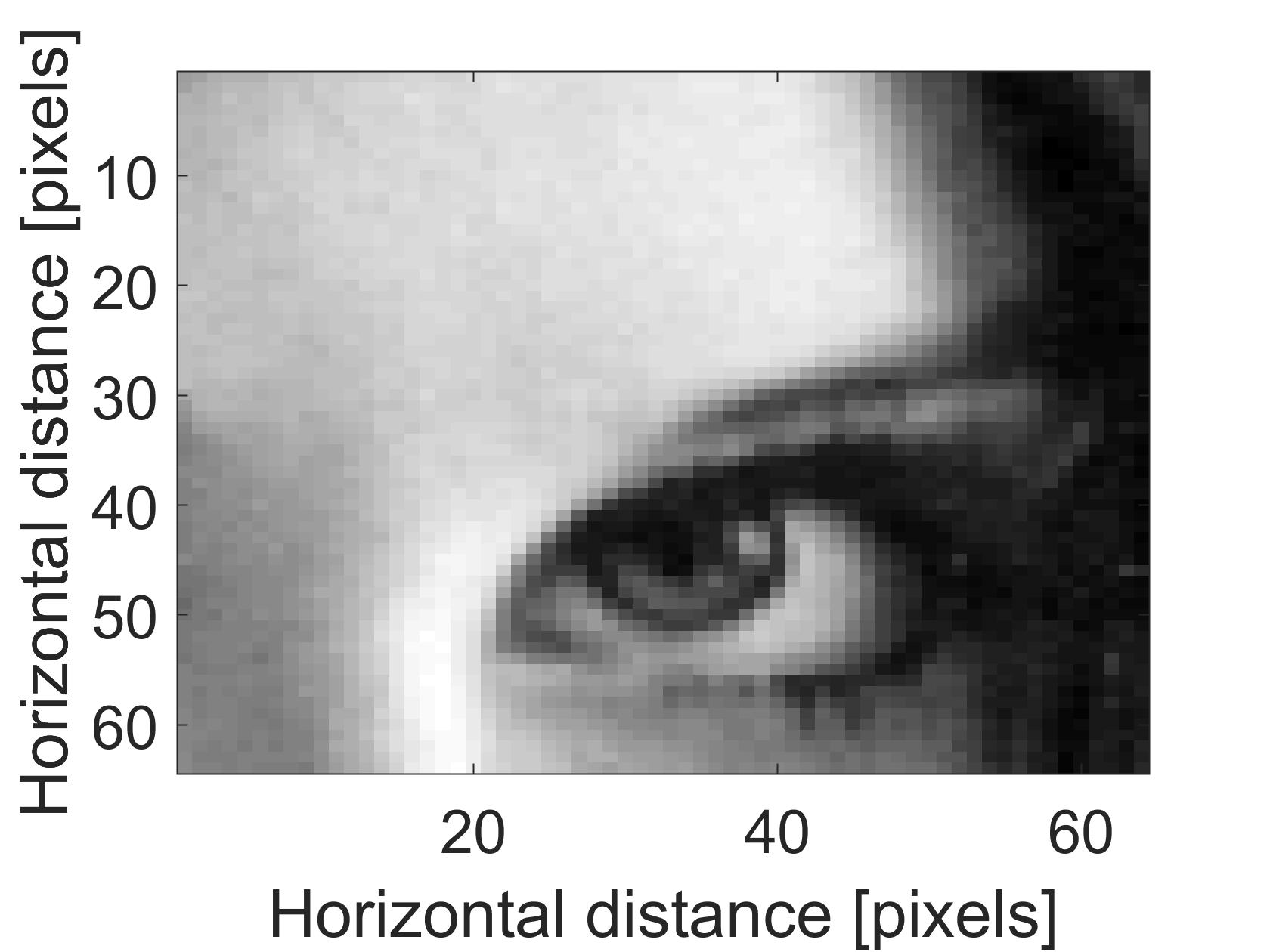}}
\subfigure[Fruits]{\includegraphics[width=0.15\textwidth]{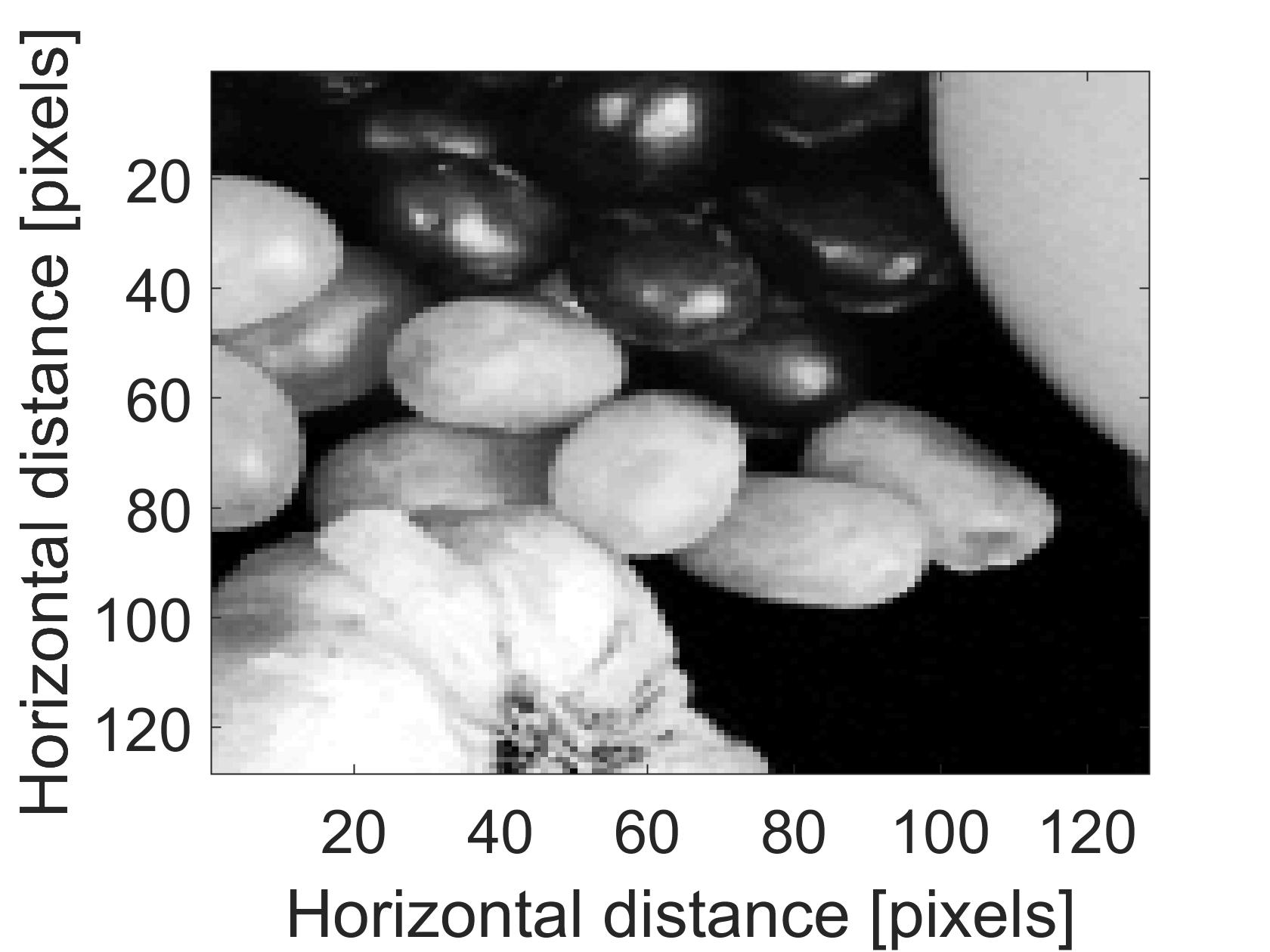}}
\subfigure[Synthetic]{\includegraphics[width=0.15\textwidth]{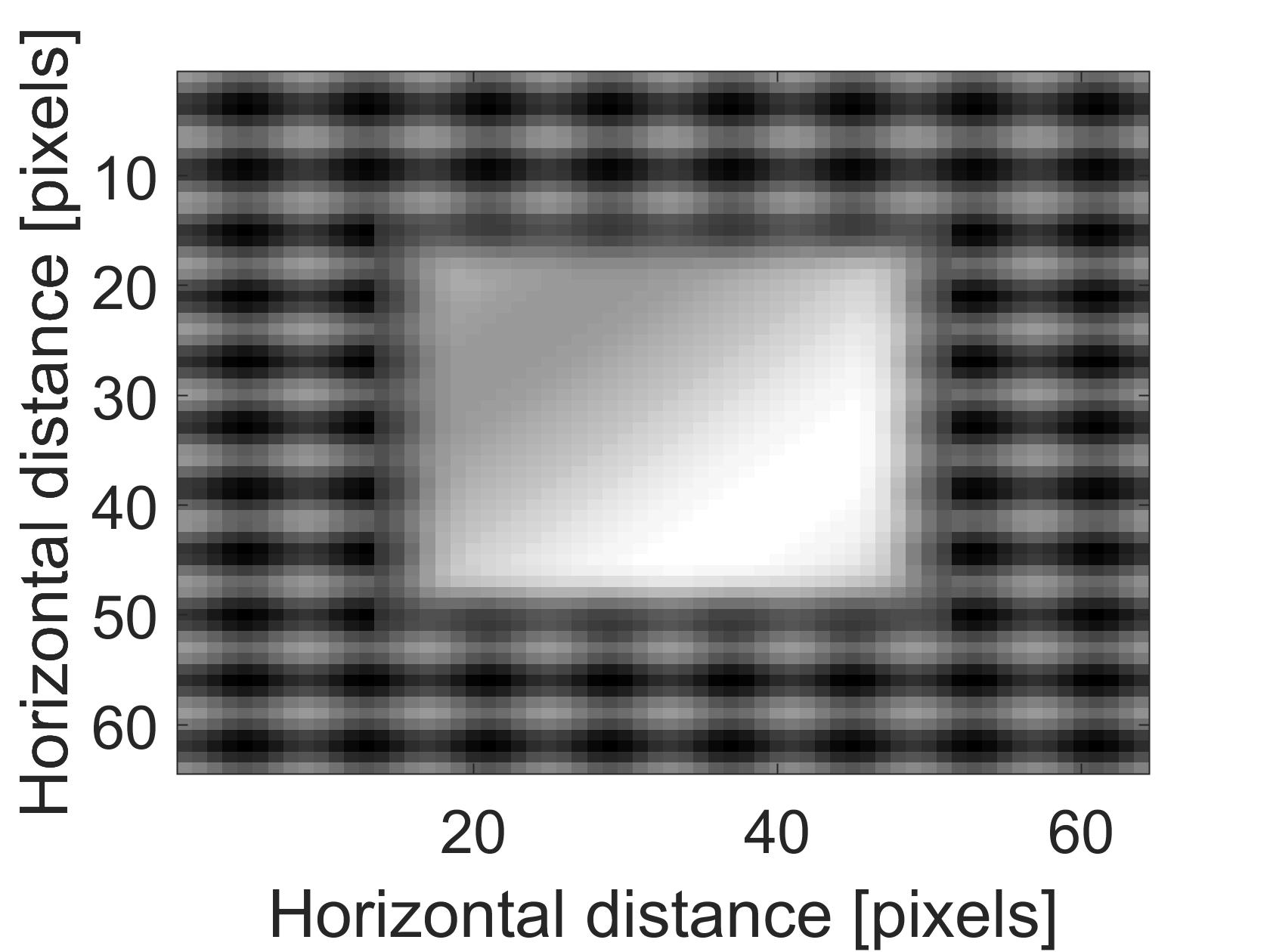}}
\caption{Images used for deconvolution simulations.}
\label{fig:sample}
\end{figure}

\begin{figure*}[h!]
\centering
\subfigure[Performed on the image in Fig.~\ref{fig:sample}(a).]
{\includegraphics[width=0.33\textwidth]{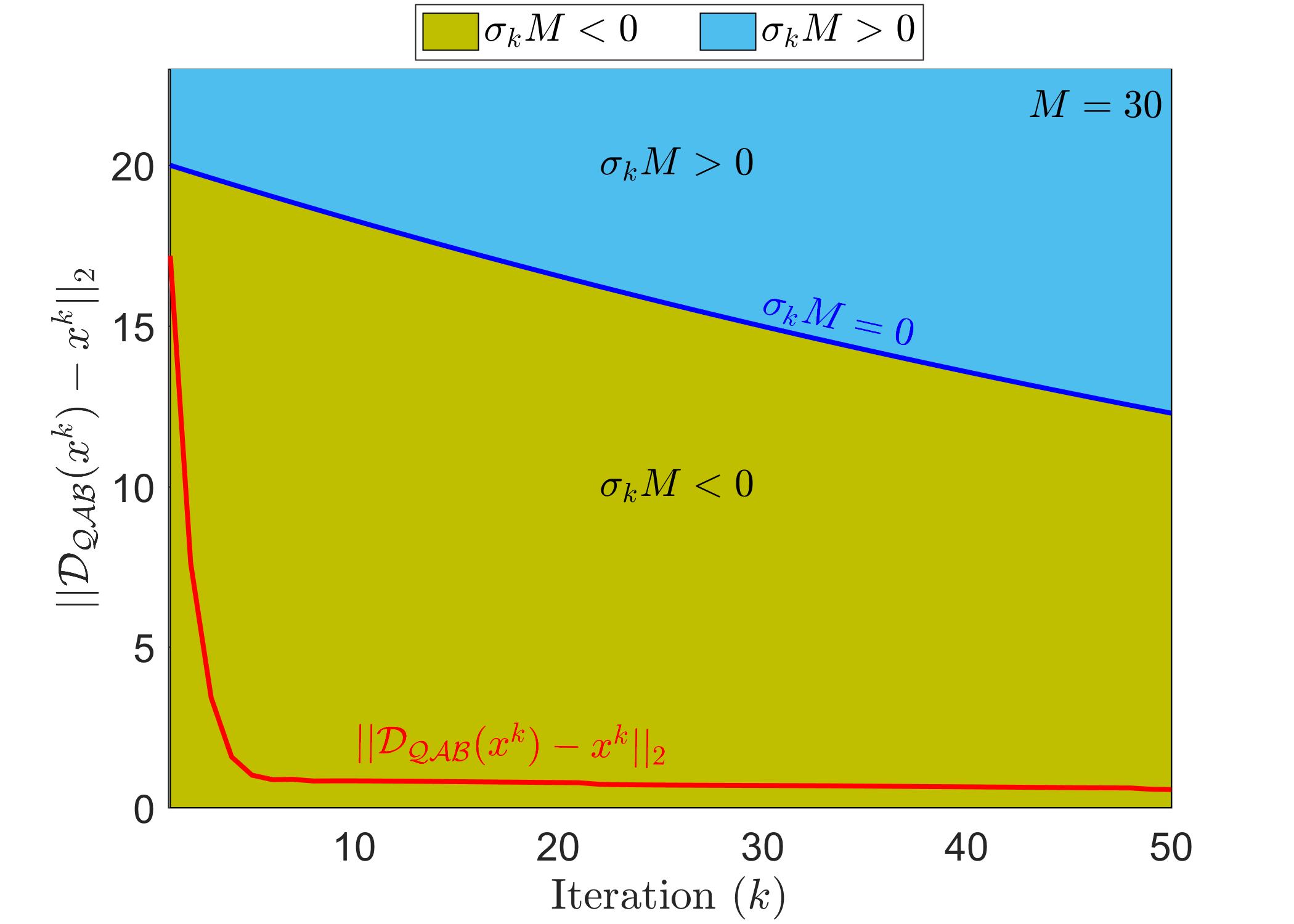}} \hfill
\subfigure[Performed on the image in Fig.~\ref{fig:sample}(b).]
{\includegraphics[width=0.33\textwidth]{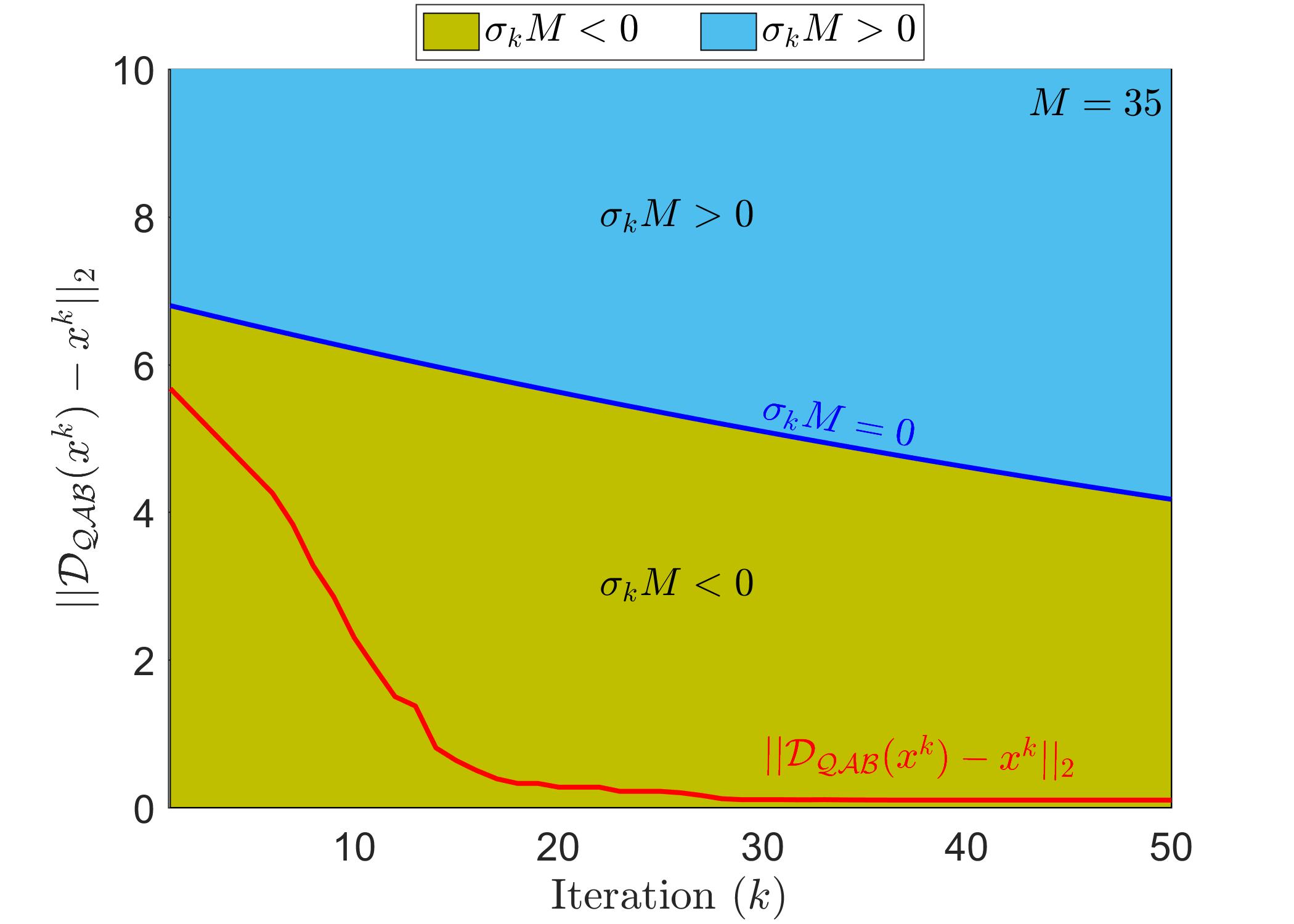}} \hfill
\subfigure[Performed on the image in Fig.~\ref{fig:sample}(c).]
{\includegraphics[width=0.33\textwidth]{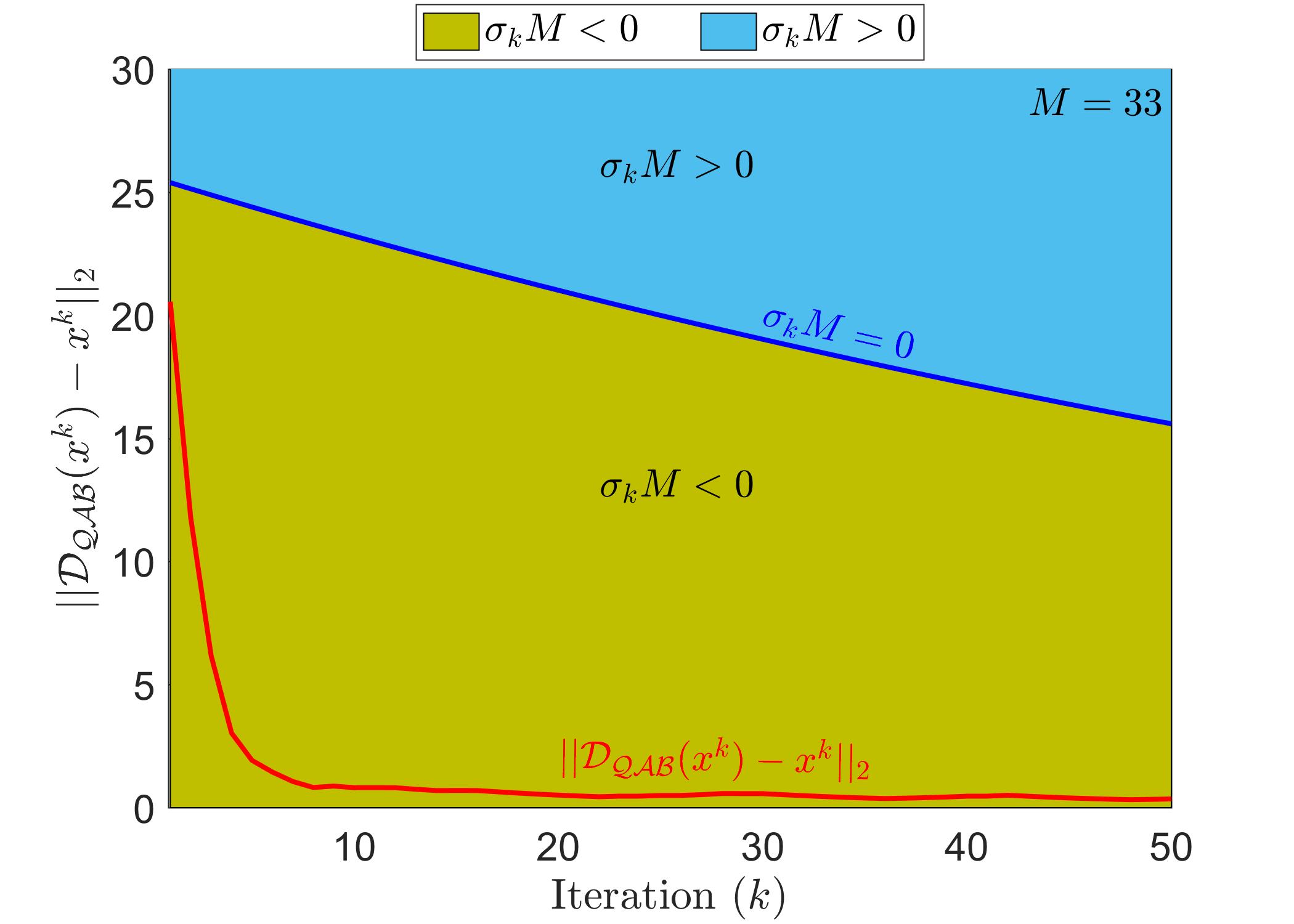}}

\caption{Numerical validation of the criteria, $ \left\| \mathcal{D_{QAB}}(\bsx^k) - \bsx^k \right\|_2 \leq \sigma_k M$ for any $ \bsx^k \in~\mathbb{R}^{n^2}$, performed on the sample images in Fig.~\ref{fig:sample}(a-c).}
\label{fig:boundenoi}
\end{figure*}


The computational complexity of the algorithm is dominated by the eigendecomposition of the high dimensional Hamiltonian matrix and the QAB image projection. For a $n \times n$ image, the Hamiltonian matrix is of size $n^2 \times n^2$. Usual textbook diagonalization methods would require $O (n^6)$ operations (time complexity) and $O (n^4) $ storage space. However, the Hamiltonian matrix is extremely sparse, and is more efficiently diagonalized by iterative methods such as the Lanczos method (as we actually did). In this case the computational complexity would be $O (n^4)$ if we compute all eigenvalues and eigenvectors (and still $O (n^4) $ in storage space). If we compute only $\mathcal{T}$ of these eigenvalues and eigenvectors (with $\mathcal{T} \leq n^2$), the time complexity becomes $O (\mathcal{T}n^2) $ and the storage space (space complexity) also $O (\mathcal{T}n^2) $. The QAB image projection is $O (n^4) $ with the simplest algorithm, and becomes $O (\mathcal{T}n^2) $ in time and space with the OMP algorithm. We thus conclude that our algorithm requires $O (\mathcal{T}n^2) $ time and space resources, with $\mathcal{T} \leq n^2$, for a $n \times n$ image. To  further decrease the complexity, a block-wise approach could be used as proposed in \cite{dutta2021quantum}, where a large image is divided into smaller patches denoised independently by the QAB denoiser. In this the complexity is $O (\mathcal{T}Pm^2) $ for $P$ patches of size $m~ (<<n)$. Moreover, such a patch-based architecture can be improved by considering the dependence between neighboring patches by borrowing tools from the quantum interaction theory as suggested in \cite{dutta2021image}.

\subsection{Convergence analysis of QAB-PnP algorithm}
\label{sec:convergence}

Despite their popularity during the last decade, the proof of convergence of PnP-ADMM algorithms may still be an issue. Some interesting developments have been proposed during the last few years on global \cite{sreehari2016plug} and fixed point \cite{chan2017plug, chan2019performance, ryu2019plug, xu2020provable, zhang2020plug, cohen2020regularization, teodoro2017scene} convergence of these algorithms, while imposing restrictions on the denoising operator. In this section, our goal is to analyse the fixed point convergence of the proposed QAB-PnP algorithm.

To enable the fixed point convergence and in particular to avoid the issue of unbounded gradient in \eqref{eq:admmPnP1Grad} for pixel values equal to $0$, i.e., to overcome the singularity problem at $\bsx=0$, we slightly modify the observation model \eqref{eq:noisemodel} by introducing a small positive constant $\epsilon \ll 1$, as suggested in \cite{harmany2012this}:
\begin{equation}
\bsy = \mathcal{P}( \bsH \bsx + \epsilon {\boldsymbol{1}}).
\label{eq:newnoisemodel}
\end{equation}
Therefore the negative Poisson log-likelihood \eqref{eq:datafidelity} becomes
\begin{equation}
f(\bsx) = - \bsy^T log(\bsH \bsx + \epsilon {\boldsymbol{1}}) + {\boldsymbol{1}}^T \bsH \bsx,
\label{eq:newdatafidelity}
\end{equation}
and the corresponding gradient 
\begin{equation}
\nabla f(\bsx) = - \bsH^T (\bsy/(\bsH \bsx + \epsilon {\boldsymbol{1}})) + \bsH^T {\boldsymbol{1}}.
\label{eq:grad}
\end{equation}
One should note that within practical experiments, $\epsilon$ is much smaller than any background value, so that its influence on the final output is negligible \cite{harmany2012this}.

\begin{rem}
\label{rem:funcboun}

For $f(\bsx) : [0,1]^{n^2}  \rightarrow   {\mathbb{R}^+}$, with nontrivial constant vector $\bsy \in \mathbb{R}^{n^2}$ and operator $\bsH \in \mathbb{R}^{n^2 \times n^2}$, the gradient $\nabla f(\bsx)$ is bounded.

\end{rem}

\begin{IEEEproof}

Since $\epsilon$ is the lower bound of $(\bsH \bsx + \epsilon {\boldsymbol{1}})$, therefore $1/\epsilon$ is the upper bound of $1/(\bsH \bsx + \epsilon {\boldsymbol{1}})$. Since $\bsy$ and $\bsH$ are constants, they are bounded. Hence one can write:

\begin{align}
\left\| \nabla f(\bsx)\right\|_2 & = \left\| - \bsH^T (\bsy/(\bsH \bsx + \epsilon {\boldsymbol{1}})) + \bsH^T {\boldsymbol{1}} \right\|_2 \nonumber \\
& \leq \left\| \bsH^T \right\|_2 \left\|\frac{\bsy}{\bsH \bsx + \epsilon {\boldsymbol{1}}} \right\|_2 + \left\| \bsH^T \right\|_2 \nonumber \\
& \leq \frac{\delta_1}{\epsilon} + \delta_2 \nonumber \\
& \leq L < \infty
\label{eq:rem2}
\end{align}
where $\delta_1, \delta_2, L \in {\mathbb{R}^+}$. 

\end{IEEEproof}

\begin{rem}
\label{rem:denoiboun}

Denoiser $\mathcal{D_{QAB}}$ is a bounded denoising operator with a parameter $\sigma_k$.
\end{rem}

We cannot offer a general proof of this statement, also it intuitively appears highly likely. The denoising process denoted by $\mathcal{D_{QAB}}$ certainly reduces the level of noise at each iteration and gets 
$\mathcal{D_{QAB}}(\bsx^k)$ closer and closer to $\bsx^k $. It is therefore fair to consider that 
 $ \left\| \mathcal{D_{QAB}}(\bsx^k) - \bsx^k \right\|_2 $ decreases with $k$. It is also bounded by 
 $ \left\| \bsx^k \right\|_2 $ since $\mathcal{D_{QAB}}$ is a projection operator.
 
The rate of decrease is not a priori easy to bound, but we offer numerical evidence that the decrease is fast. Indeed, in all three examples shown in Fig.~\ref{fig:boundenoi} the decrease is very fast. In particular, it is much faster  that the rate of decrease of $\sigma_k \overset{\textrm{def}}{=} 1/ \lambda^k$. We thus generalize this result and take as generic that $ \left\| \mathcal{D_{QAB}}(\bsx^k) - \bsx^k \right\|_2 \leq \sigma_k M$ where $M$ is a system-dependent constant.

\begin{rem}[Fixed Point Convergence of QAB-PnP algorithm]
If
\begin{enumerate}
\item $f(\bsx) : [0,1]^{n^2}  \rightarrow   {\mathbb{R}^+}$ is analytic and has bounded gradient, \textit{i.e.}, for all $ \bsx \in [0,1]^{n^2} $, there exists $ L < \infty $ such that $ \left\| \nabla f(\bsx) \right\|_2 \leq L$, and
\item $\mathcal{D_{QAB}}$ is a bounded denoising operator with a parameter $\sigma_k$,
\end{enumerate}
then QAB-PnP converges to a fixed point. That is, there exists $( \bsx^*, \bsz^*, \bsu^* )$ such that $ \left\| \bsx^k - \bsx^* \right\|_2  \rightarrow {\boldsymbol{0}} $, $ \left\| \bsz^k - \bsz^* \right\|_2  \rightarrow {\boldsymbol{0}}$, $ \left\| \bsu^k - \bsu^* \right\|_2  \rightarrow {\boldsymbol{0}} $ as $ k  \rightarrow \infty $.
\label{rem:theorem}
\end{rem}

\begin{IEEEproof}

\textit{$\ast$ First condition:}
The first condition holds as shown in Remark \ref{rem:funcboun}.

\textit{$\ast$ Second condition:}
The second condition should hold generically as discussed in Remark~\ref{rem:denoiboun}.


Given that the two conditions are satisfied within the proposed framework, let us move to the proof of the fixed point convergence in Remark~\ref{rem:theorem}. We start by proving the following statements:
\begin{align}
& \left\| \bsz^{k+1} - \bsz^k \right\| \leq \frac{C_2}{\lambda^k}
\label{eq:con2} \\
& \left\| \bsx^{k+1} - \bsx^k \right\| \leq \frac{C_1}{\lambda^k}
\label{eq:con1} \\
& \left\| \bsu^{k+1} - \bsu^k \right\| \leq \frac{C_3}{\lambda^k}
\label{eq:con3}
\end{align}
where $C_1$, $C_2$ and $C_3$ are constants and $\lambda^k$ is the penalty parameter with $\lambda^{k+1} = \gamma \lambda^k$, where $\gamma > 1$.

\textit{$\ast$ First step:} Proof of condition \eqref{eq:con2}.

From \eqref{eq:admm1}, we have
\begin{equation}
\bsx^{k+1} = \underset{\bsx} {\textrm{arg min}}~~ f(\bsx) + \dfrac{\lambda^k}{2} \left\| \bsx - \bsz^k + \bsu^k  \right\|_2^2.
\label{eq:th1}
\end{equation}

The first order optimality implies
\begin{equation}
\bsx - (\bsz^k - \bsu^k) = - \dfrac{\nabla f(\bsx)}{\lambda^k}.
\label{eq:th2}
\end{equation}

Since the minimizer is obtained in $\bsx = \bsx^{k+1}$, replacing $\bsx$ by $\bsx^{k+1}$ and using the boundedness property of $\nabla f(\bsx)$, we have
\begin{equation}
\left\| \bsx^{k+1} - (\bsz^k - \bsu^k) \right\|_2 = \frac{\left\| \nabla f(\bsx^{k+1}) \right\|_2}{\lambda^k} \leq \frac{L}{\lambda^k}.
\label{eq:th3}
\end{equation}

Furthermore, since the denoiser $\mathcal{D_{QAB}}$ is bounded and $\bsz^{k+1} =  \mathcal{D_{QAB}} (\bsx^{k+1} + \bsu^k)$, one can write
\begin{align}
& \left\| \bsz^{k+1} - (\bsx^{k+1} + \bsu^k) \right\|_2 \nonumber \\
& = \left\| \mathcal{D_{QAB}} (\bsx^{k+1} + \bsu^k) - (\bsx^{k+1} + \bsu^k) \right\|_2 \nonumber \\
& \leq \sigma_k M = \frac{M}{\lambda^k}.
\label{eq:th4}
\end{align}

One also has
\begin{align}
\left\| \bsz^{k+1} - \bsz^k \right\|_2 & \leq \left\| \bsz^{k+1} - (\bsx^{k+1} + \bsu^k) \right\|_2 \nonumber \\
& ~~~ + \left\| (\bsx^{k+1} + \bsu^k) - \bsz^k \right\|_2.
\label{eq:th5}
\end{align}

Finally, using \eqref{eq:th3} and \eqref{eq:th4}, we obtain
\begin{equation}
\left\| \bsz^{k+1} - \bsz^k \right\|_2 \leq \frac{L}{\lambda^k} + \frac{M}{\lambda^k} = \frac{C_2}{\lambda^k}.
\label{eq:th6}
\end{equation}

\textit{$\ast$ Second step:} Proof of condition \eqref{eq:con3}.

From \eqref{eq:admm3}, we get
\begin{align}
\left\| \bsu^{k+1} \right\|_2 & = \left\| \bsu^k + \bsx^{k+1} - \bsz^{k+1} \right\|_2 \nonumber \\
& = \left\| (\bsx^{k+1} + \bsu^k) - \mathcal{D_{QAB}} (\bsx^{k+1} + \bsu^k) \right\|_2 \nonumber \\
& \leq \frac{M}{\lambda^k}.
\label{eq:th7}
\end{align}

Using \eqref{eq:th7}, we have
\begin{equation}
\left\| \bsu^{k+1} - \bsu^k \right\|_2  \leq  \left\| \bsu^{k+1} \right\| + \left\| \bsu^k \right\|_2 \leq  \frac{M}{\lambda^k} + \frac{M}{\lambda^k} = \frac{C_3}{\lambda^k}.
\label{eq:th8}
\end{equation}

\textit{$\ast$ Third step:} Proof of condition \eqref{eq:con1}.

\eqref{eq:admm3} can be written as
\begin{equation}
\bsx^{k+1} = \bsu^{k+1} - \bsu^k + \bsz^{k+1}.
\label{eq:th9}
\end{equation}

Using \eqref{eq:th9}, we have
\begin{align}
& \left\| \bsx^{k+1} - \bsx^k \right\|_2 \nonumber \\
& = \left\| (\bsu^{k+1} - \bsu^k + \bsz^{k+1}) - (\bsu^k - \bsu^{k-1} + \bsz^k ) \right\|_2 \nonumber \\
& \leq \left\| \bsu^{k+1} - \bsu^k \right\|_2 + \left\| \bsz^{k+1} - \bsz^k \right\|_2 + \left\| \bsu^k - \bsu^{k-1} \right\|_2 \nonumber \\
& \leq \frac{C_3}{\lambda^k} + \frac{C_2}{\lambda^k} + \frac{C_3}{\lambda^{k-1}} \leq \frac{C_3}{\lambda^k} + \frac{C_2}{\lambda^k} + \frac{\gamma C_3}{\lambda^k} = \frac{C_1}{\lambda^k}
\label{eq:th10}
\end{align}

Hence all three conditions \eqref{eq:con2}, \eqref{eq:con1} and \eqref{eq:con3} are true.


Next, we aim at proving that $\lbrace \bsx^k \rbrace_{k = 1}^\infty$ is a Cauchy sequence. Therefore, one has to show that for all integer $n > k$, $ \left\| \bsx^n - \bsx^k \right\|_2  \rightarrow {\boldsymbol{0}} $ as $ n \rightarrow \infty$ and $ k  \rightarrow \infty $.

For any finite $n$ and $k$, one can write using the condition \eqref{eq:con1}
\begin{equation}
\left\| \bsx^n - \bsx^k \right\|_2 \leq \sum_{l = k}^{n-1} \frac{C_1}{\lambda^l} = C_1 \sum_{l = k}^{n-1} \frac{1}{\lambda_0 \gamma^l} = \frac{C_1}{\lambda_0 \gamma^k} \sum_{l = 0}^{n-k-1} \frac{1}{\gamma^l}.
\label{eq:th11}
\end{equation}

Therefore, as $ n \rightarrow \infty$ and $ k  \rightarrow \infty $, $ \left\| \bsx^n - \bsx^k \right\|_2  \rightarrow {\boldsymbol{0}} $, since $\gamma > 1$, so $\lbrace \bsx^k \rbrace_{k = 1}^\infty$ is a Cauchy sequence. Hence, the sequence $\lbrace \bsx^k \rbrace_{k = 1}^\infty$ is convergent, thus there exits $\bsx^* \in [0,1]^{n^2}$ such that $ \left\| \bsx^k - \bsx^* \right\|_2  \rightarrow {\boldsymbol{0}} $ as $ k  \rightarrow \infty $.

Similarly, one can show that the sequence $\lbrace \bsz^k \rbrace_{k = 1}^\infty$ and $\lbrace \bsu^k \rbrace_{k = 1}^\infty$ are convergent, so there exit $\bsz^*, \bsu^* \in [0,1]^{n^2}$ such that $ \left\| \bsz^k - \bsz^* \right\|_2  \rightarrow {\boldsymbol{0}} $ and $ \left\| \bsu^k - \bsu^* \right\|_2  \rightarrow {\boldsymbol{0}} $ as $ k  \rightarrow \infty $.

Therefore we can conclude that the proposed QAB-PnP algorithm converges to a fixed point.

\end{IEEEproof}

The proof we propose is not a convergence proof in the mathematical sense, since it reposes on Remark \ref{rem:denoiboun} for which we only have plausibility arguments and numerical evidence. Nevertheless, the discussion above and the numerical results in Fig.~\ref{fig:boundenoi} for three very different images, indicate that with high confidence the algorithm should converge in practice for any image.

\section{Simulation results}
\label{sec:simu}

This section illustrates the efficiency of the proposed QAB-PnP algorithm for Poisson image deconvolution. An analysis of the influence of the hyperparameters on the deconvolution accuracy is first provided in Subsection~\ref{sec:parameters}, before comparing its performance to several state-of-the-art methods in Subsection~\ref{sec:application}. In \cite{dutta2021quantum} we already performed a detailed analysis of the hyperparameters $\sigma_{\mathcal{QAB}}$, $s$ and $\rho$ for the efficiency of the denoiser. We recall that these hyperparameters control respectively the smoothing of the potential to avoid localization effects in the expansion basis, and the cutoff in energy which leads to denoising. We therefore chose these hyperparameters to be optimal according to the study in \cite{dutta2021quantum}. However, the computational method used in the present work (OMP algorithm) introduces a new hyperparameter $\mathcal{E}$ which controls the accuracy and efficiency of the OMP process. The accuracy of OMP increases for increasing $\mathcal{E}$, but at the cost of higher computational time. A trade-off is thus necessary, and we will show that the optimal value of $\mathcal{E}$ is also influenced by the value of the hyperparameter $\hbar ^2/2m$, which fixes how the local frequencies of the basis vectors vary as a function of pixels' amplitudes.

The simulations are conducted on three images, shown in Fig.~\ref{fig:sample}. Two of them represent cropped versions of the standard Lena and fruits images. The third one was synthetically constructed so that it contains high frequencies for low gray levels and, vice versa, low frequencies for high intensity pixels. Its purpose is to illustrate the ability of the proposed deconvolution method, and in particular of the embedded quantum-based denoiser, to handle such images. All the sample images are distorted with two Gaussian blurring kernel $h_{\sigma}^{4\times 4}$ of size $4 \times 4$ and standard deviation $\sigma = 3$ and $\sigma = 5$ respectively. The study was conducted with three different Poisson noise levels corresponding to SNRs of 20, 15 and 10 dB. Note that the noise was image-dependent Poisson distributed and that the SNRs of the observations was computed a posteriori to emphasize the amount of noise. Finally, Subsection~\ref{sec:applifluo} shows the abbility of the the proposed method to enhance experimental fluorescence microscopy images.

\subsection{Hyperparameter analysis}
\label{sec:parameters}

This subsection presents a detailed analysis on the influence of the hyperparameters on the proposed method. In particular, the role of the hyperparameter $\mathcal{E}$ will be evaluated, given its important impact on the compromise between accuracy and computational time, and its relationship with the hyperparameter $\hbar ^2/2m$  will be assessed. It is important to mention that in general  the hyperparameter $\hbar ^2/2m$ and the number of significant wave vectors $\mathcal{T}$ vary in an opposite way, one of them increasing when the other one decreases. In addition,  there is a linear relation between $\mathcal{T}$ and the processing time. Therefore, to achieve an optimal behaviour of the algorithm, a good balance between the hyperparameters $\hbar ^2/2m$ and $\mathcal{E}$ needs to be achieved. We will also discuss the choice of the hyperparameter $\lambda_0$ which controls the iterations of the ADMM algorithm described in Section \ref{sec:background}.

\begin{figure}[b!]
\centering
\includegraphics[width=0.45\textwidth]{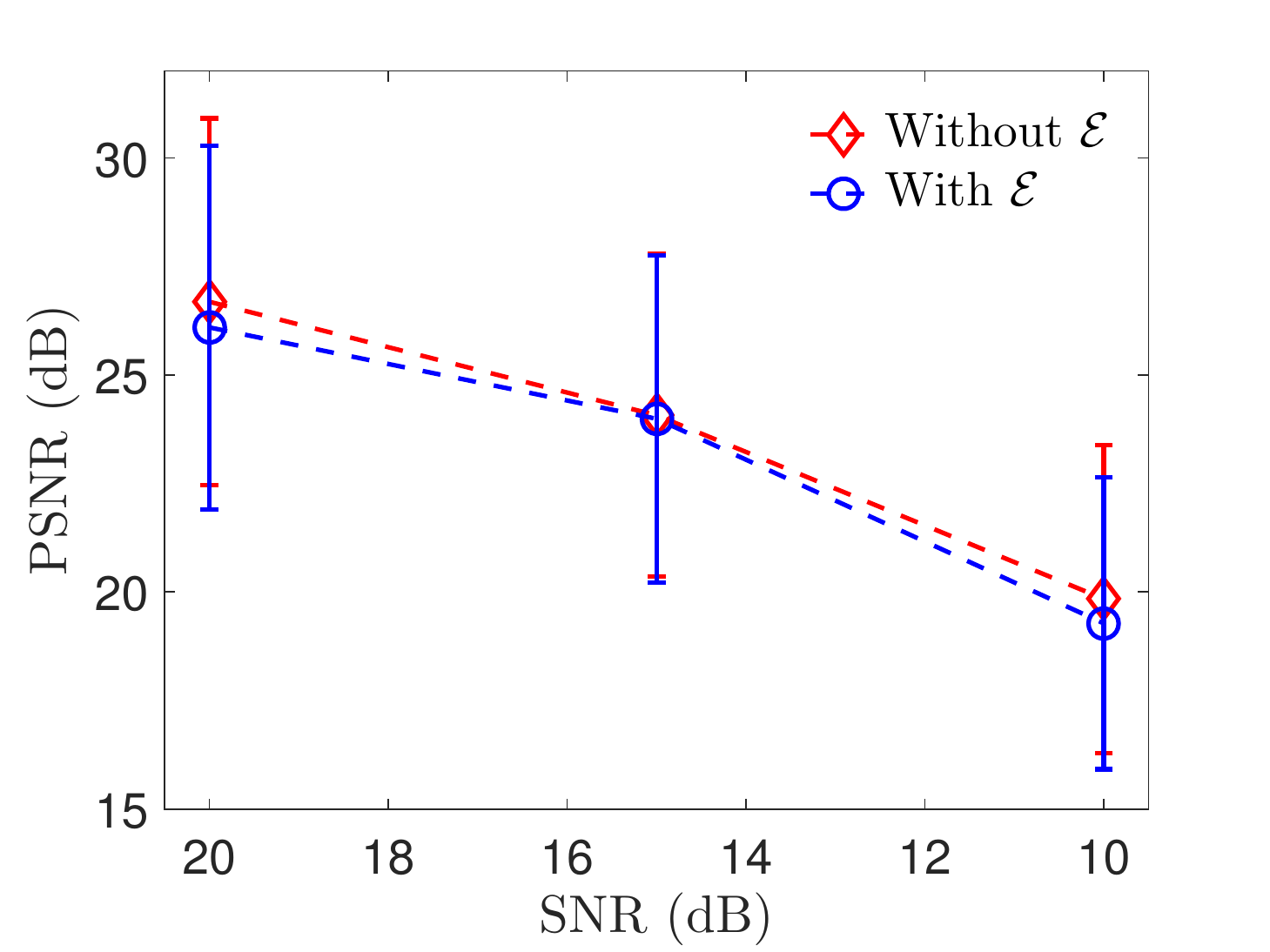}

\caption{PSNR mean and standard deviation values for all the three sample images in Fig. \ref{fig:sample} as a function of Poisson noise level.}
\label{fig:Ewow}
\end{figure}

\begin{figure*}[b!]
\centering
\subfigure[Influence of the hyperparameters $\mathcal{E}$ and $\hbar ^2/2m$ on the proposed method in terms of PSNR (dB)]{\includegraphics[width=0.47\textwidth]{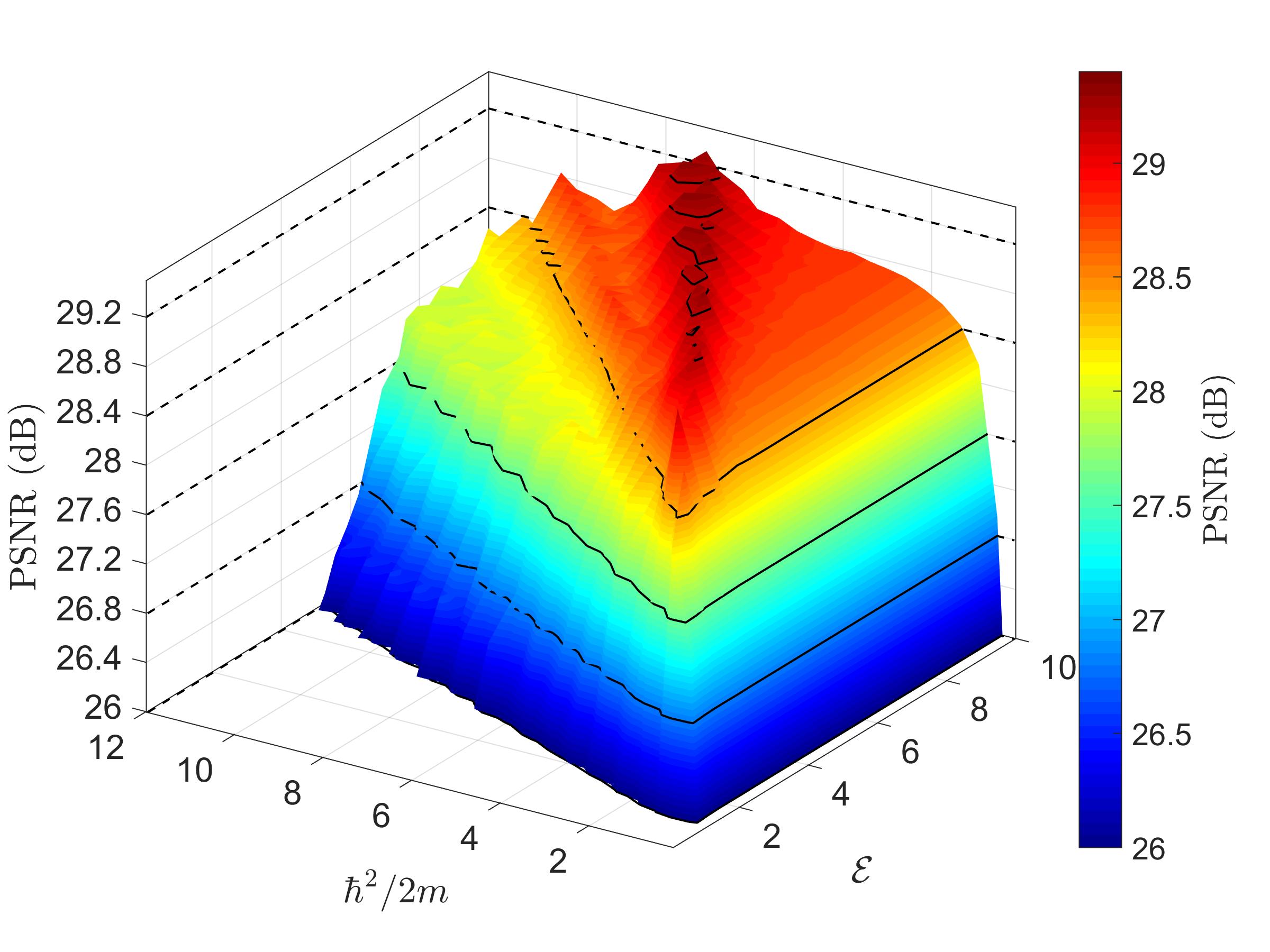}\hfill
\includegraphics[width=0.47\textwidth]{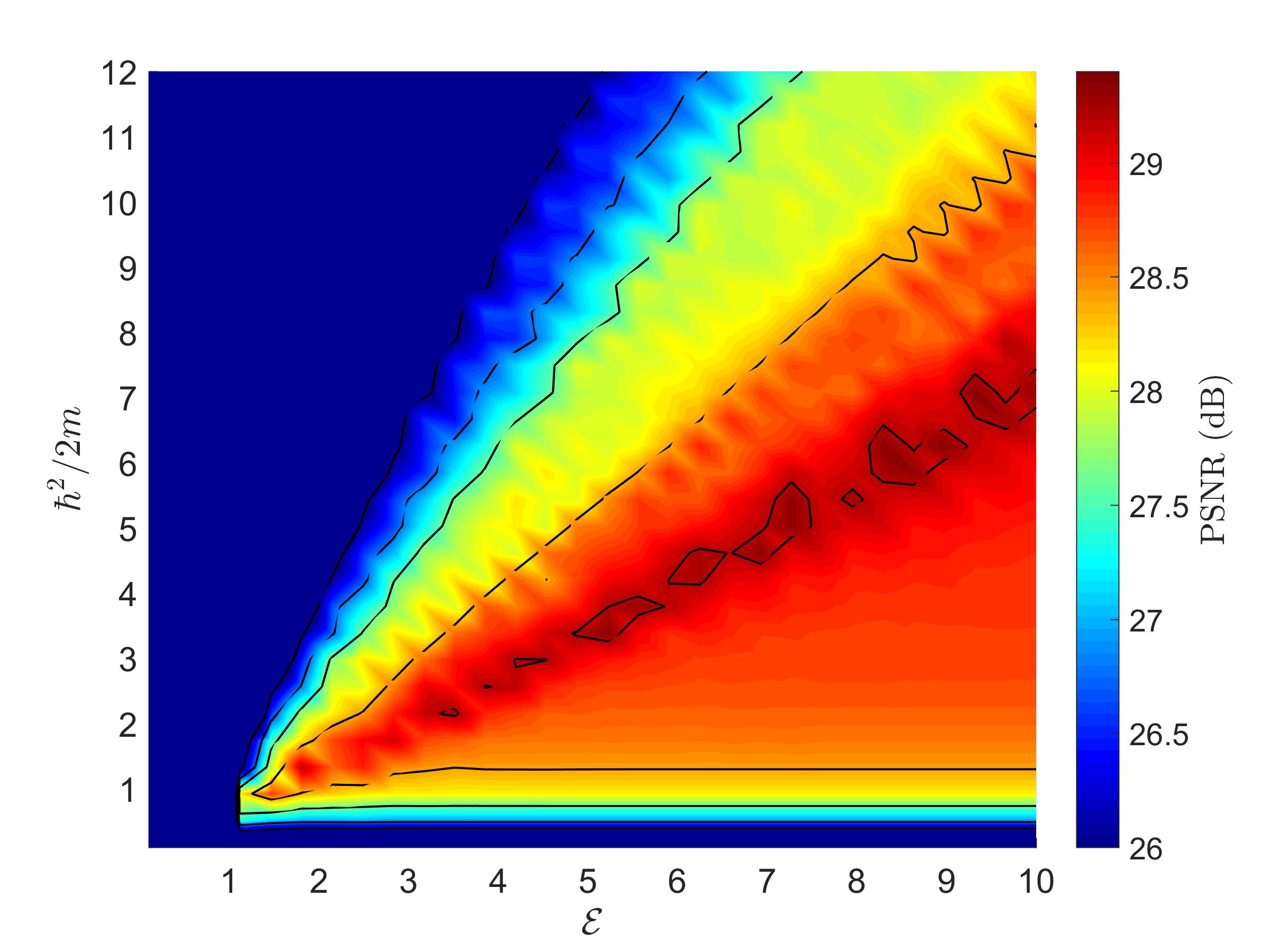}}

\subfigure[Number of significant wave vectors $\mathcal{T}$ for different values of the hyperparameters $\mathcal{E}$ and $\hbar ^2/2m$]{\includegraphics[width=0.47\textwidth]{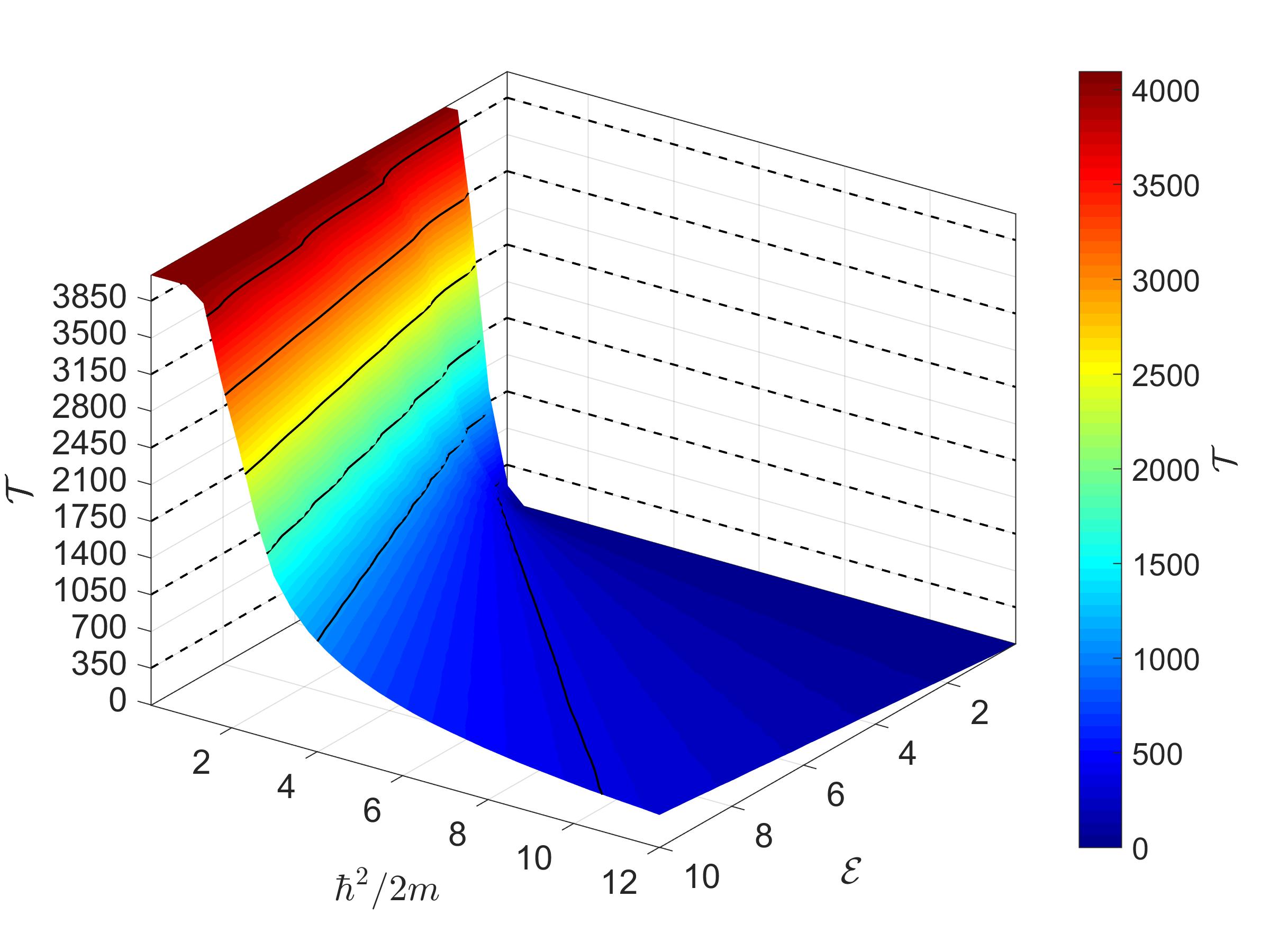}\hfill
\includegraphics[width=0.47\textwidth]{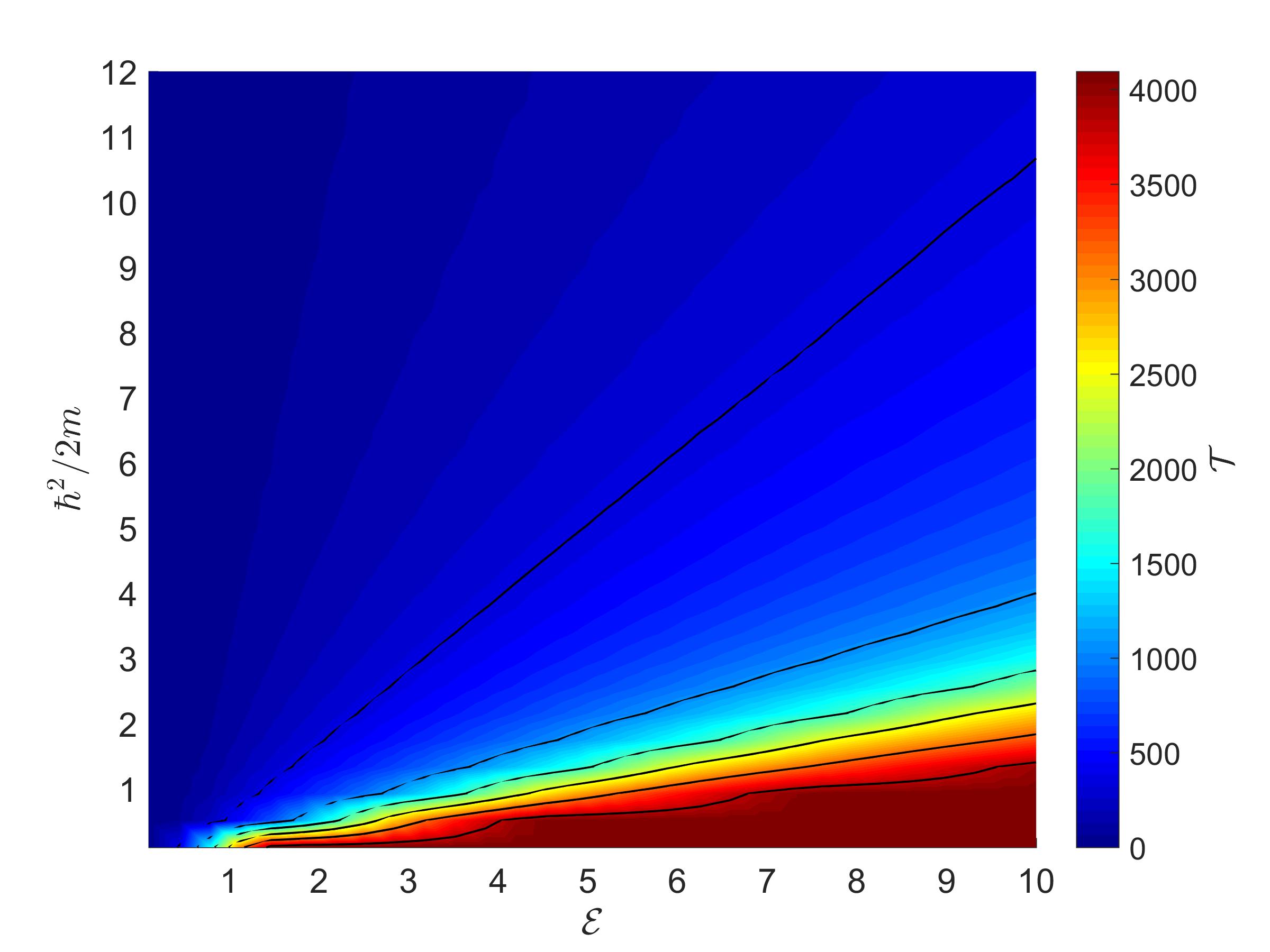}}

\subfigure[Computation time for different values of the hyperparameters $\mathcal{E}$ and $\hbar ^2/2m$]{\includegraphics[width=0.47\textwidth]{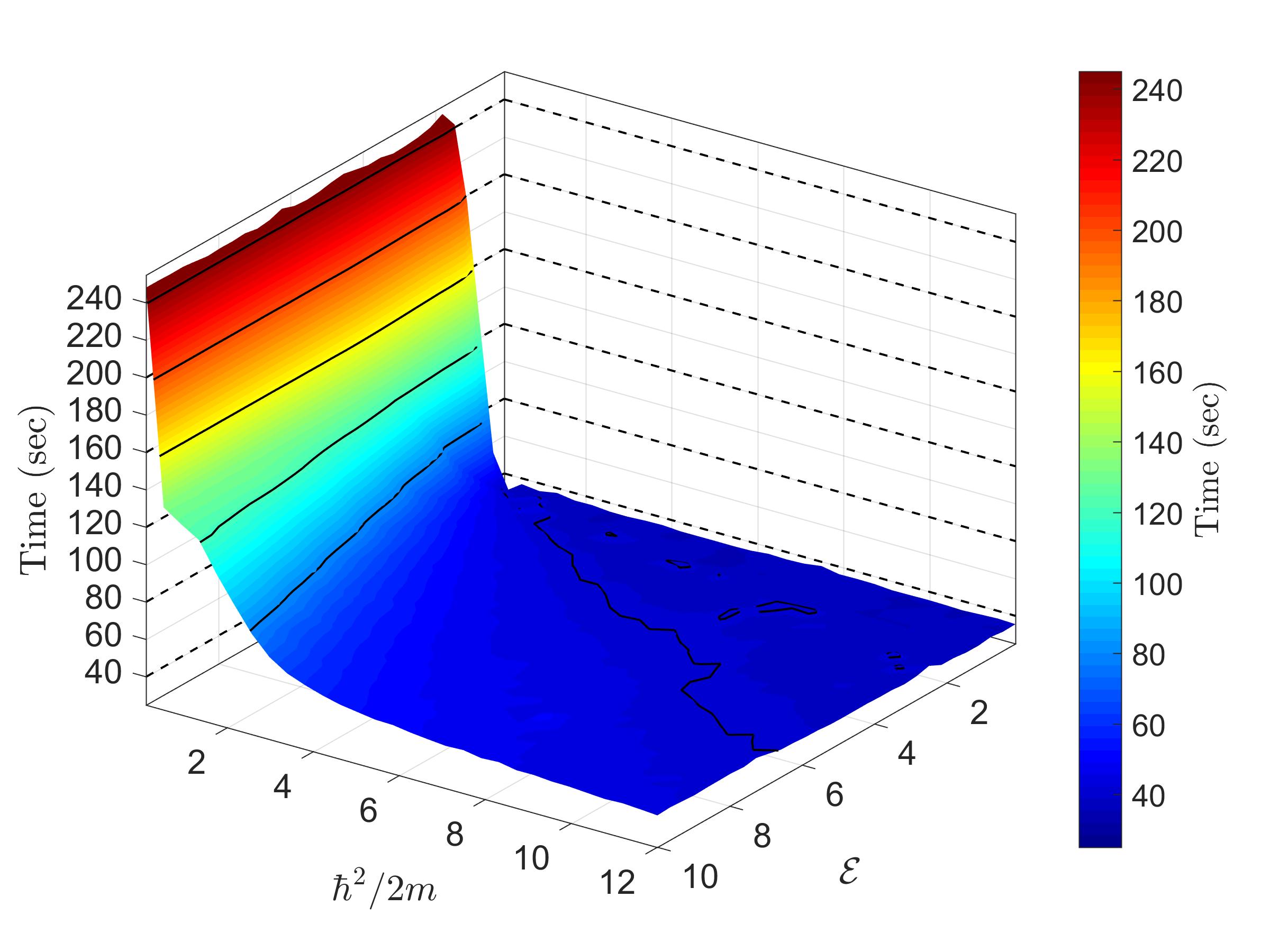}\hfill
\includegraphics[width=0.47\textwidth]{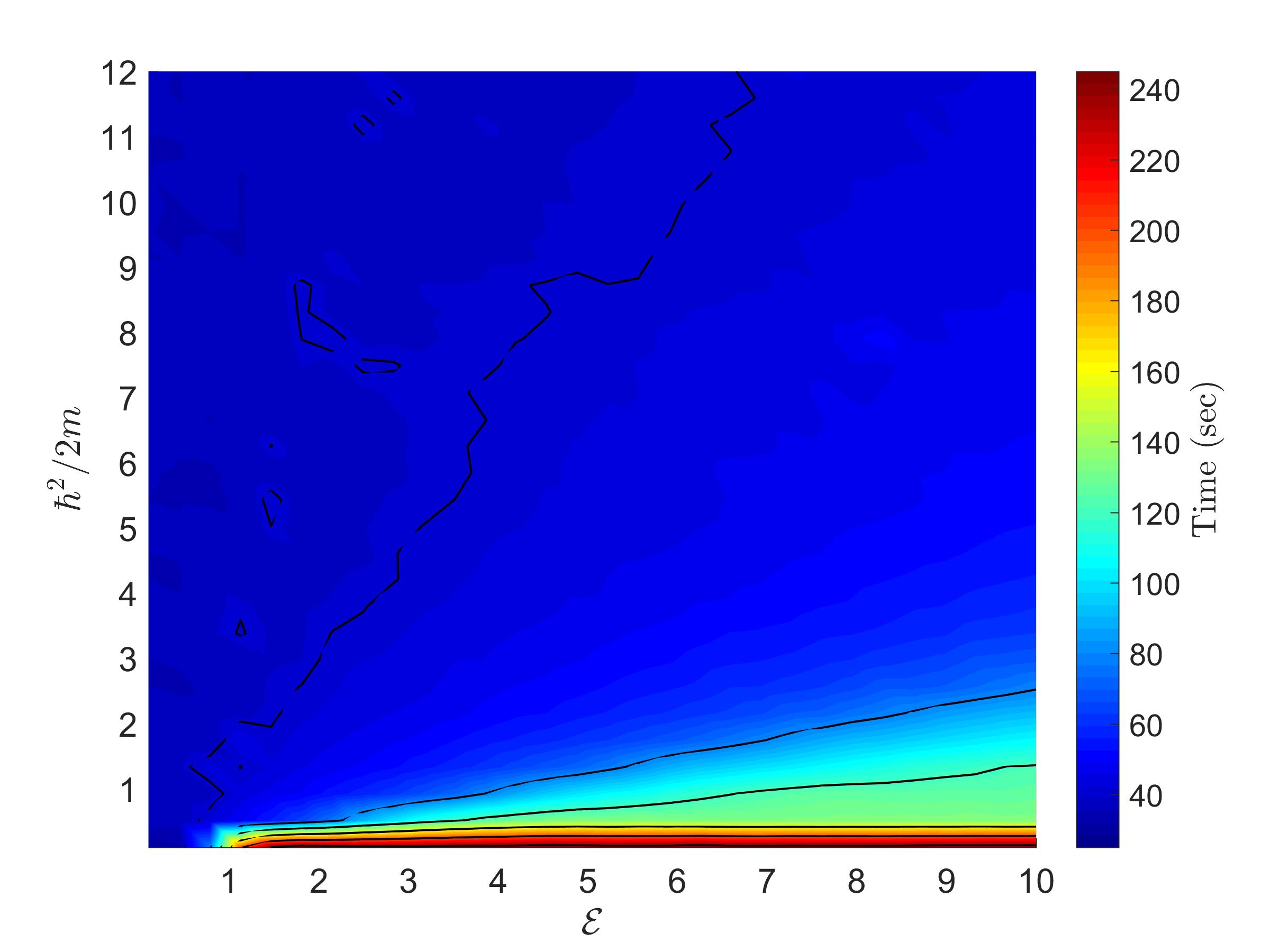}}

\caption{Experiment performed on the image in Fig. \ref{fig:sample}(a) blurred by a Gaussian kernel $h_{\sigma}^{4\times 4}$ of size $4\times 4$ with standard deviation $\sigma = 3$, and corrupted by Poisson noise corresponding to a SNR of 20 dB. QAB-PnP was performed with $\lambda_0 = 1.5$, and $\gamma$, $\sigma_{\mathcal{QAB}}$, $s$ and $\rho$ manually tuned to their best possible values for each set of experiments.}
\label{fig:hypar}
\end{figure*}

\begin{figure*}[h!]
\centering
\subfigure[Performed on the image in Fig. \ref{fig:sample}(a) with  $\mathcal{E} = 3.9$, $\hbar ^2/2m = 4$]
{\includegraphics[width=0.32\textwidth]{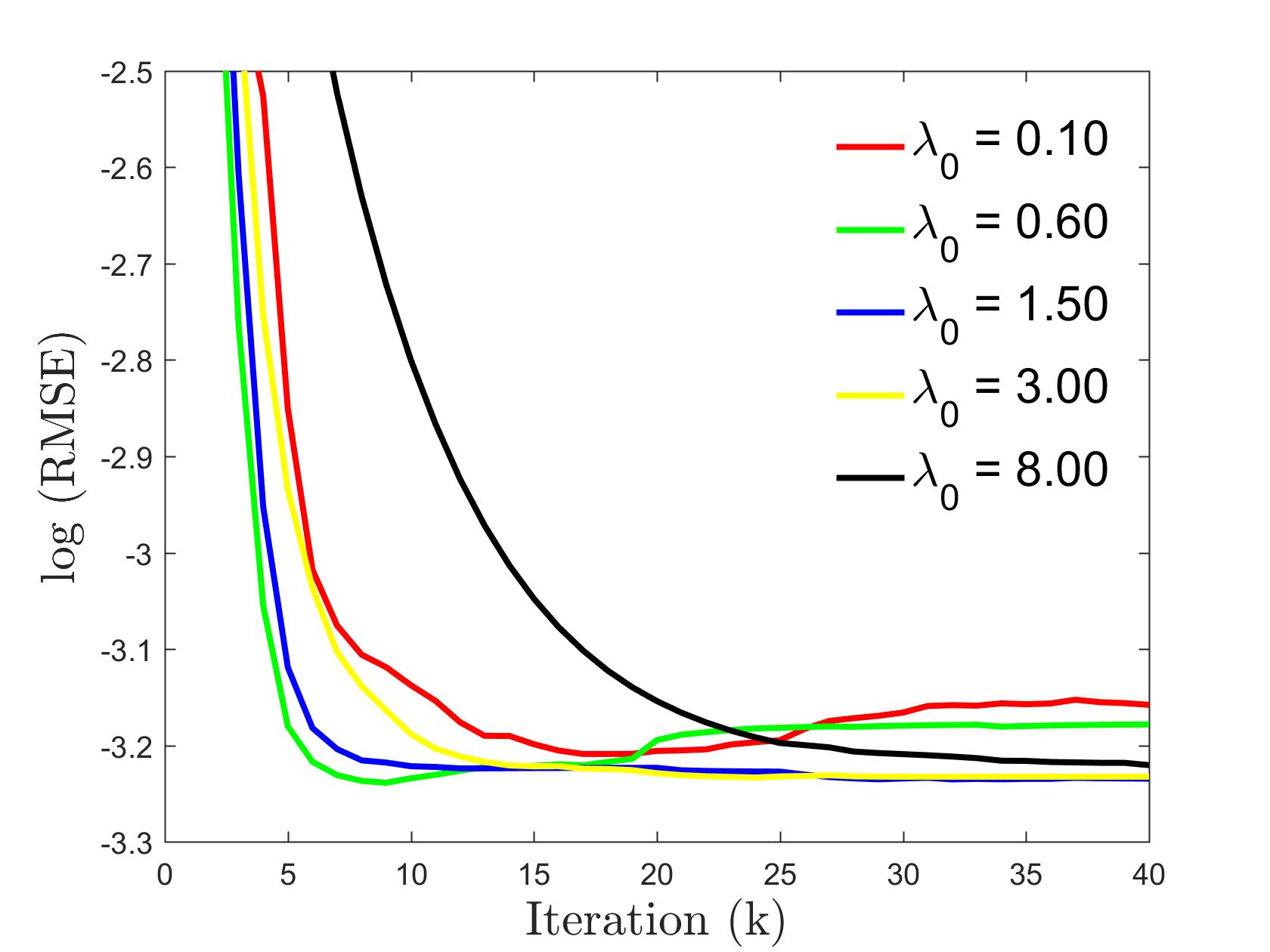}} \hfill
\subfigure[Performed on the image in Fig. \ref{fig:sample}(b) with  $\mathcal{E} = 4.1$, $\hbar ^2/2m = 4$]
{\includegraphics[width=0.32\textwidth]{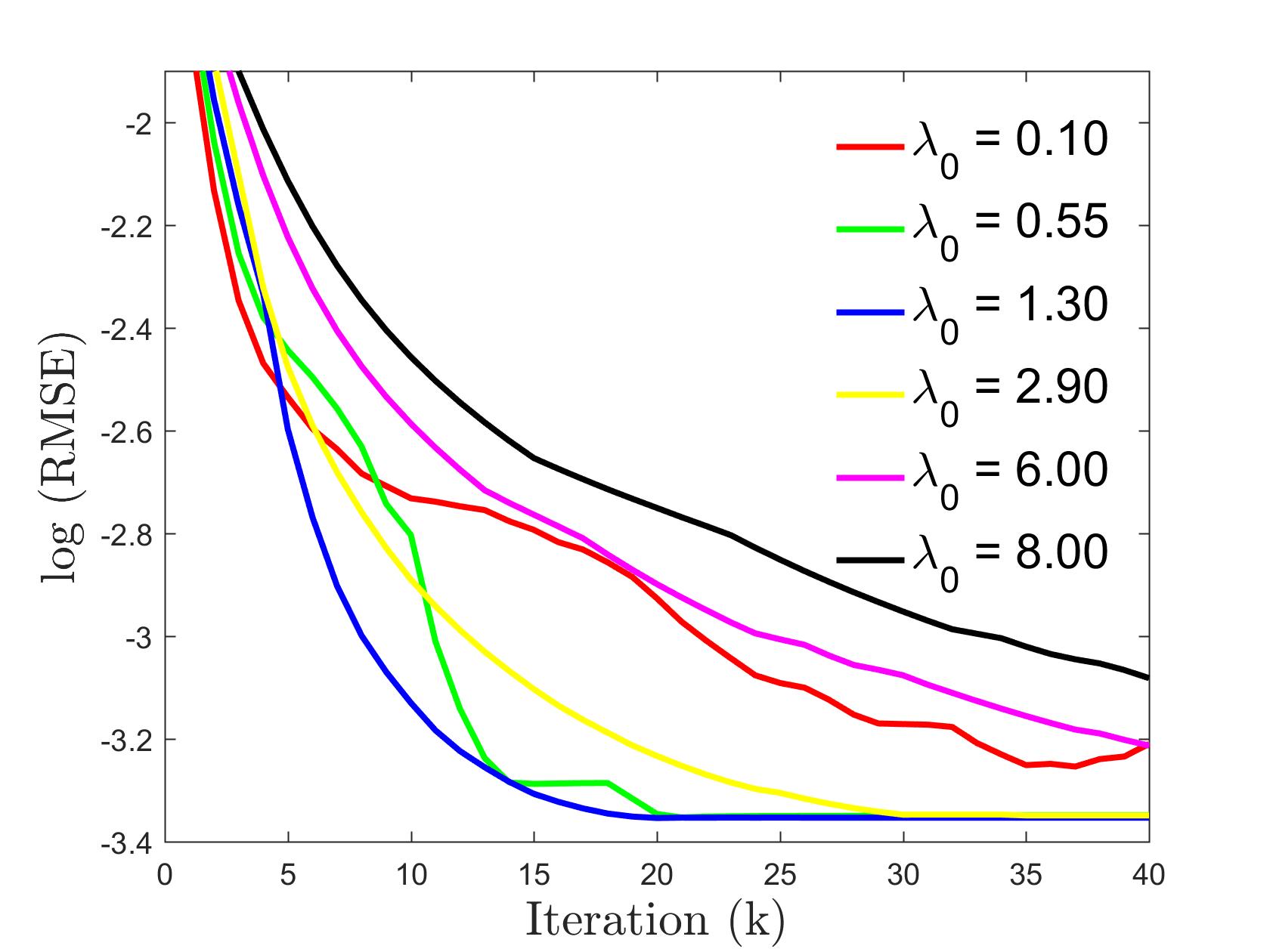}} \hfill
\subfigure[Performed on the image in Fig. \ref{fig:sample}(c) with  $\mathcal{E} = 4.5$, $\hbar ^2/2m = 4.3$]
{\includegraphics[width=0.32\textwidth]{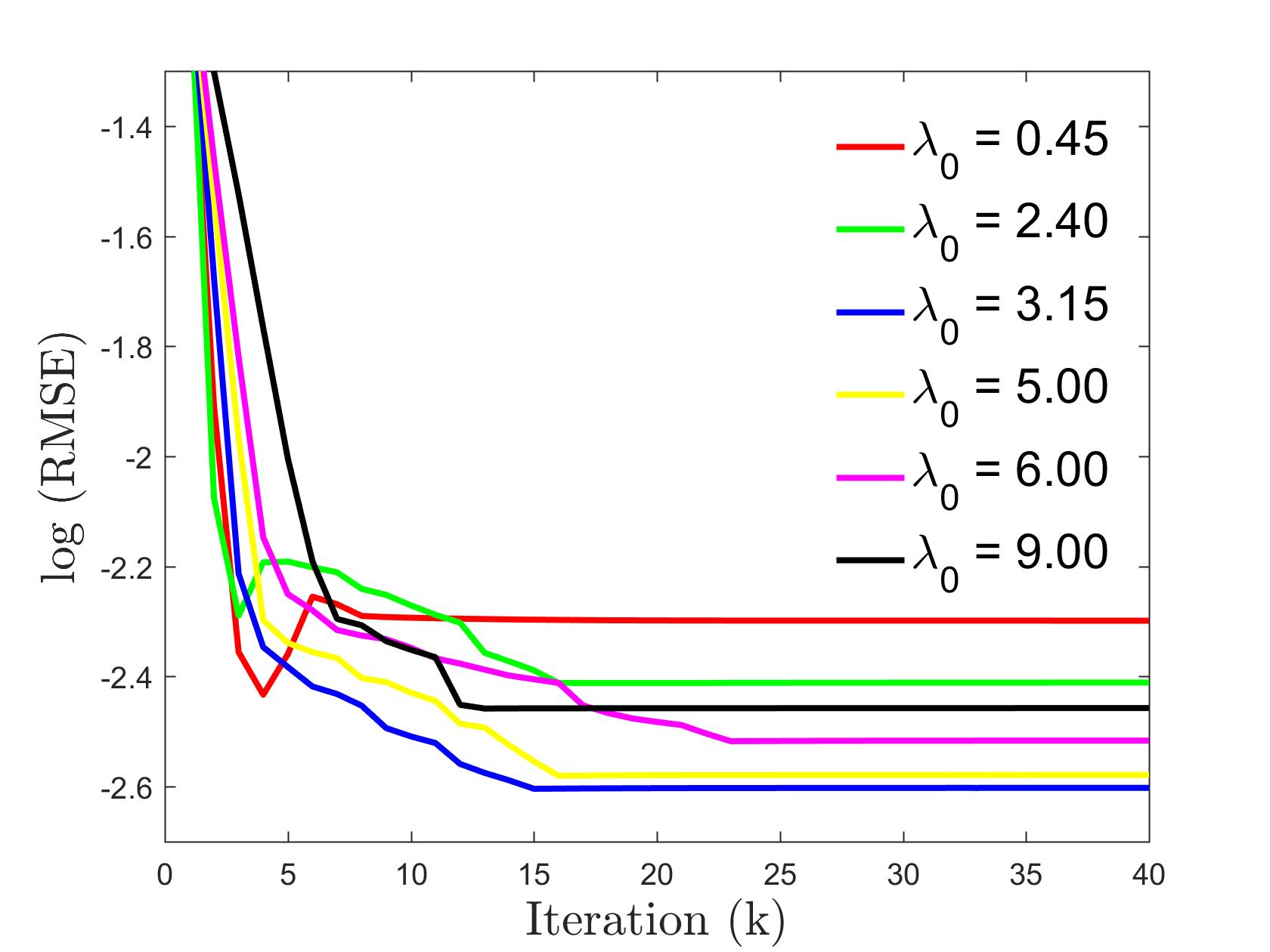}}

\caption{Evolution of the RMSE (logarithmic scale) for different values of the hyperparameter $\lambda_0$, for a Gaussian blurring kernel $h_{\sigma}^{4\times 4}$ of size $4\times 4$ with standard deviation $\sigma = 3$ and Poisson noise corresponding to a SNR of 20 dB. The other hyperparameters $\gamma$, $\sigma_{\mathcal{QAB}}$, $s$ and $\rho$ have been manually tuned to their best possible values for each set of experiments.}
\label{fig:lambda}
\end{figure*}

\begin{table}[h!]
\begin{footnotesize}

\begin{center}
\caption{Quantitative measurements obtained using the proposed QAB-PnP algorithm with and without modified OMP}
\label{tab:tab_E}
\begin{tabular}{c c c c c c}

\thickhline

\multirow{2}{*}{Sample} & \multirow{2}{*}{Noise} & \multicolumn{2}{c}{Without OMP} & \multicolumn{2}{c}{With OMP, best $\mathcal{E}$}\\
 &  & PSNR (dB) & SSIM & PSNR (dB) & SSIM\\
\thickhline

\multirow{3}{*}{Synthetic}
			
	& 20 dB & \textbf{30.1724} & \textbf{0.9179} & 29.9497 & 0.8934\\
	& 15 dB & \textbf{26.8101} & 0.8604 & 26.7300 & \textbf{0.8620}\\
	
\vspace{5pt}
	
	& 10 dB & \textbf{23.1674} & 0.7489 & 23.1006 & \textbf{0.7493}\\
	

\multirow{3}{*}{Lena}
			
		& 20 dB & \textbf{29.1330} & \textbf{0.8112} & 28.9842 & 0.8091\\
		& 15 dB & \textbf{26.5853} & \textbf{0.7712} & 26.5805 & 0.7709\\
		
\vspace{5pt}

		& 10 dB & \textbf{21.4328} & \textbf{0.6989} & 19.8070 & 0.6942\\
		

\multirow{3}{*}{Fruits}
			
		& 20 dB & \textbf{20.7366} & \textbf{0.6908} & 20.1657 & 0.6817\\
		& 15 dB & \textbf{18.8144} & 0.6471 & 18.6564 & \textbf{0.6474}\\
		& 10 dB & \textbf{14.9236} & 0.6114 & 14.9200 & \textbf{0.6117}\\

\thickhline

\end{tabular}
\end{center}
\end{footnotesize}
\end{table}



\begin{table*}[b!]
\begin{center}
\caption{Average computation time (all the algorithms have been implemented in Matlab and tested on a computer with an Intel(R) Core(TM) i7-10510U CPU of 4 cores each with 1.80 GHz, 16 GB memory and using Windows 10 Pro version 20H2 as operating system) and required number of iterations for different images.}
\label{tab:iterVStime}
\begin{tabular}{c  c c c   c  c c c}
\thickhline

\multirow{2}{*}{Method}
			 & \multicolumn{3}{c}{ Run time (sec)} && \multicolumn{3}{c}{ Number of iterations}\\
			 &  Synthetic &  Lena &  Fruits  &&  Synthetic &  Lena &  Fruits \\
\thickhline
 TV-ADMM 		&  0.111 &  0.107 &  0.130 &&  26 &  17 &  23 \\

 ADMM+BM3D	&  0.017 &  0.017 &  0.022 &&  27 &  20 &  26 \\

 ADMM+TNRD &  78.375 &  81.980 &  104.179 &&  17 &  22 &  25 \\

 ADMM+VST+TNRD	&  77.310 &  82.630 &  112.070 &&  20 &  19 &  17 \\

 P$^4$IP		&  0.037 &  0.039 &  0.049 &&  18 &  8 &  19 \\

 QAB-PnP (Without OMP)		&  190.284 &  186.677 &  266.221 &&  17 &  7 &  14 \\

 QAB-PnP (With OMP, best $\mathcal{E}$)		&  37.425 &  35.732 &  48.568 &&  18 &  7 &  15 \\

\thickhline	
\end{tabular} \end{center} 
\end{table*}

From this perspective, we first show that considering the wave vectors up to the energy level $\mathcal{E}$ and evaluating only the corresponding coefficients $\alpha_i$ following the modified OMP algorithm in Algo.~\ref{Algo:OMP} helps reducing the computation time with minimal accuracy loss. Quantitative results showing the influence of $\mathcal{E}$ on the simulations performed over the three sample images in Fig.~\ref{fig:sample}, distorted by a Gaussian blurring kernel $h_{\sigma}^{4\times 4}$ of size $4\times 4$ and standard deviation $\sigma = 3$, and corrupted by Poisson noise corresponding to a SNR of 20 dB, 15 dB, and 10 dB, have been regrouped in Table~\ref{tab:tab_E}, where the best results have been highlighted in bold. Similarly, the average peak signal to noise ratios (PSNR) values for different SNR, obtained with the proposed deconvolution method with and without the modified OMP algorithm, are shown in Fig.~\ref{fig:Ewow}. The results in Fig.~\ref{fig:Ewow} and Table~\ref{tab:tab_E} prove that the accuracy loss, caused by the use of the parameter $\mathcal{E}$ within the modified OMP algorithm, is very limited. This accuracy loss is caused by the denoising process that reconstructs the denoised image only from the wave functions associated with an energy level lower than $\mathcal{E}$. Indeed, although wave functions associated with higher energies are dominated by noise, they may still carry information about certain features of the clean image. The average computation time for different images obtained with a Matlab implementation on a desktop computer, with and without $\mathcal{E}$, given in Table~\ref{tab:iterVStime}, confirms the computational efficiency gain enabled by the modified OMP algorithm embedded in QAB-PnP method.

In addition to $\mathcal{E}$, as stated previously, $\hbar ^2/2m$ is also an important hyperparameter of the proposed deconvolution technique. The hyperparameter $\hbar ^2/2m$ dictates how the local frequencies of the basis vectors vary with the amplitude of the image pixel values. On the other hand, $\mathcal{E}$ is associated with the sparsity. Given their mutual dependence, Fig.~\ref{fig:hypar}(a) shows the accuracy of QAB-PnP algorithm for different couple values of these two hyperparameters over an acceptable range. This experiment consisted in recovering the image in Fig.~\ref{fig:sample}(a) from a degraded version blurred by a $4 \times 4$ Gaussian kernel with standard deviation equal to $3$ and Poisson noise corresponding to a SNR of 20 dB.

Similarly, Figs.~\ref{fig:hypar}(b) and (c) show the variation of the number of the significant wave vectors $\mathcal{T}$ and of the computation time. These results also justify the linear proportionality of $\mathcal{T}$ and processing time. Note that as explained previously, the other hyperparameters, $\sigma_{\mathcal{QAB}}$, $s$ and $\rho$, were chosen as suggested in \cite{dutta2021quantum}.

Finally, the choice of the hyperparameter $\lambda_0$ used within the iterations of the ADMM algorithm described in Section \ref{sec:background} is important to accelerate the convergence. The curves in Fig.~\ref{fig:lambda} show, within a logarithmic scale, the evolution of the root mean square error (RMSE) over the iterations of the  proposed deconvolution method, for different values of $\lambda_0$. These simulations were performed for the three images in Fig.~\ref{fig:sample}, distorted by a Gaussian blurring kernel $h_{\sigma}^{4\times 4}$ of size $4\times 4$ and standard deviation $\sigma = 3$, and corrupted by Poisson process corresponding to a SNR of 20 dB.

The studies performed in this subsection show that a certain range of optimal choice of the hyperparameters considered is possible. Without a priori knowledge, it should be possible to use values in this range for arbitrary images, taking care to choose $\mathcal{E}$ and $\hbar ^2/2m$ in a correlated way. As a further note, keeping the hyperparameters constant to the same values for all the images considered hereafter leads to a very low PSNR degradation of about $0.1$ dB. From the discussions above, one may note that the hyperparameters $\hbar ^2/2m$ and $\mathcal{E}$ are primarily associated with the construction of the quantum adaptive basis and the sparsity of the clean image in this basis, both related to the denoising process. In contrast, $\lambda_0$, the penalty parameter, regulates the restoration process by accelerating the convergence. Therefore, the optimal choice of $\hbar ^2/2m$ and $\mathcal{E}$ discussed above is independent of the value of $\lambda_0$.

\begin{figure}[b!]
\centering
\includegraphics[width=0.5\textwidth]{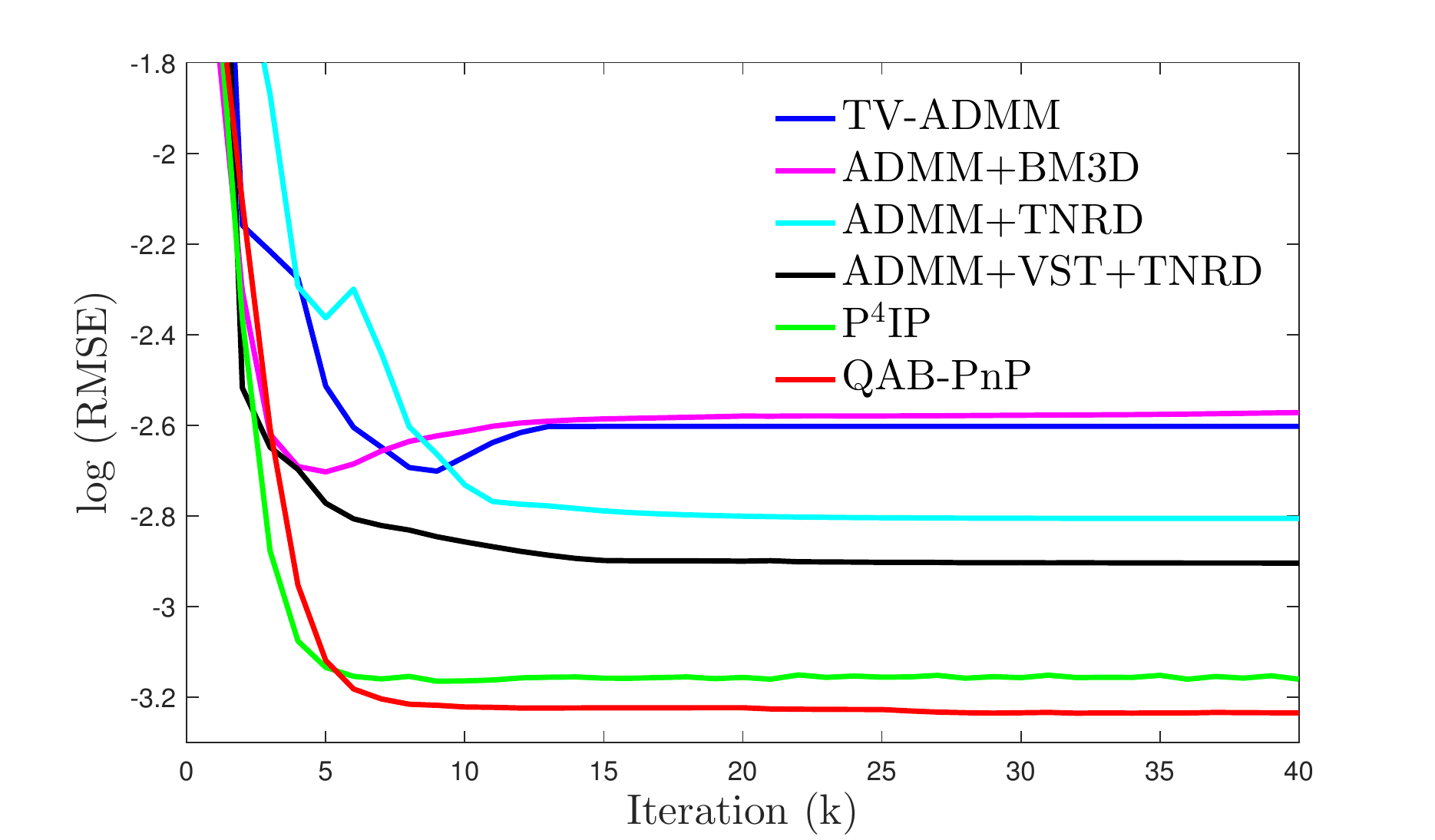}

\caption{RMSE in logarithmic scale as a function of iteration number for TV-ADMM, ADMM+BM3D, ADMM+TNRD, ADMM+VST+TNRD, P$^4$IP and proposed QAB-PnP methods. The results correspond to the restoration of the image in Fig.~\ref{fig:sample}(a) from a degraded image by a Gaussian blurring kernel $h_{\sigma}^{4\times 4}$ of size $4\times 4$ and standard deviation $\sigma = 3$, and Poisson noise corresponding to a SNR of 20 dB. All hyperparameters were manually tuned to their best possible values for all the methods.}
\label{fig:p4ipvsprop}
\end{figure}

\begin{figure*}[h!]
\centering
\subfigure[Clean image]{\includegraphics[width=0.17\textwidth]{Figure/simu/CleanLena.jpg}}
\subfigure[Corrupted image]{\includegraphics[width=0.17\textwidth]{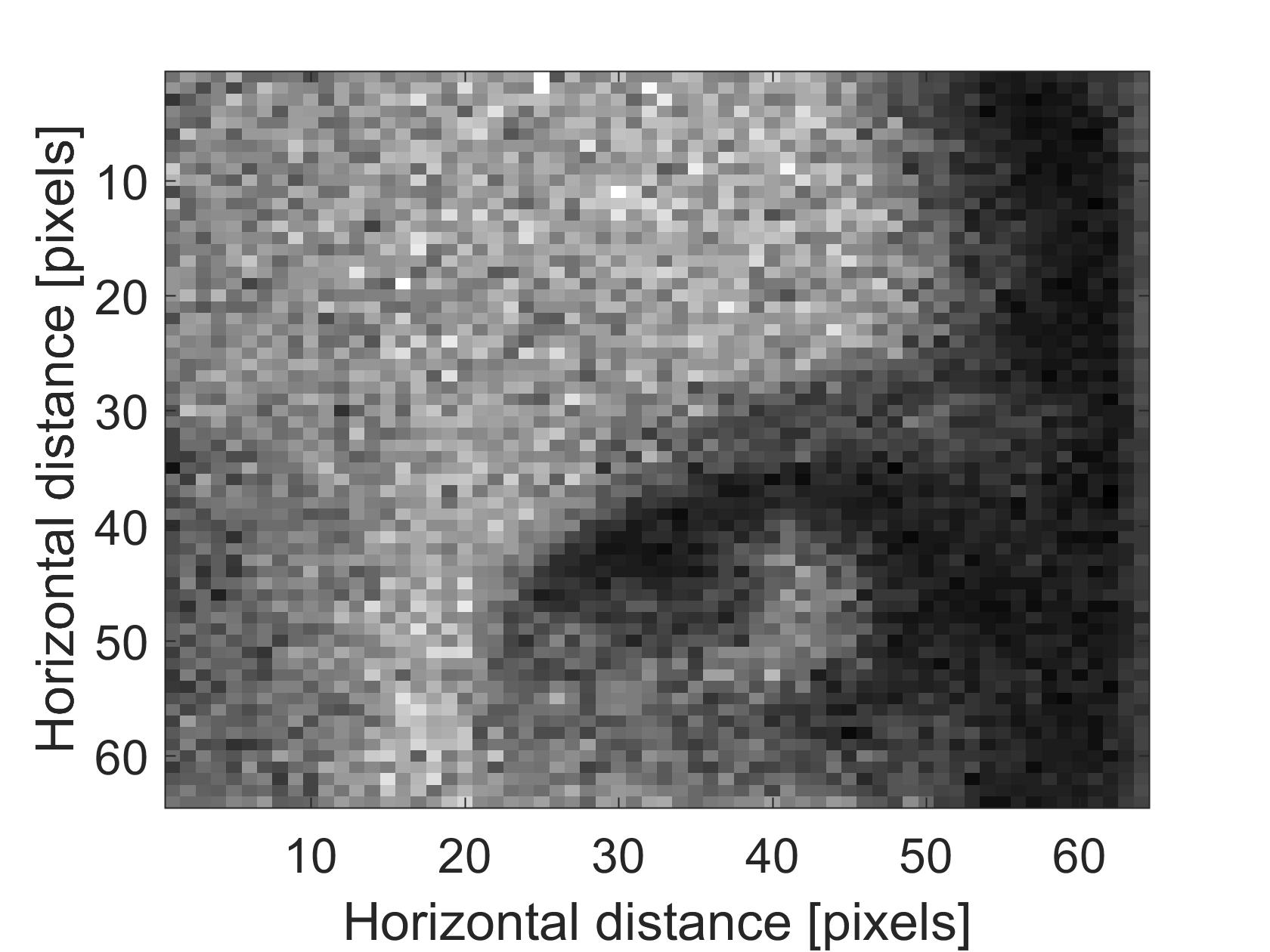}}
\subfigure[TV-ADMM]{\includegraphics[width=0.17\textwidth]{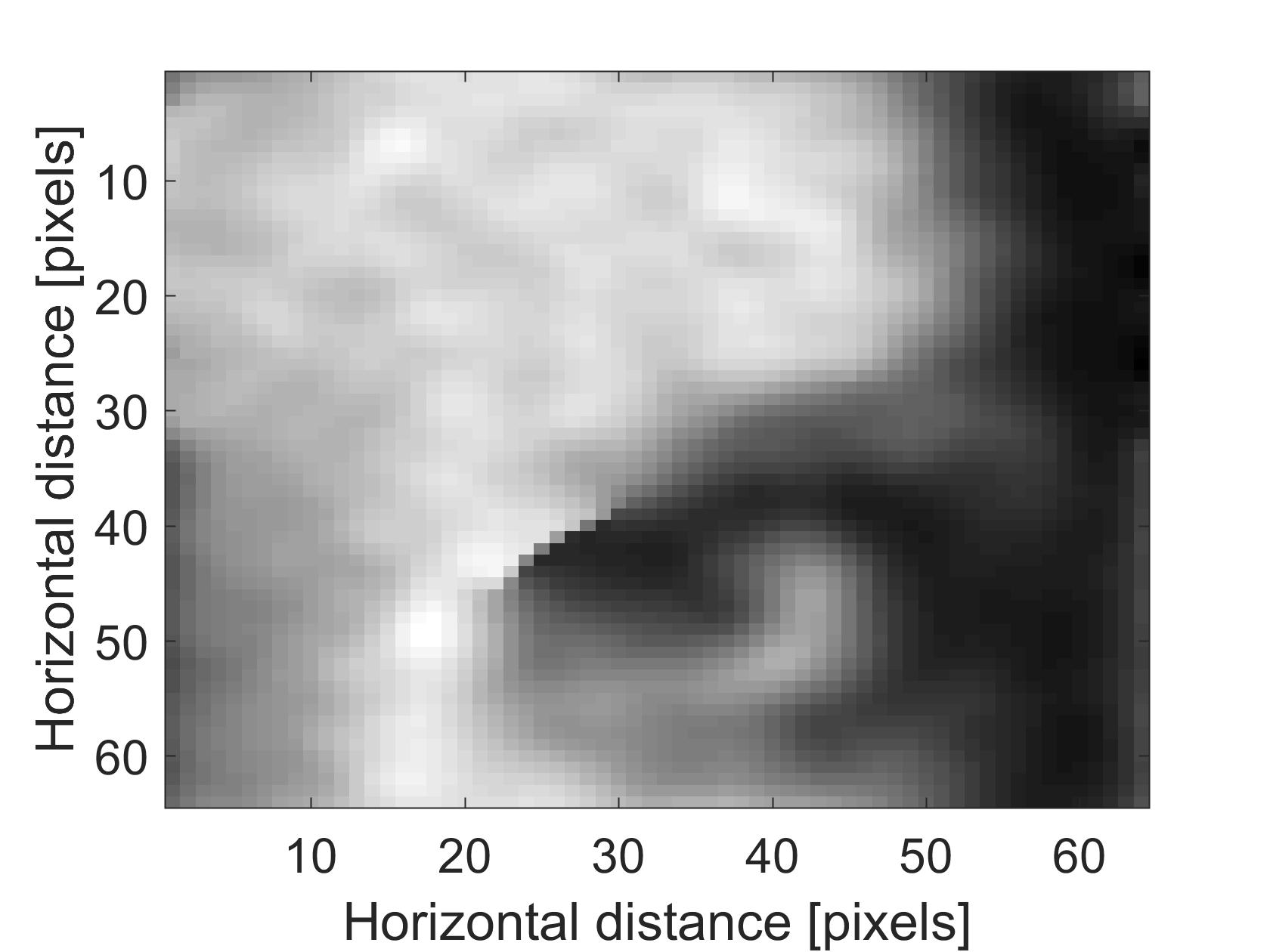}}
\subfigure[ADMM+BM3D]{\includegraphics[width=0.17\textwidth]{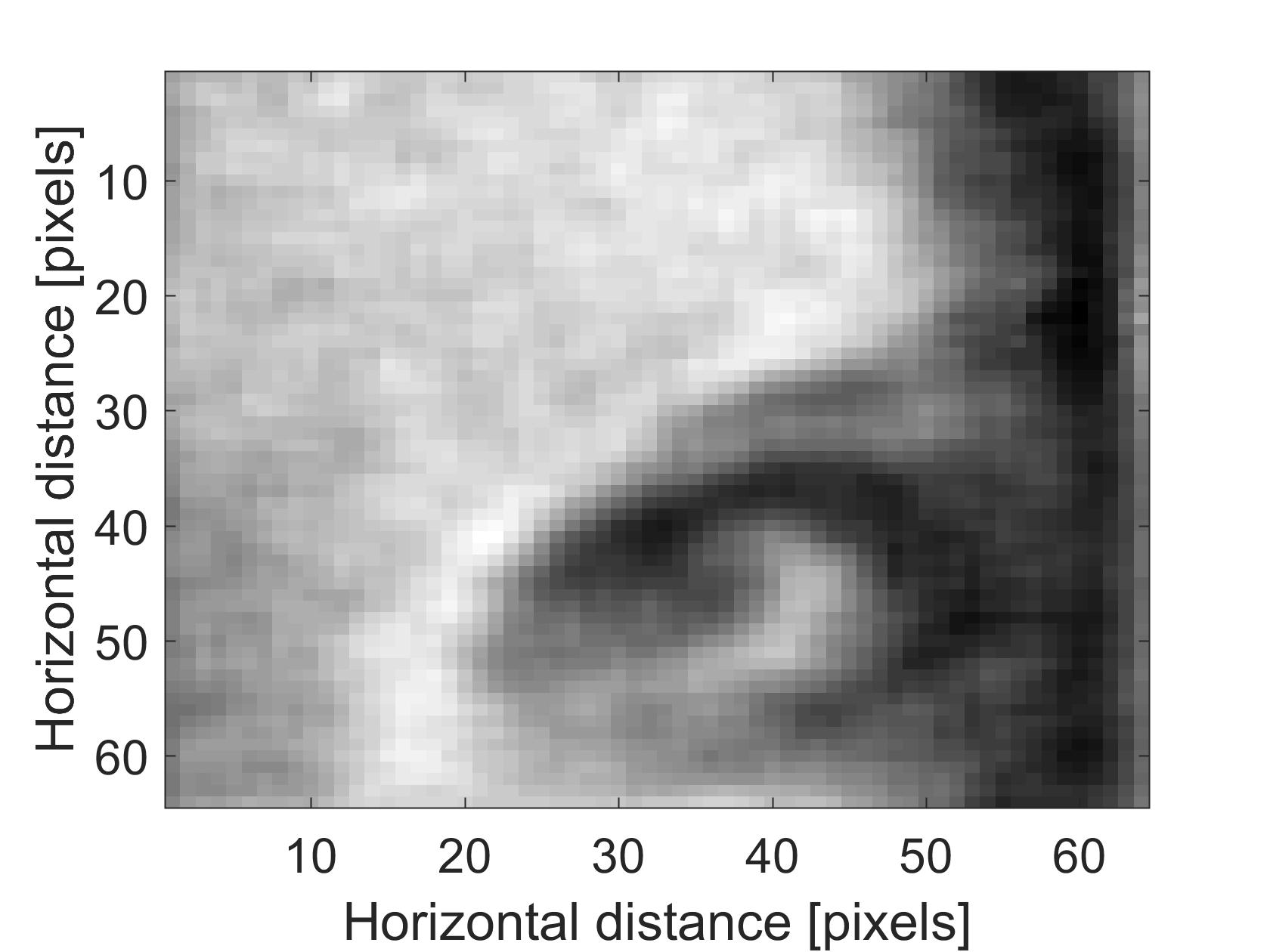}}
\subfigure[ADMM+TNRD]{\includegraphics[width=0.17\textwidth]{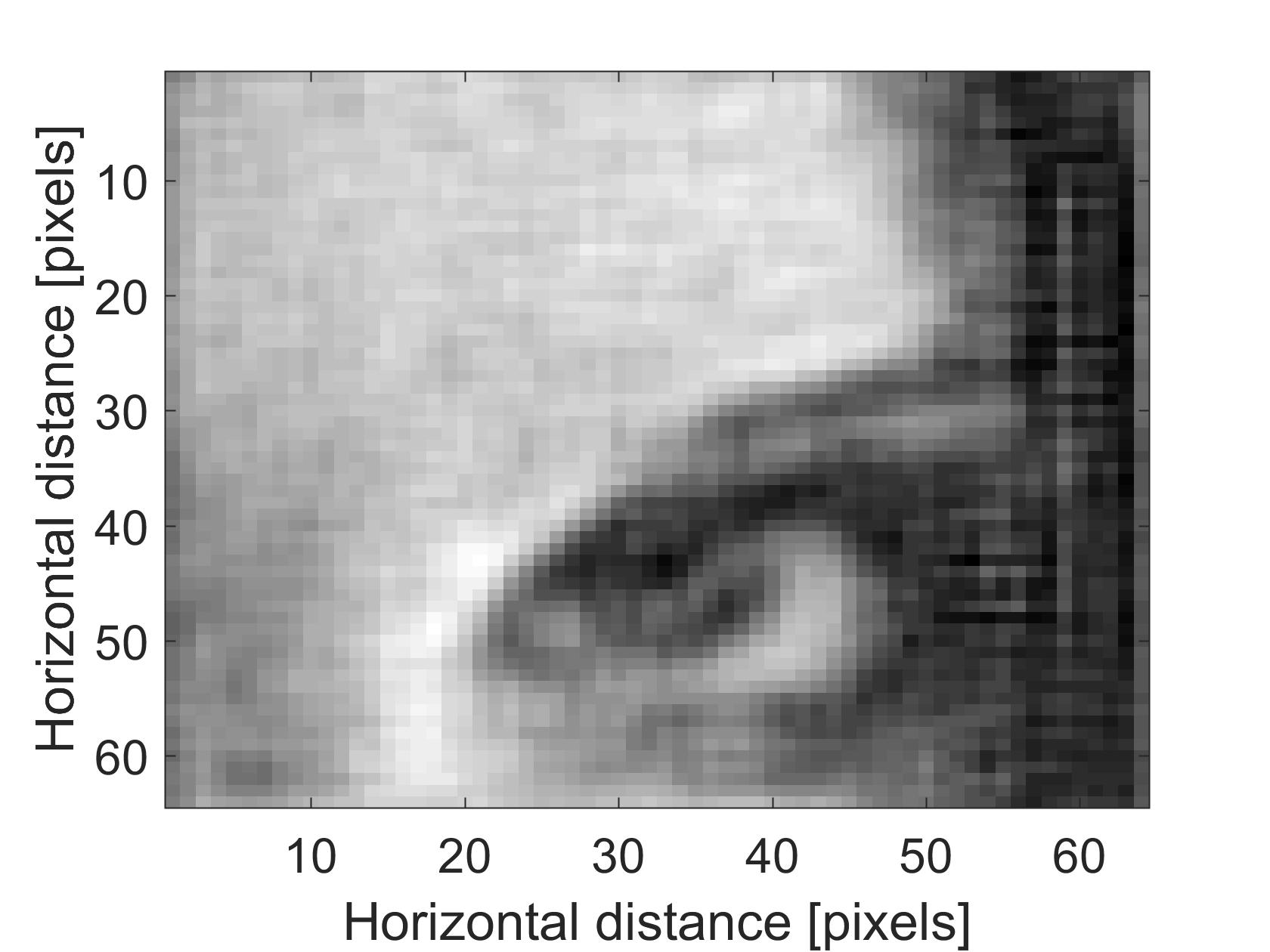}}
\subfigure[ADMM+VST+TNRD]{\includegraphics[width=0.17\textwidth]{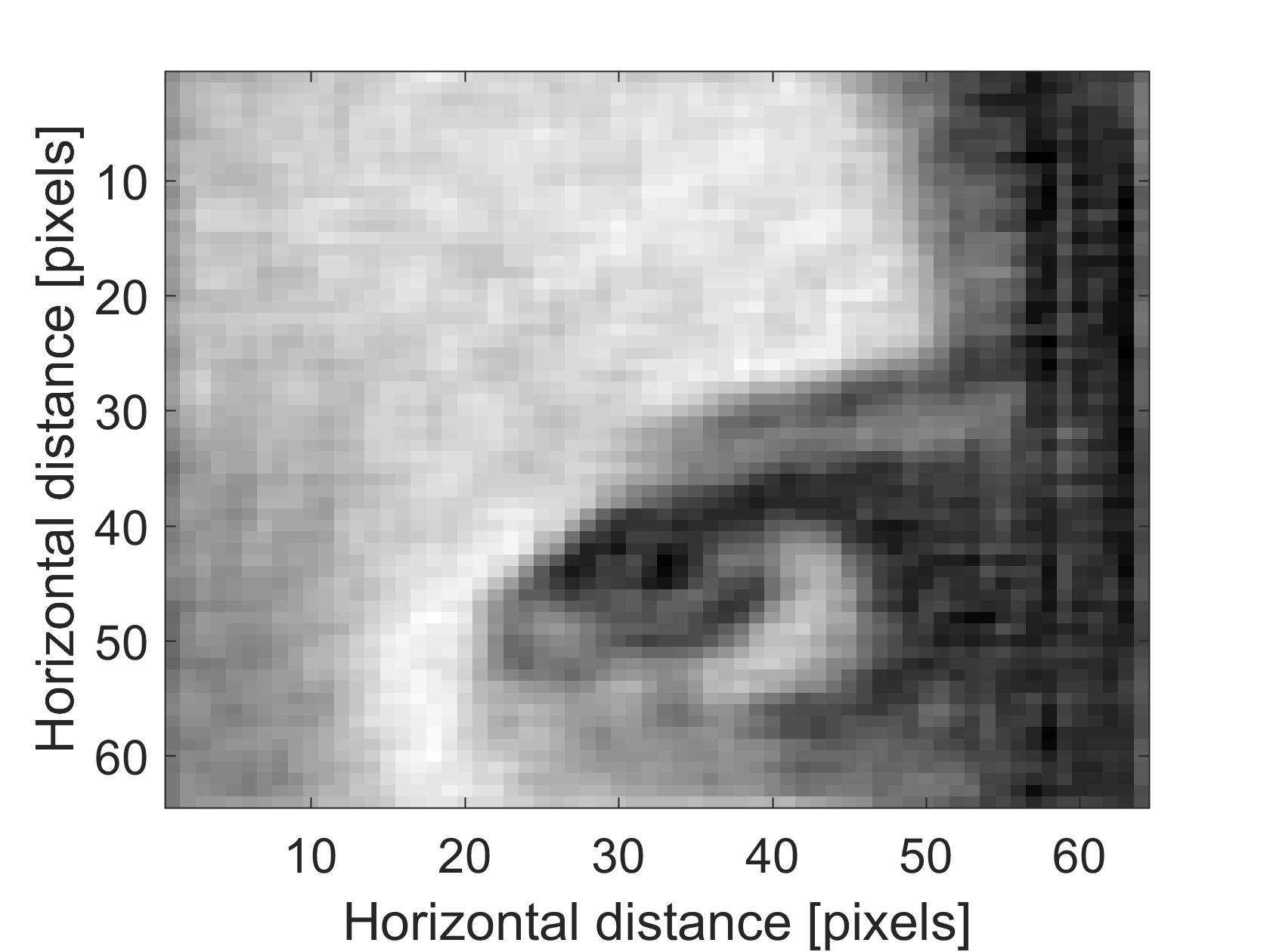}}
\subfigure[P$^4$IP]{\includegraphics[width=0.17\textwidth]{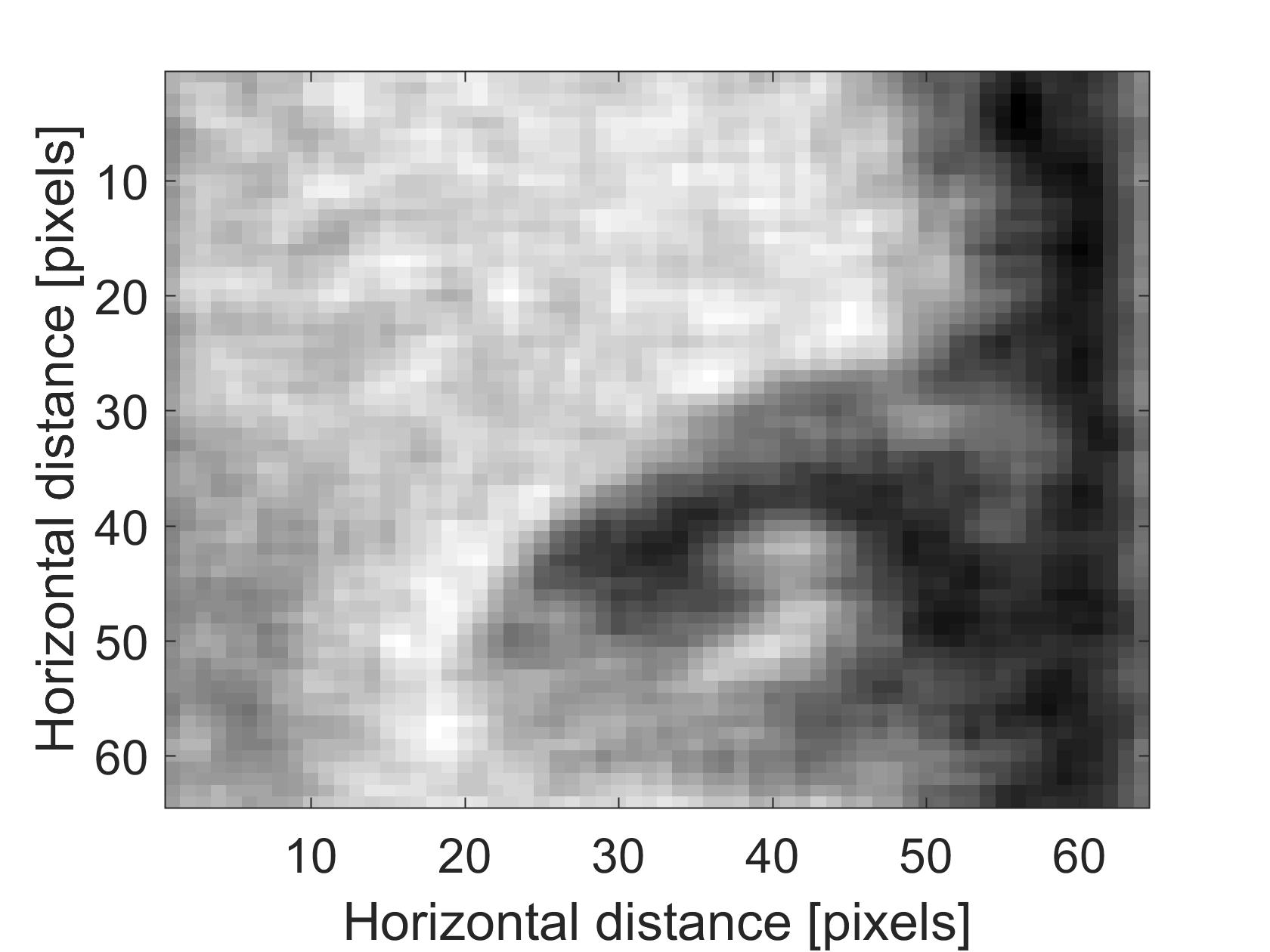}}
\subfigure[QAB-PnP]{\includegraphics[width=0.17\textwidth]{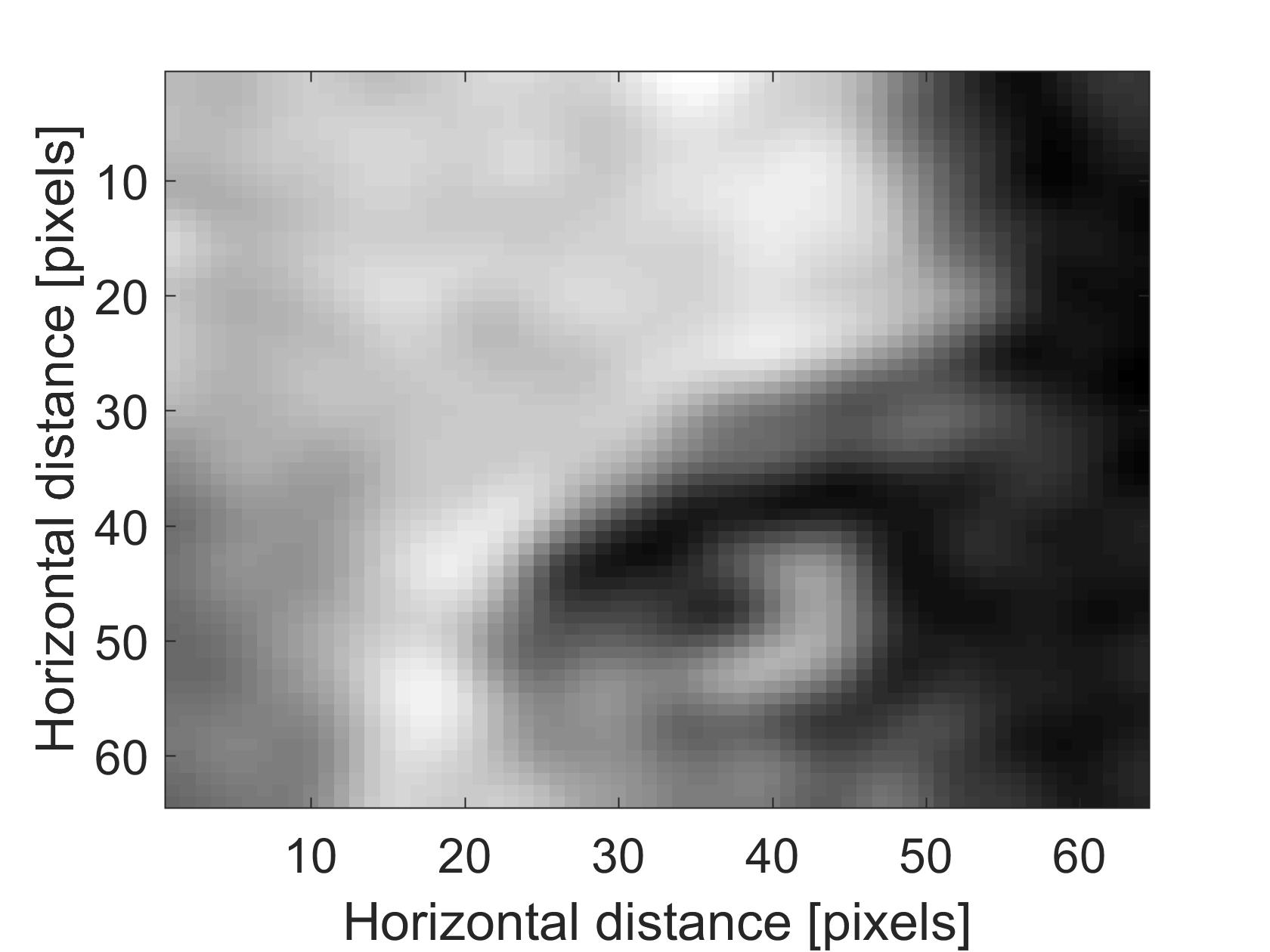}}

\caption{Deconvolution result for Lena image, blurred by a Gaussian kernel $h_{\sigma = 3}^{4\times 4}$ and corrupted by Poisson noise corresponding to a SNR of 10 dB. The proposed QAB-PnP algorithm used $\mathcal{E} = 3.9$, $\lambda_0 = 1.5$, $\hbar ^2/2m = 4$ and $\gamma = 1.01$, $\sigma_{\mathcal{QAB}} = 7$.}
\label{fig:lena10}
\vspace*{15pt}
\end{figure*}

\begin{figure*}[h!]
\centering
\subfigure[Clean image]{\includegraphics[width=0.17\textwidth]{Figure/simu/CleanSinImage2.jpg}}
\subfigure[Corrupted image]{\includegraphics[width=0.17\textwidth]{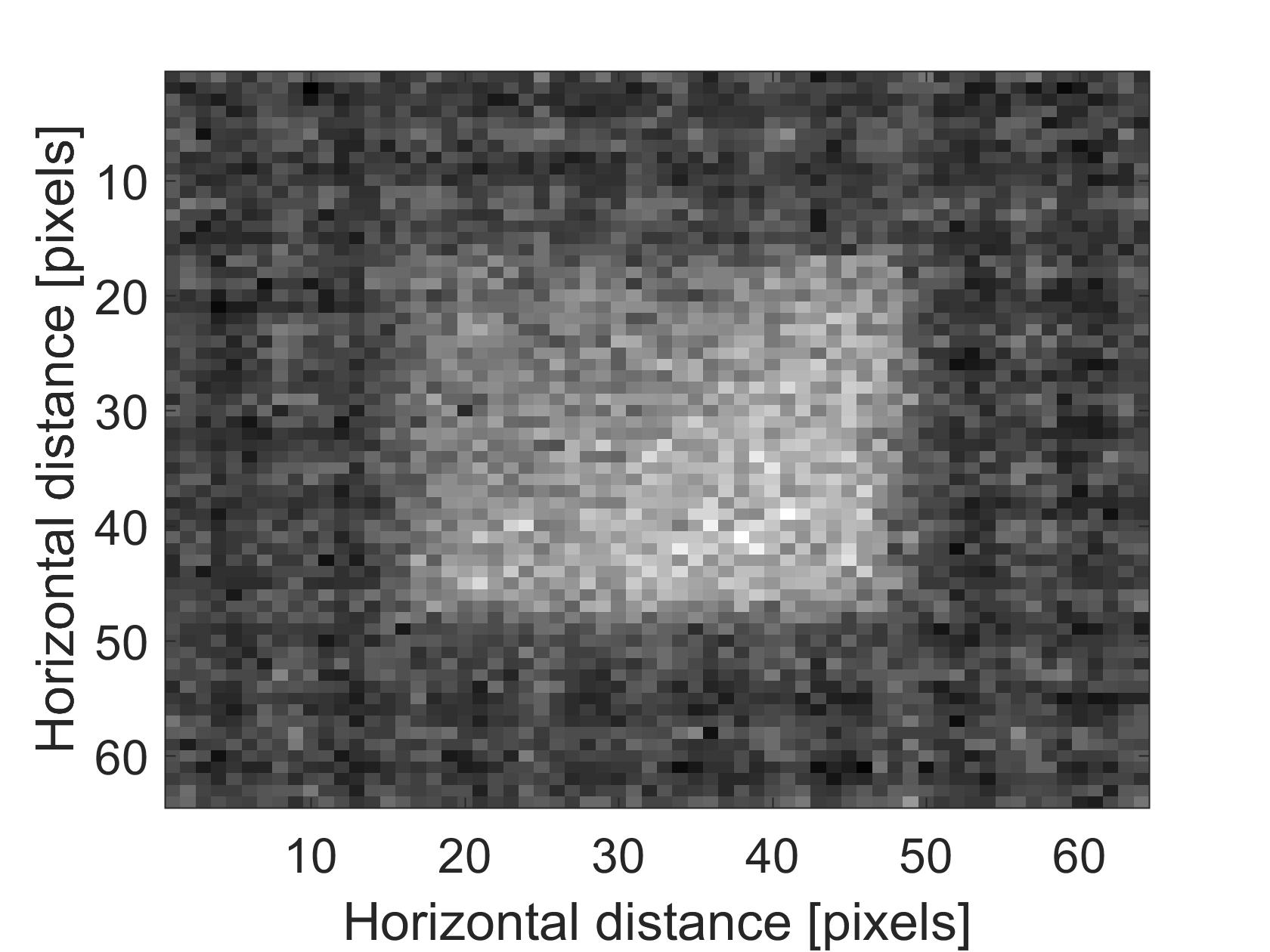}}
\subfigure[TV-ADMM]{\includegraphics[width=0.17\textwidth]{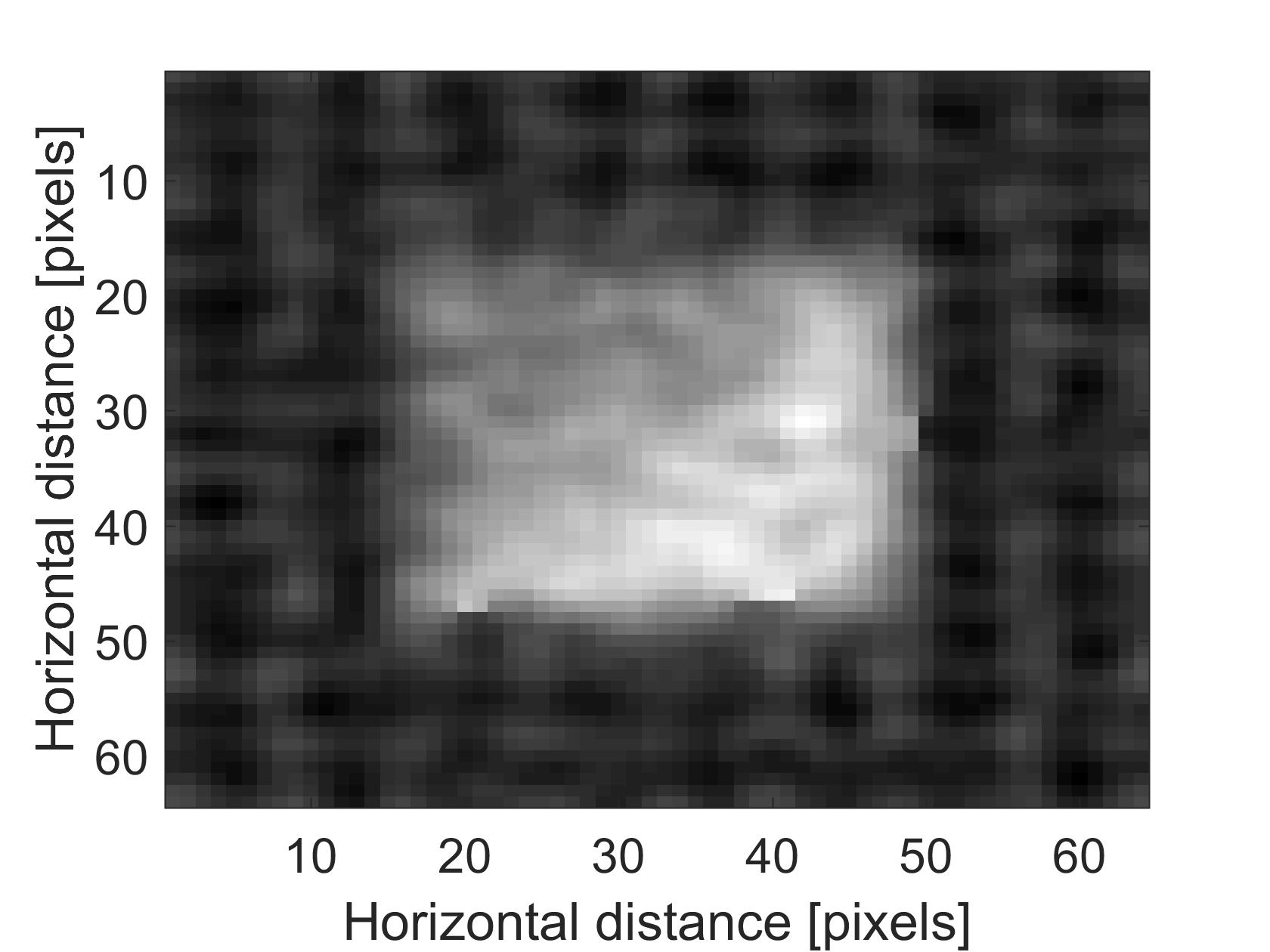}}
\subfigure[ADMM+BM3D]{\includegraphics[width=0.17\textwidth]{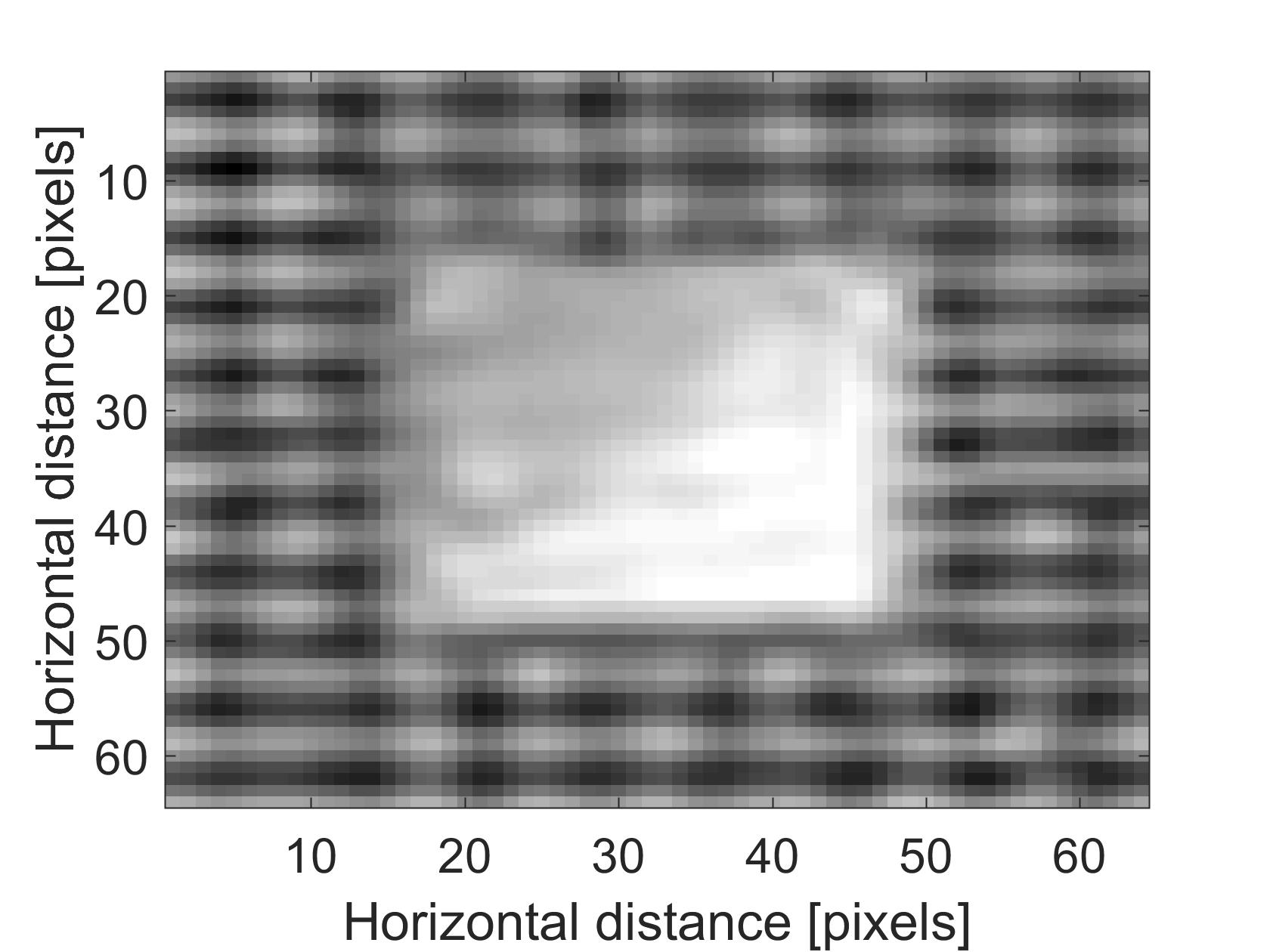}}
\subfigure[ADMM+TNRD]{\includegraphics[width=0.17\textwidth]{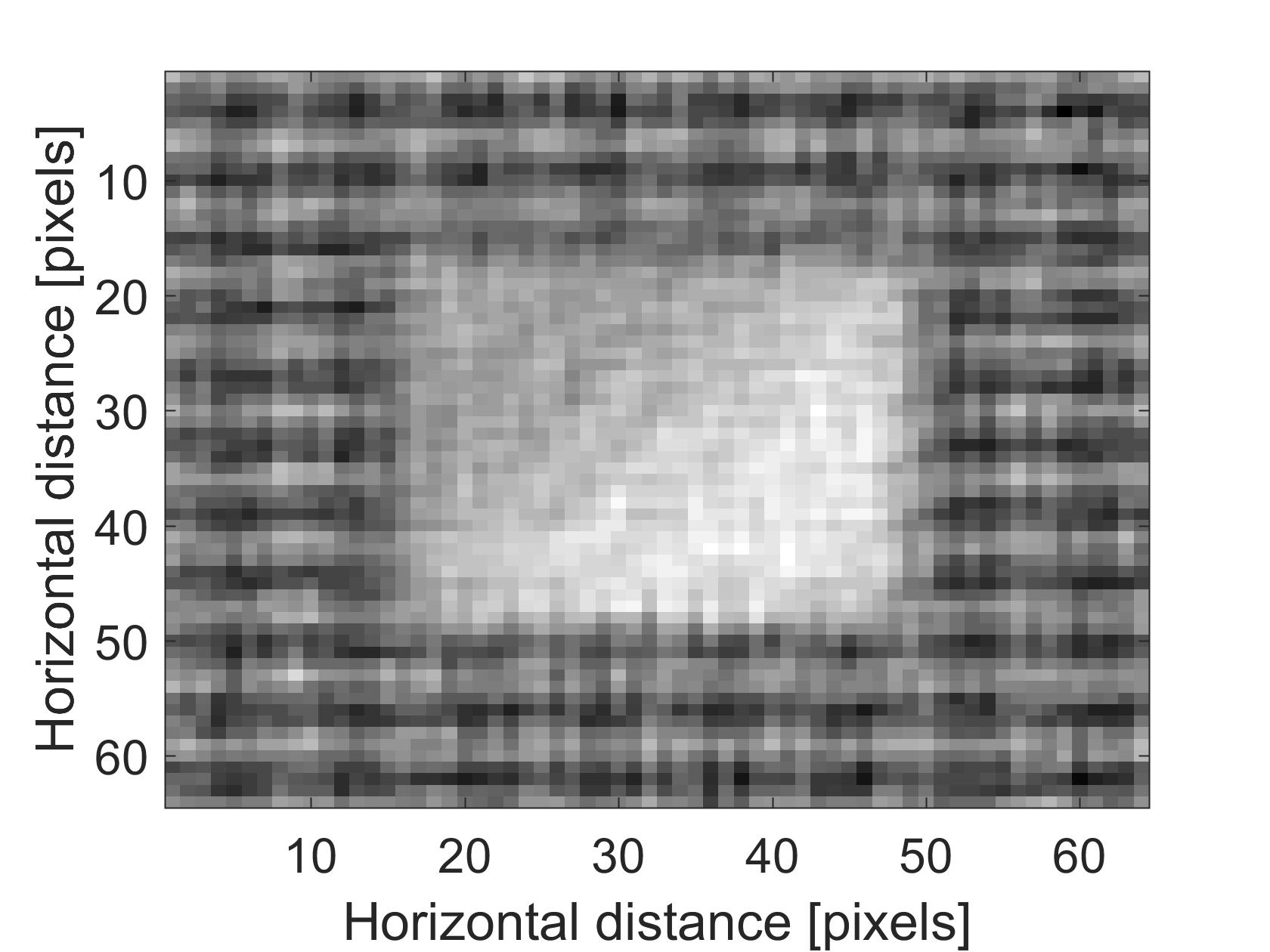}}
\subfigure[ADMM+VST+TNRD]{\includegraphics[width=0.17\textwidth]{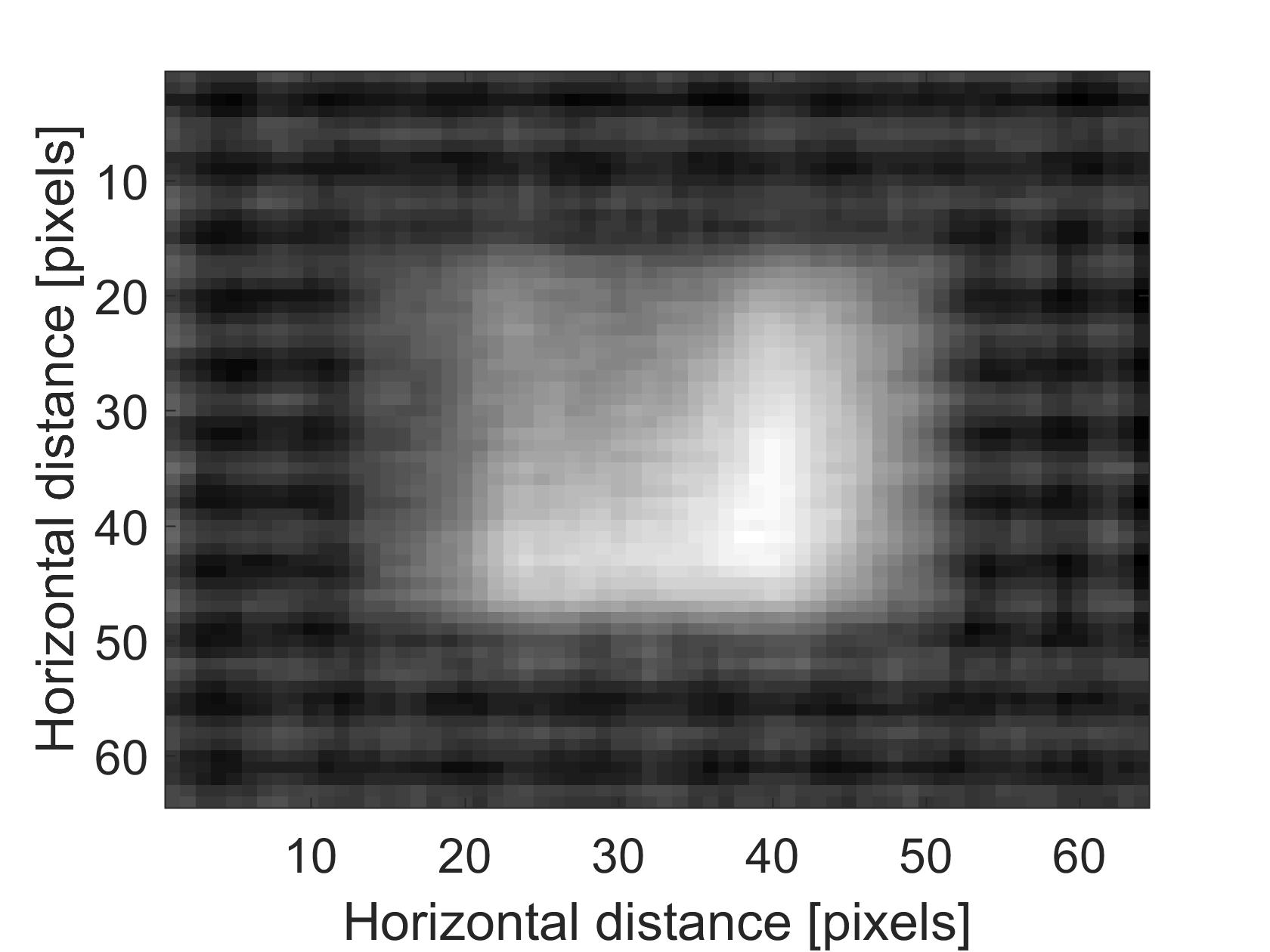}}
\subfigure[P$^4$IP]{\includegraphics[width=0.17\textwidth]{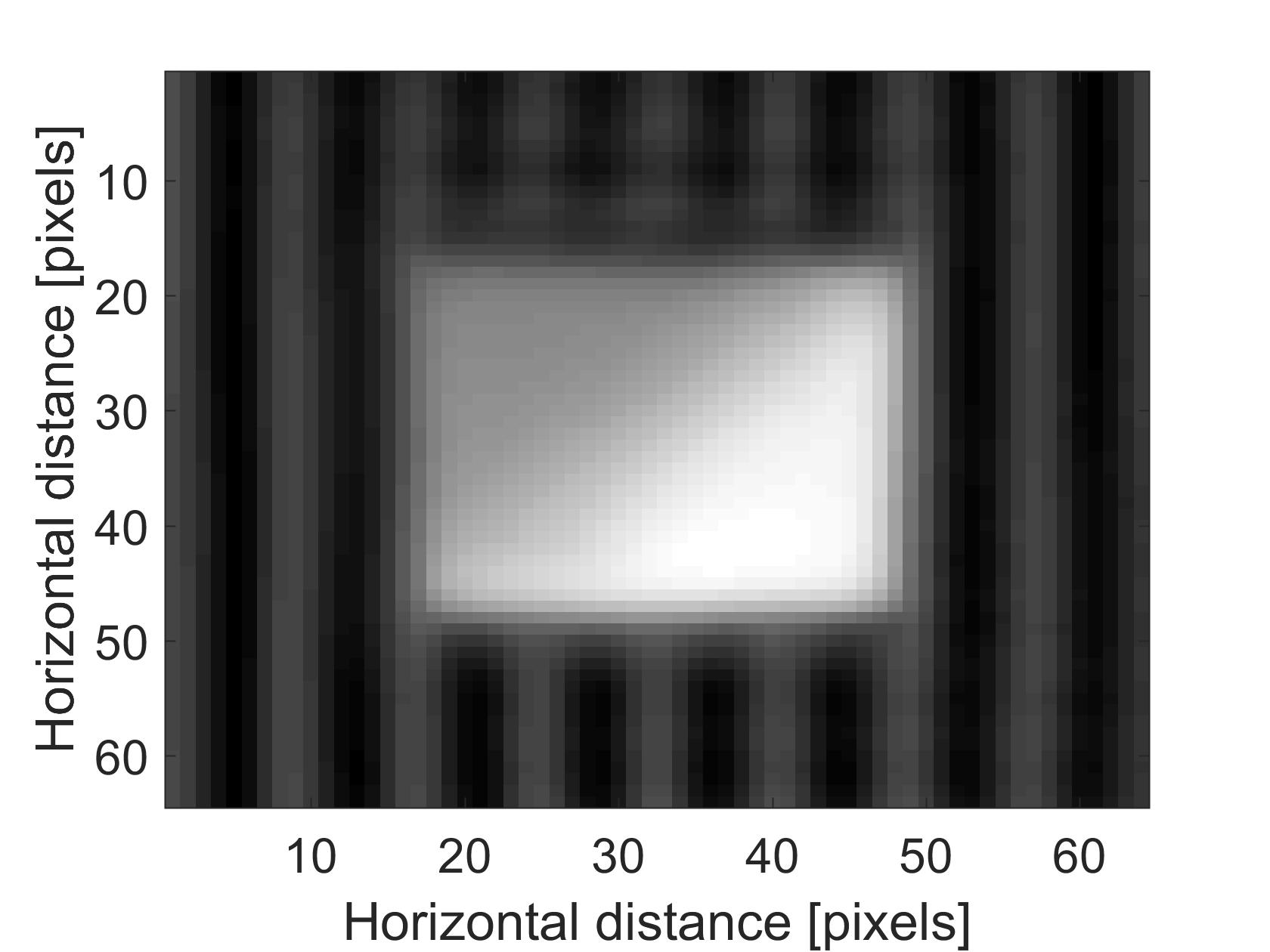}}
\subfigure[QAB-PnP]{\includegraphics[width=0.17\textwidth]{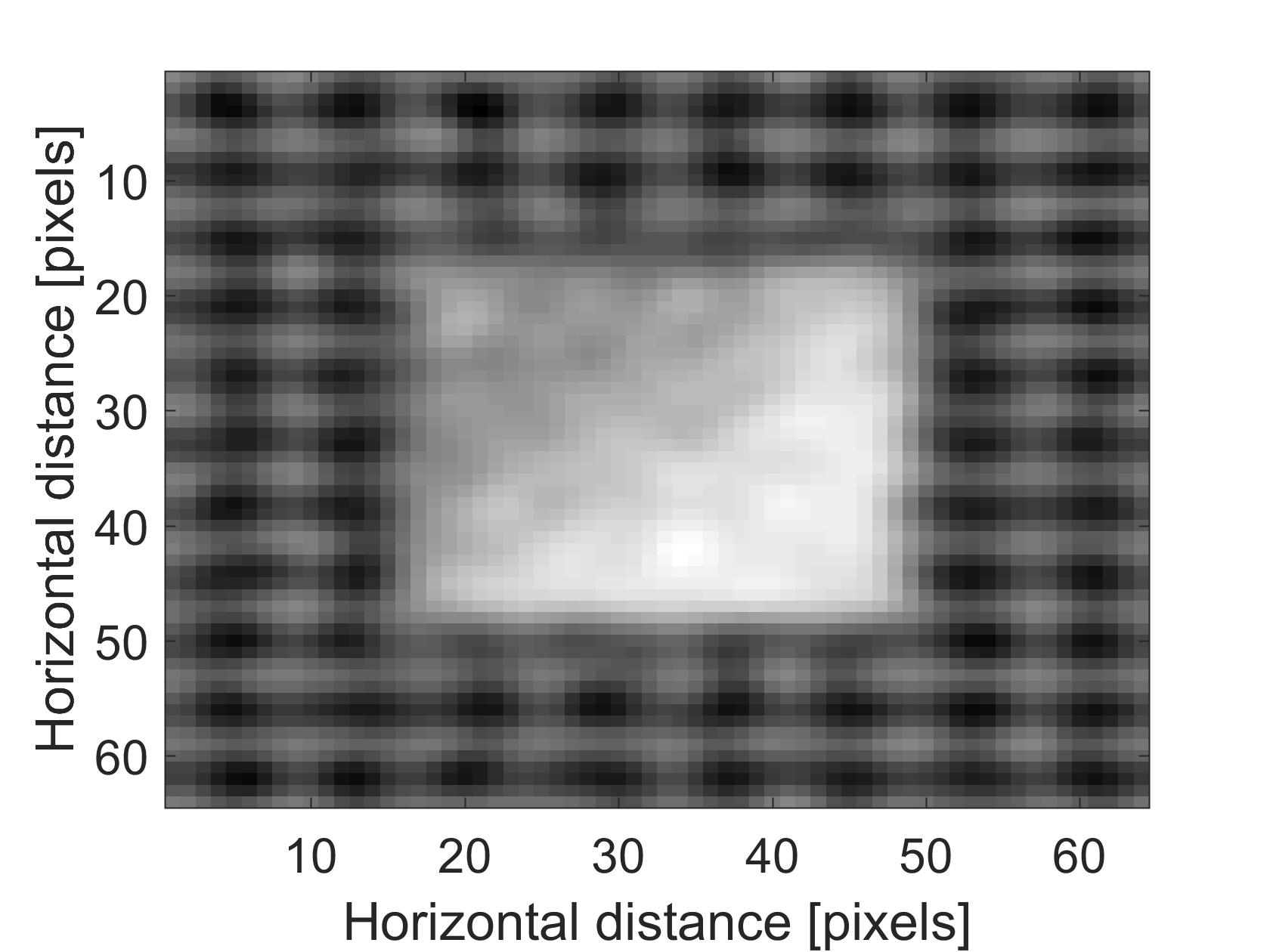}}

\caption{Deconvolution result for Synthetic image, blurred by a Gaussian kernel $h_{\sigma = 5}^{4\times 4}$ and corrupted by Poisson noise corresponding to a SNR of 15 dB. The proposed QAB-PnP algorithm used $\mathcal{E} = 4.1$, $\lambda_0 = 1.3$, $\hbar ^2/2m = 4$ and $\gamma = 1.01$, $\sigma_{\mathcal{QAB}} = 7$.}
\label{fig:sinimag15}
\vspace*{15pt}
\end{figure*}

\begin{figure*}[h!]
\centering 
\subfigure[Clean image]{\includegraphics[width=0.17\textwidth]{Figure/simu/CleanFruits.jpg}}
\subfigure[Corrupted image]{\includegraphics[width=0.17\textwidth]{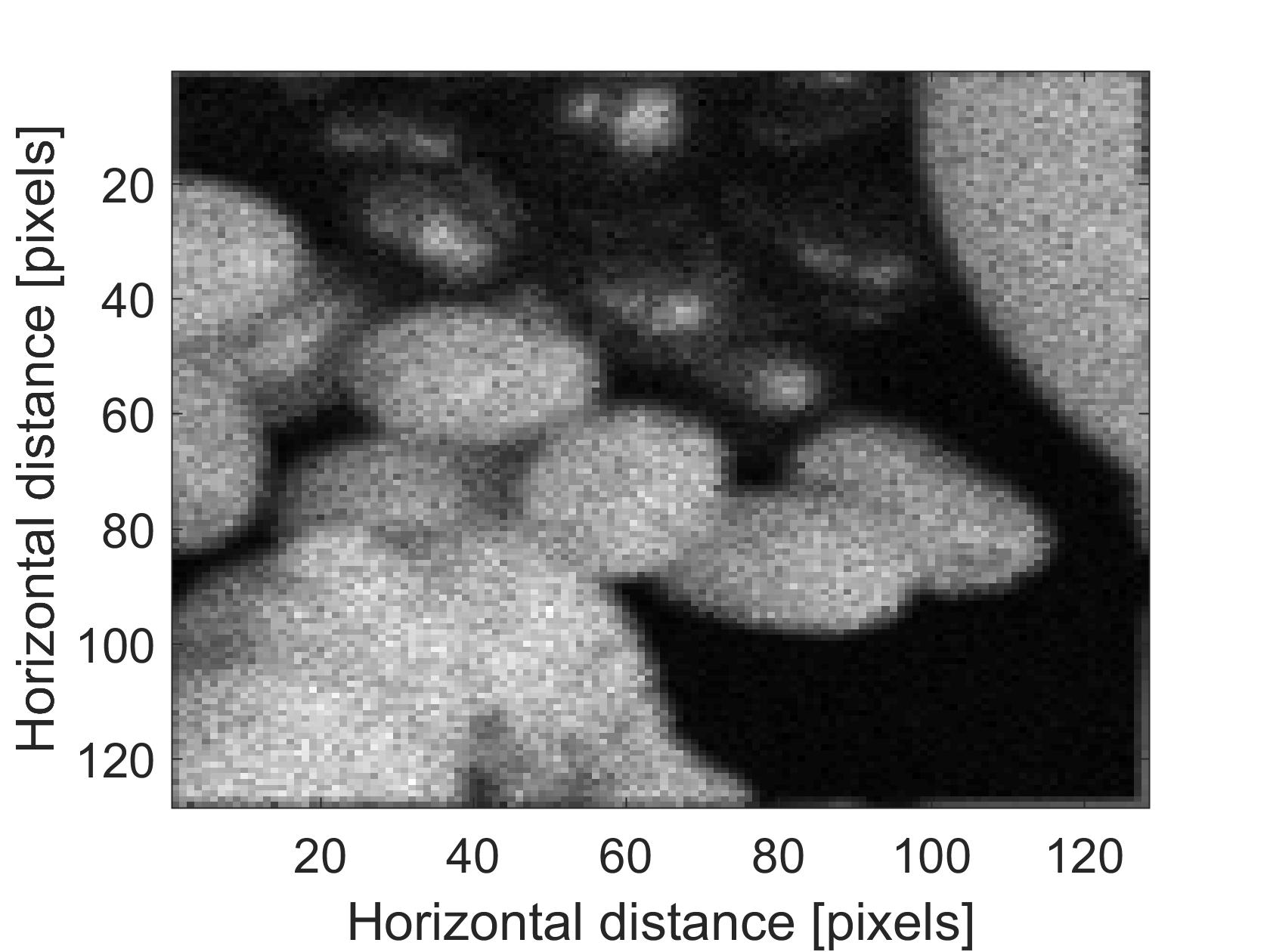}}
\subfigure[TV-ADMM]{\includegraphics[width=0.17\textwidth]{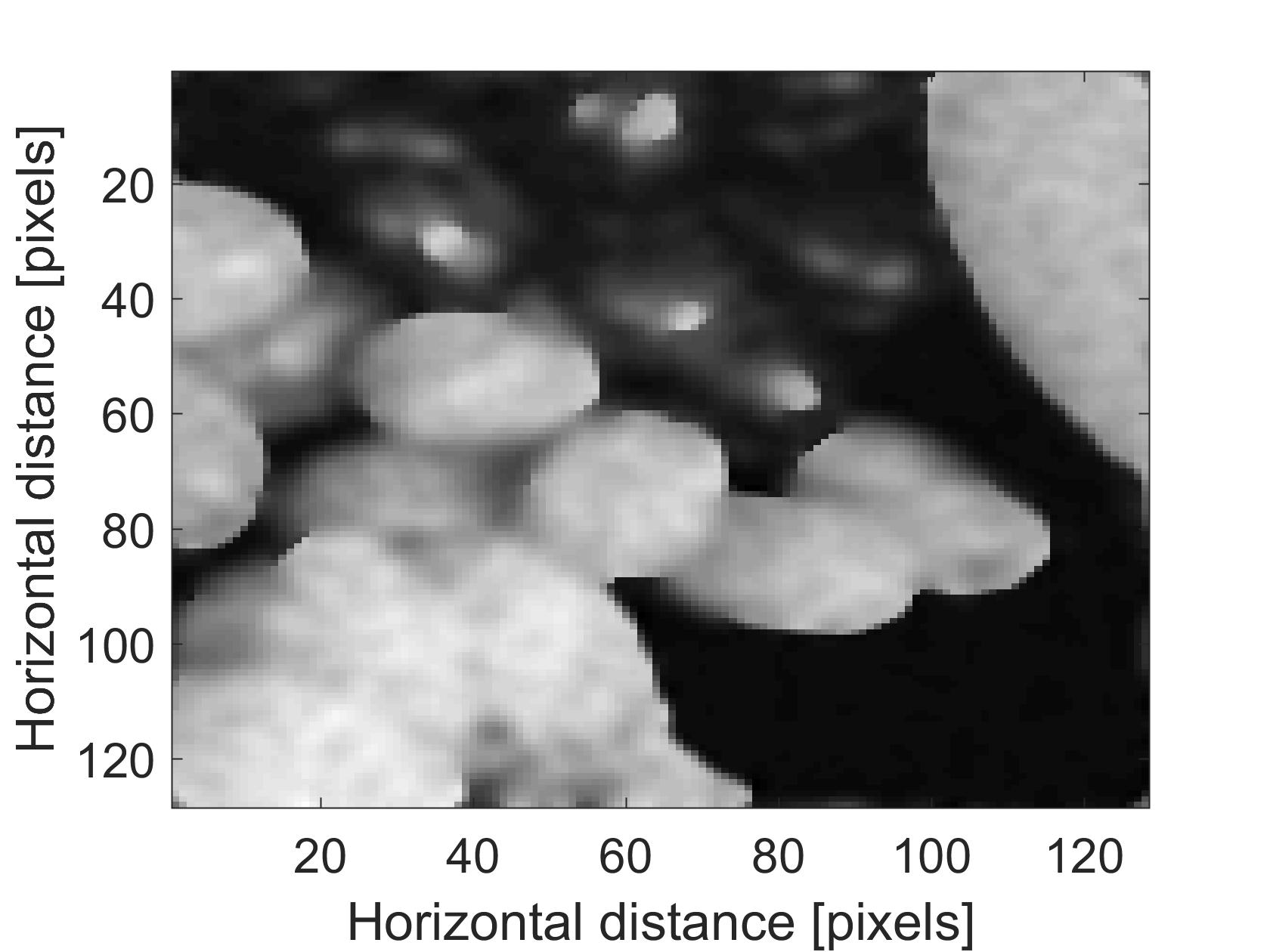}}
\subfigure[ADMM+BM3D]{\includegraphics[width=0.17\textwidth]{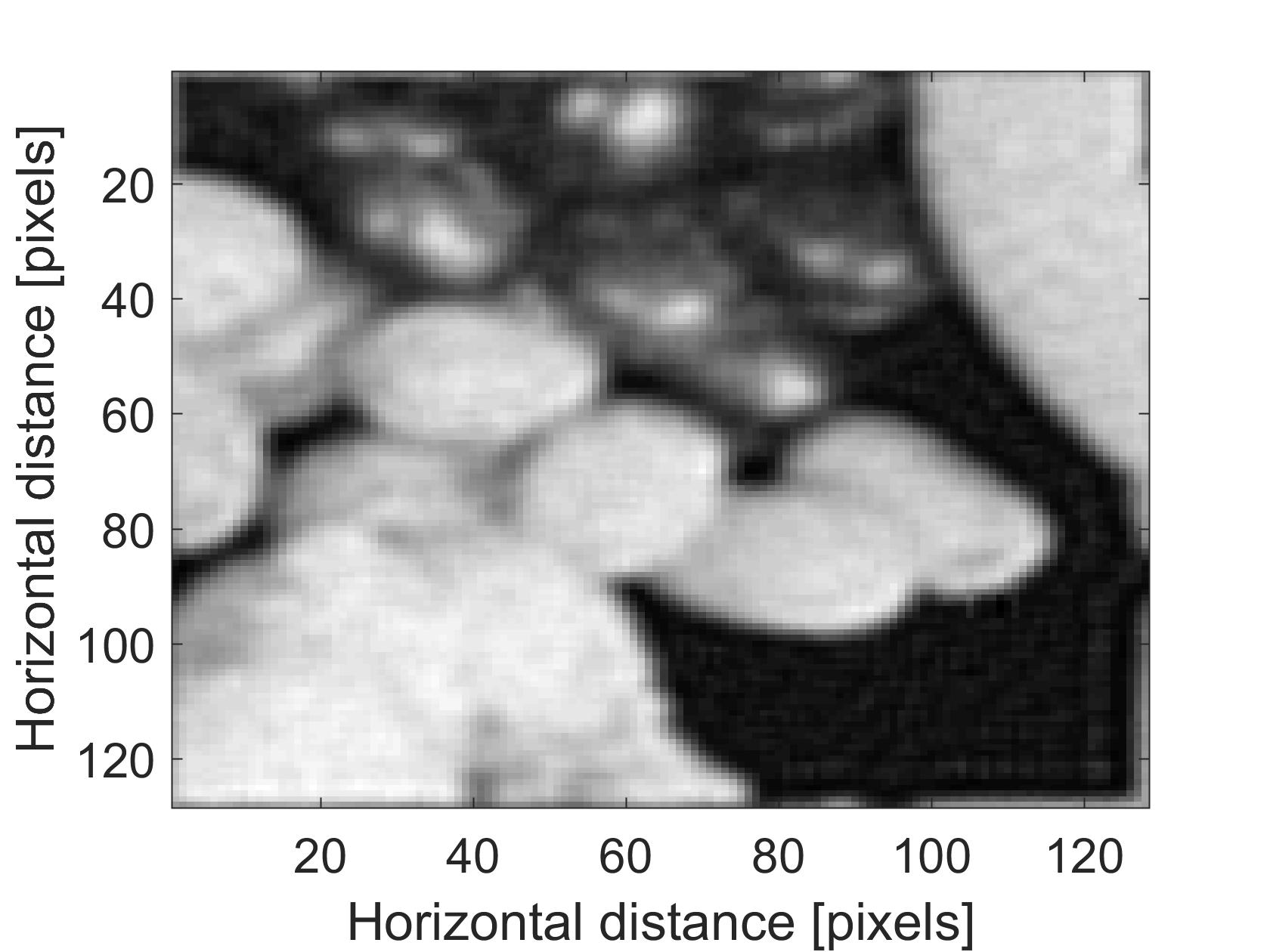}}
\subfigure[ADMM+TNRD]{\includegraphics[width=0.17\textwidth]{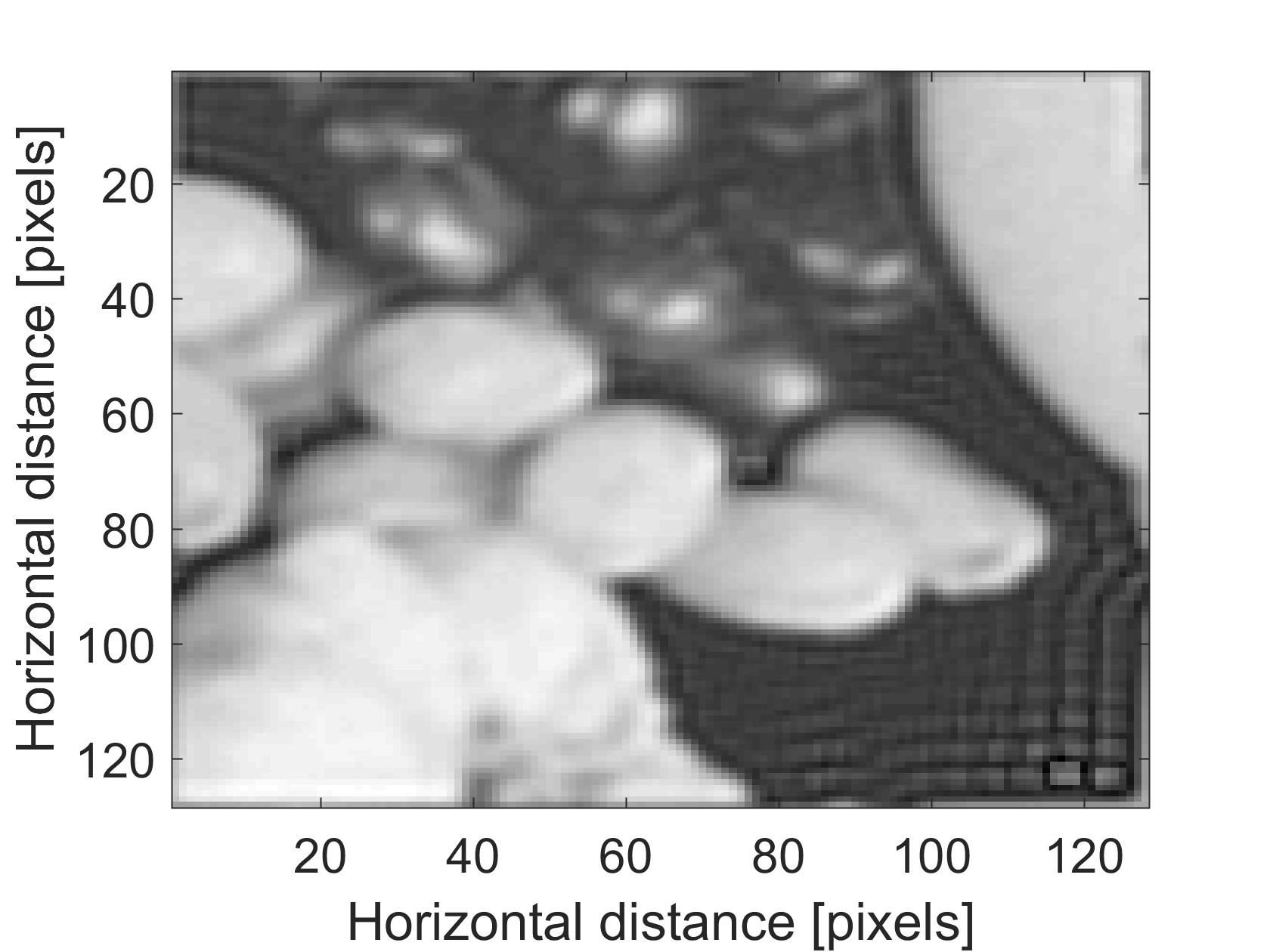}}
\subfigure[ADMM+VST+TNRD]{\includegraphics[width=0.17\textwidth]{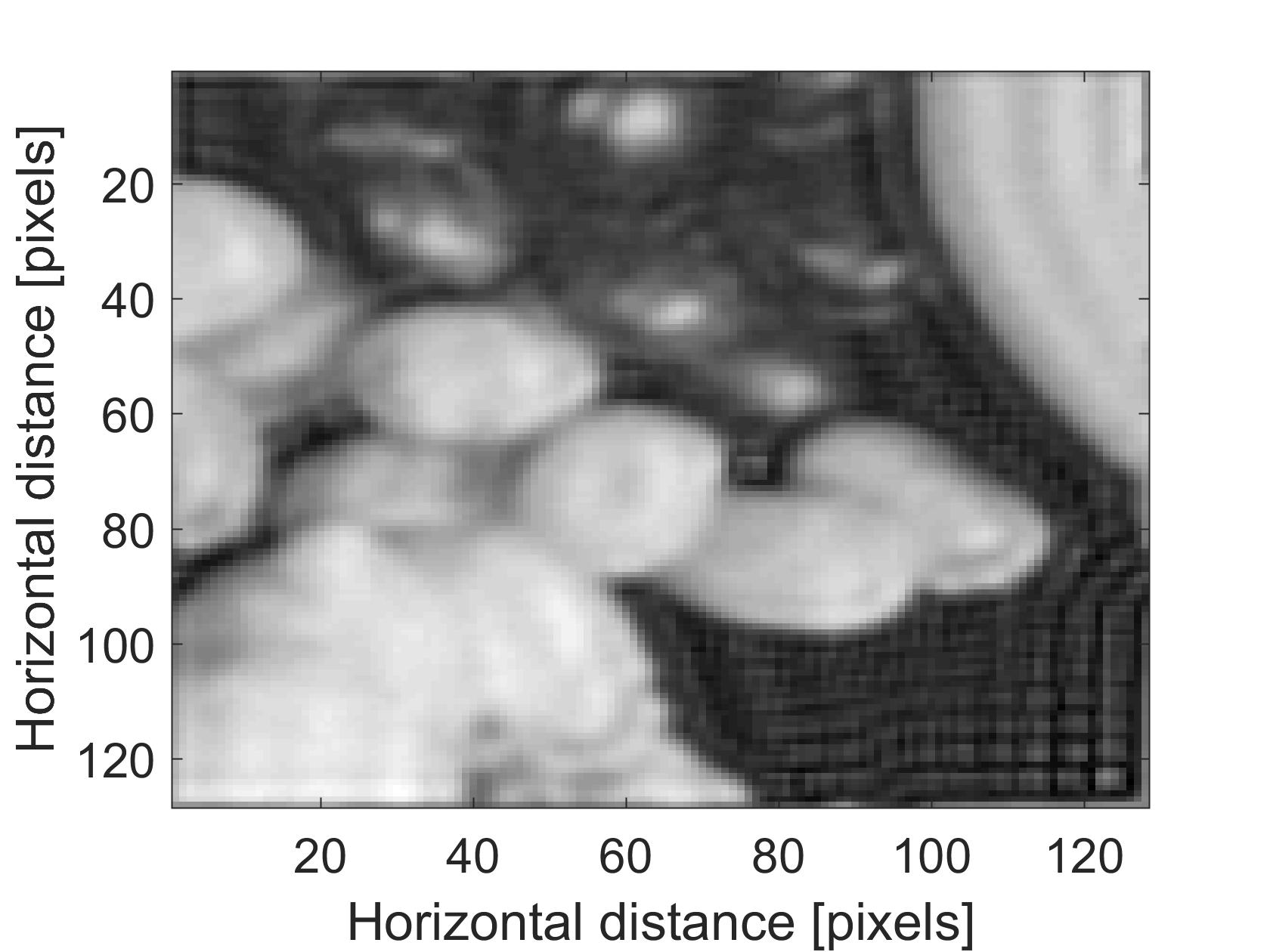}}
\subfigure[P$^4$IP]{\includegraphics[width=0.17\textwidth]{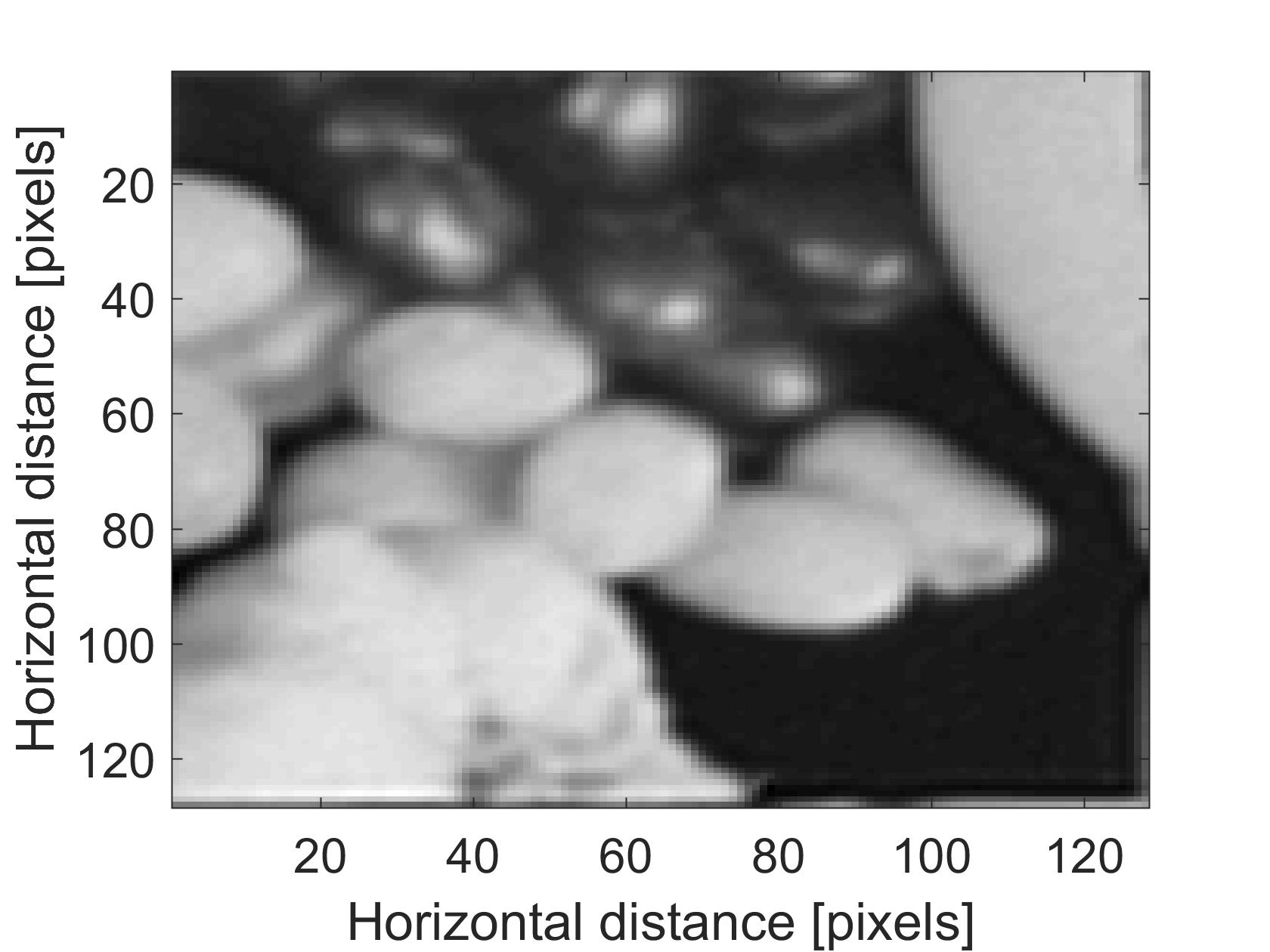}}
\subfigure[QAB-PnP]{\includegraphics[width=0.17\textwidth]{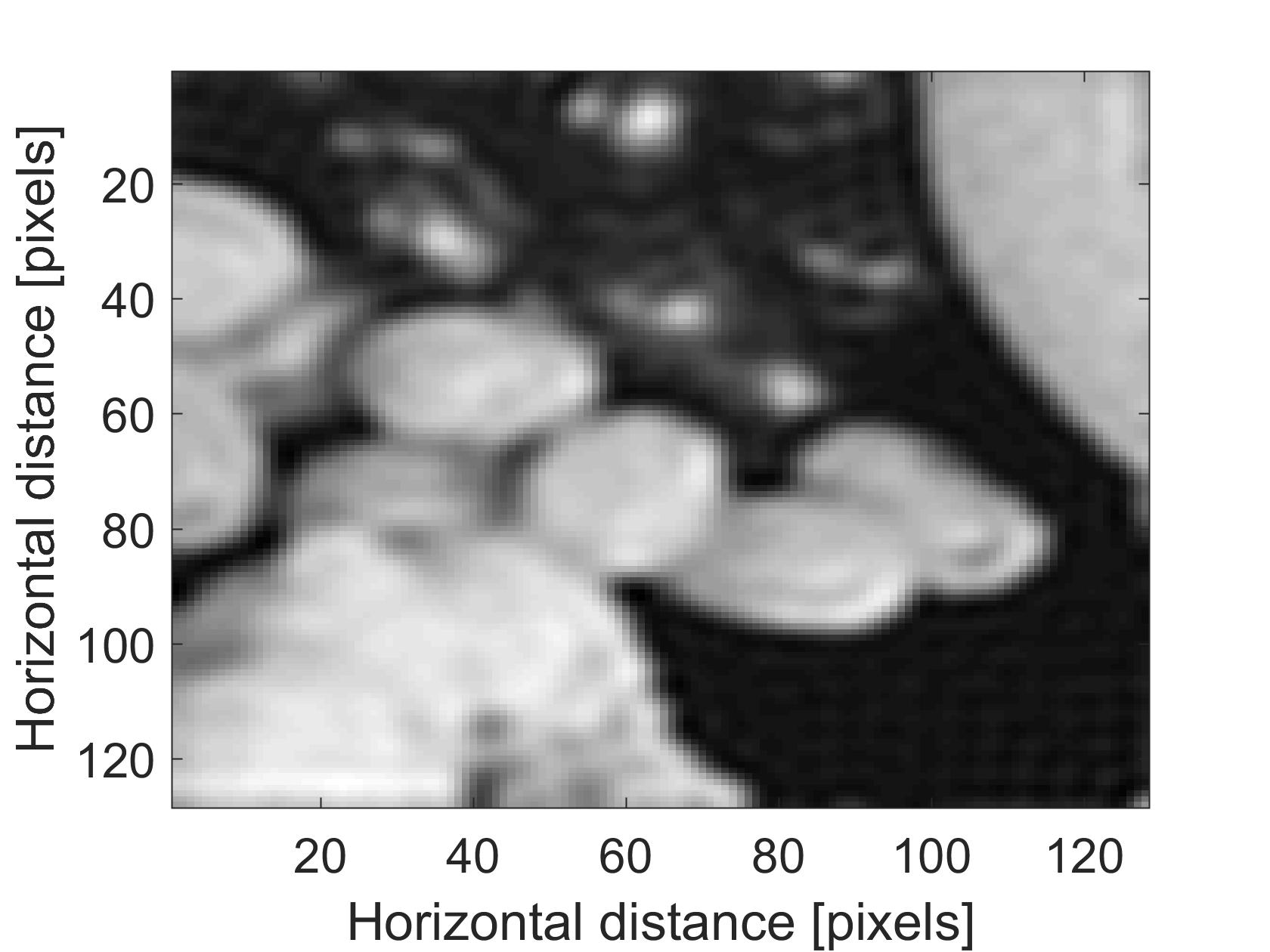}}

\caption{Deconvolution result for Fruits image, blurred by a Gaussian kernel $h_{\sigma = 3}^{4\times 4}$ and corrupted by Poisson noise corresponding to a SNR of 20 dB. The proposed QAB-PnP algorithm used $\mathcal{E} = 4.5$, $\lambda_0 = 3.15$, $\hbar ^2/2m = 4.3$ and $\gamma = 1.01$, $\sigma_{\mathcal{QAB}} = 8$.}
\label{fig:fruimag20}
\end{figure*}

\begin{figure*}[h!]
\centering

\begin{turn}{90}$\overbrace{~~~~~~~~~~~~~~~~~~~~~~~~~~~~~~~~~}^{\textrm{TV-ADMM}}$\end{turn}
$\overbrace{
\subfigure[\scriptsize Clean Lena image]{\includegraphics[width=0.21\textwidth]{Figure/simu/CleanLena.jpg}}
}^{\textrm{Clean image}}$ \hfill
$\overbrace{
\subfigure[\scriptsize PSNR = 26.05 dB, SSIM = 0.68]{\includegraphics[width=0.21\textwidth]{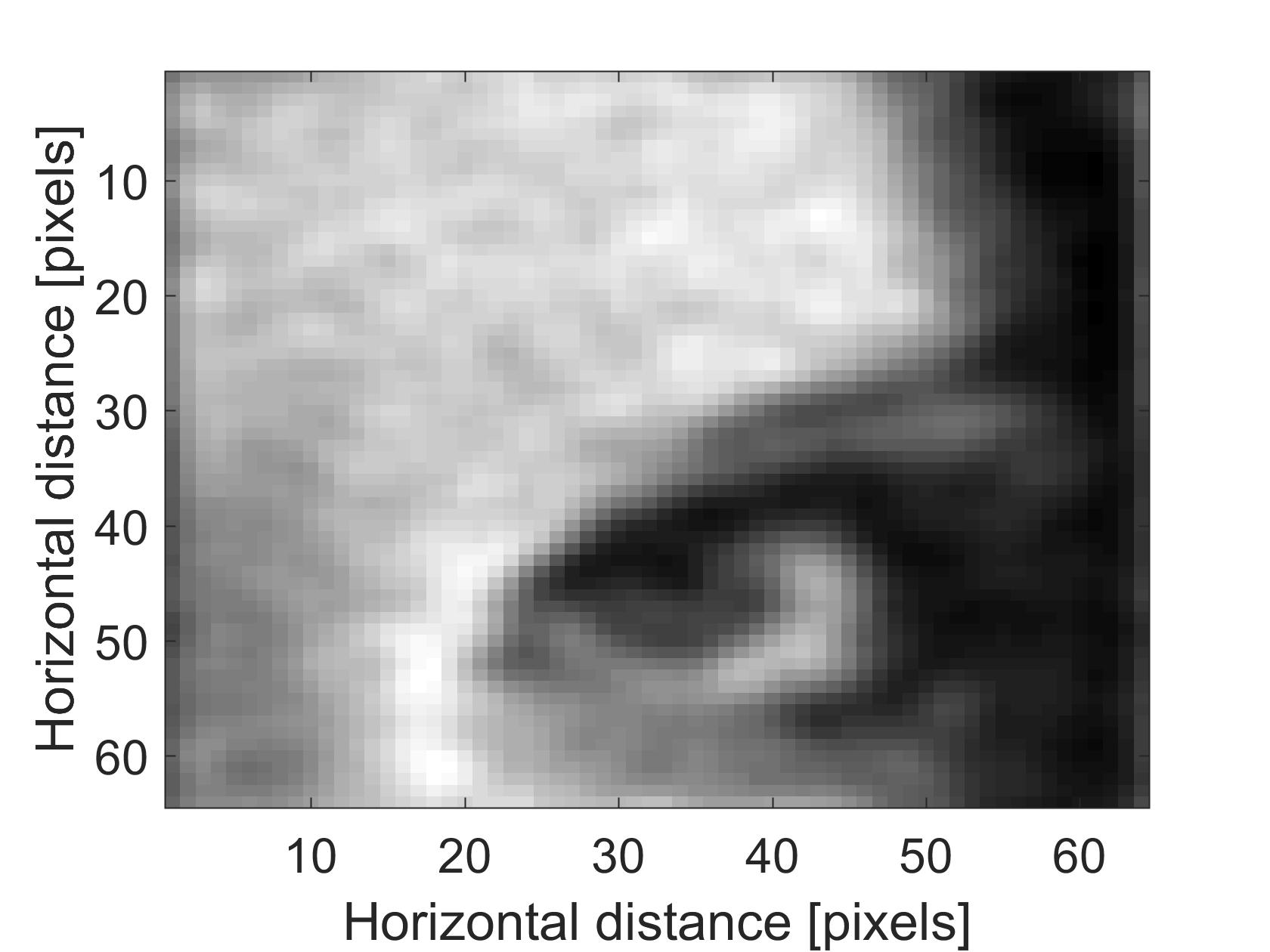}}
}^{\textrm{Best result}}$ \hfill
$\overbrace{
\subfigure[\scriptsize PSNR = 23.68 dB, SSIM = 0.61]{\includegraphics[width=0.21\textwidth]{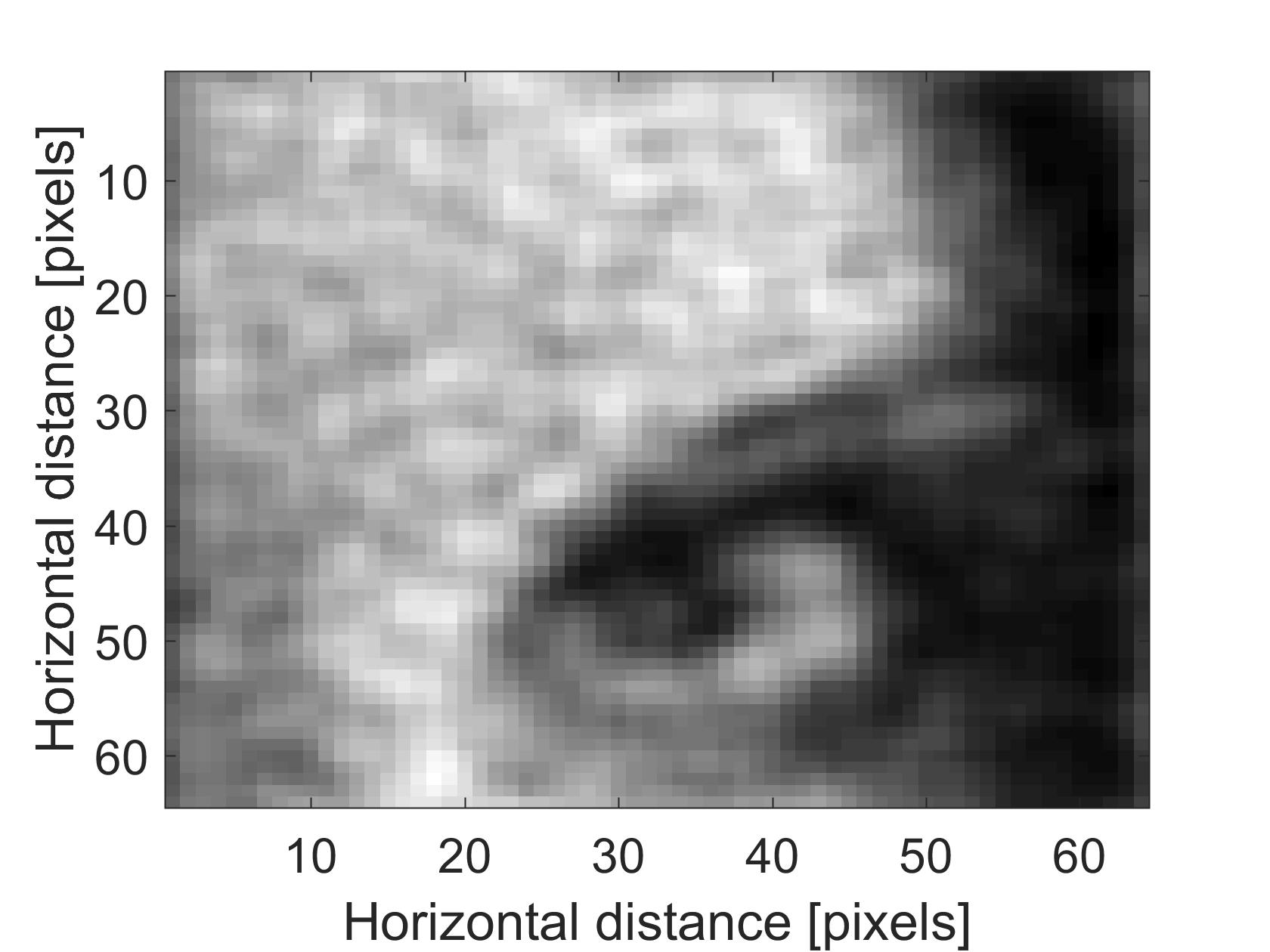}}
}^{\textrm{Worst result}}$ \hfill
$\overbrace{
\subfigure[\scriptsize PSNR = 24.88 dB, SSIM = 0.64]{\includegraphics[width=0.21\textwidth]{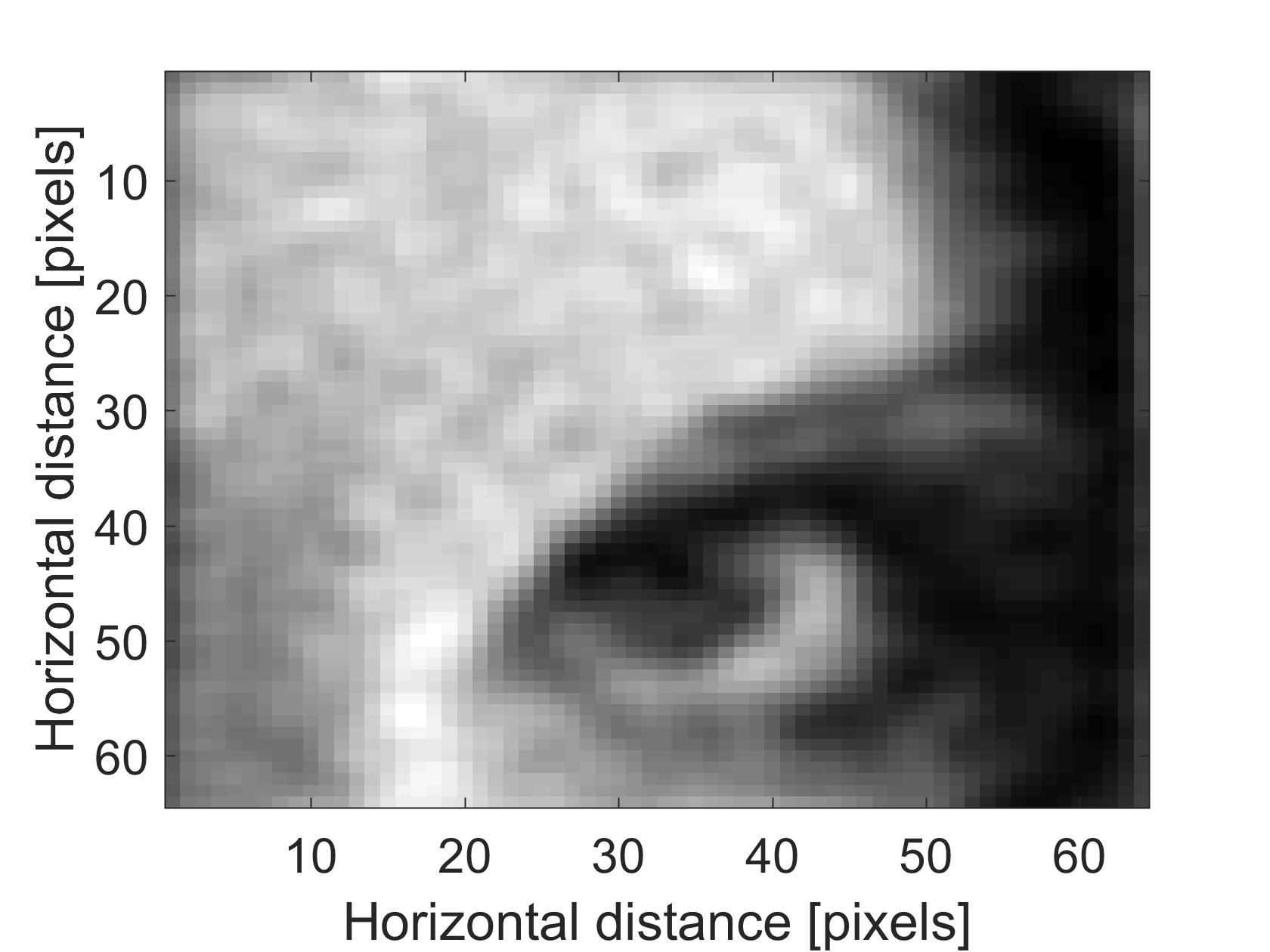}}
}^{\textrm{Intermediate result}}$

\begin{turn}{90}$\overbrace{~~~~~~~~~~~~~~~~~~~~~~~~~~~~~~~~~}^{\textrm{ADMM+BM3D}}$\end{turn}
\subfigure[\scriptsize Clean Lena image]{\includegraphics[width=0.21\textwidth]{Figure/simu/CleanLena.jpg}} \hfill
\subfigure[\scriptsize PSNR = 24.35 dB, SSIM = 0.69]{\includegraphics[width=0.21\textwidth]{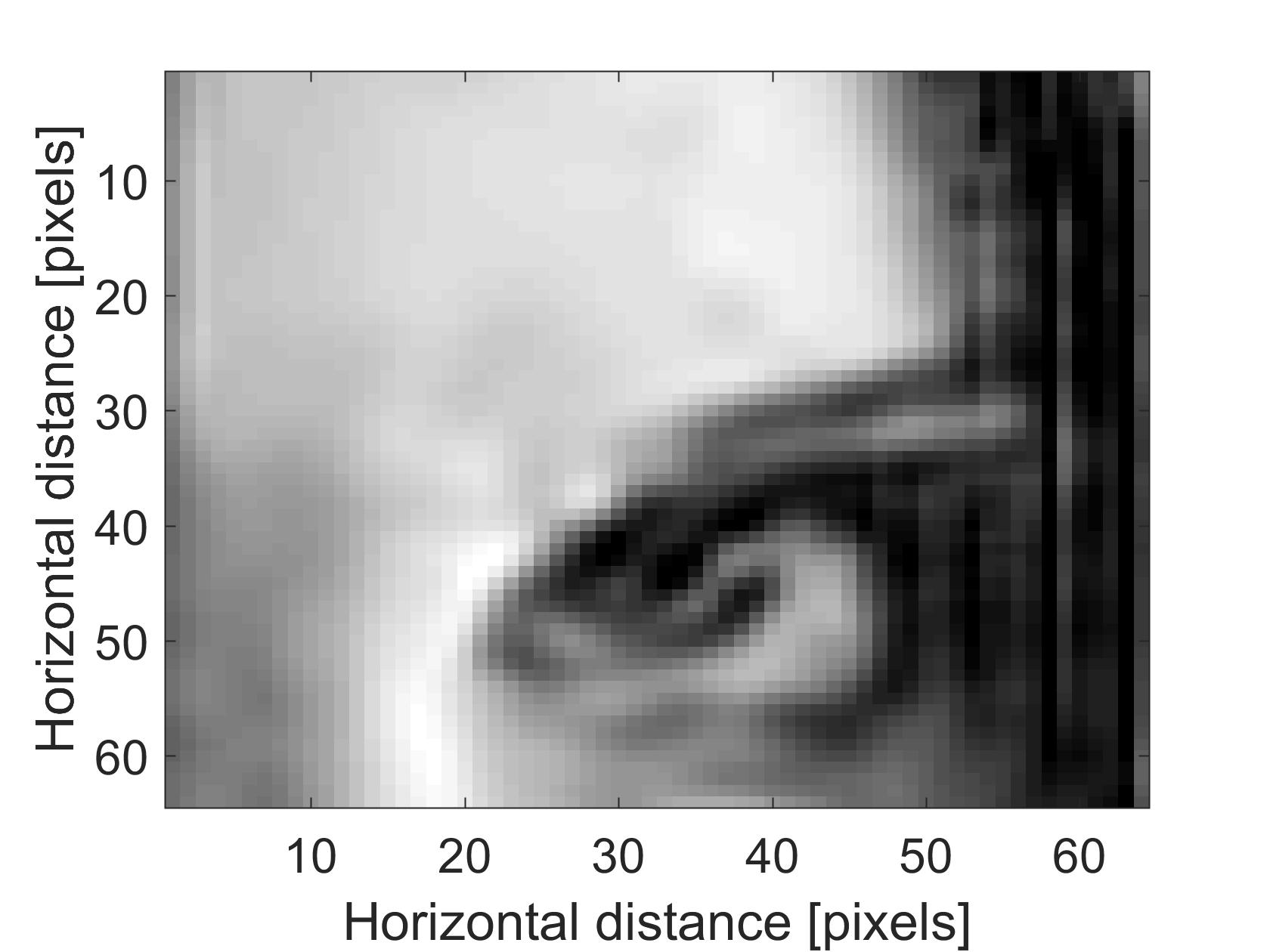}}  \hfill
\subfigure[\scriptsize PSNR = 22.42 dB, SSIM = 0.58]{\includegraphics[width=0.21\textwidth]{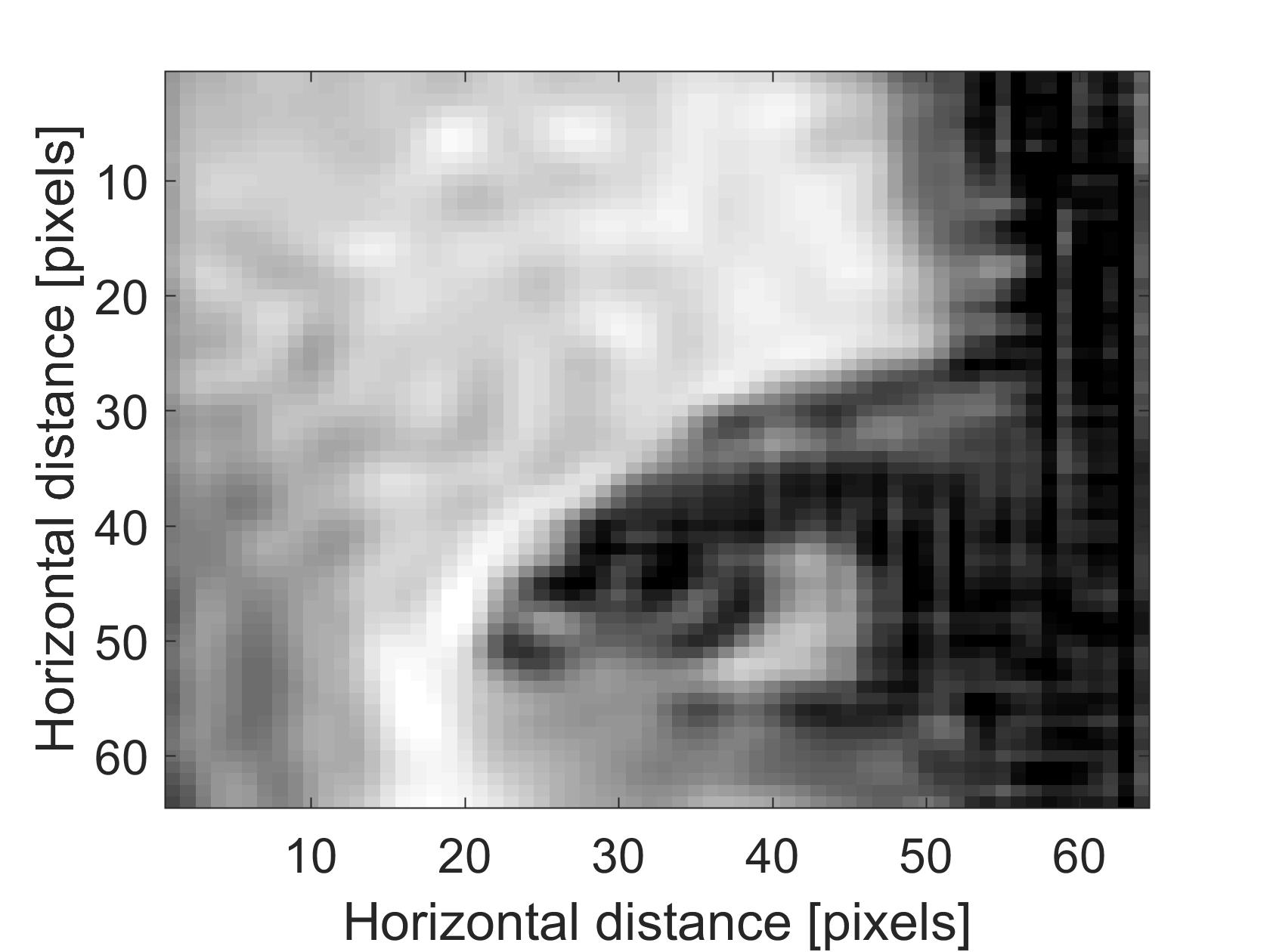}} \hfill
\subfigure[\scriptsize PSNR = 23.87 dB, SSIM = 0.65]{\includegraphics[width=0.21\textwidth]{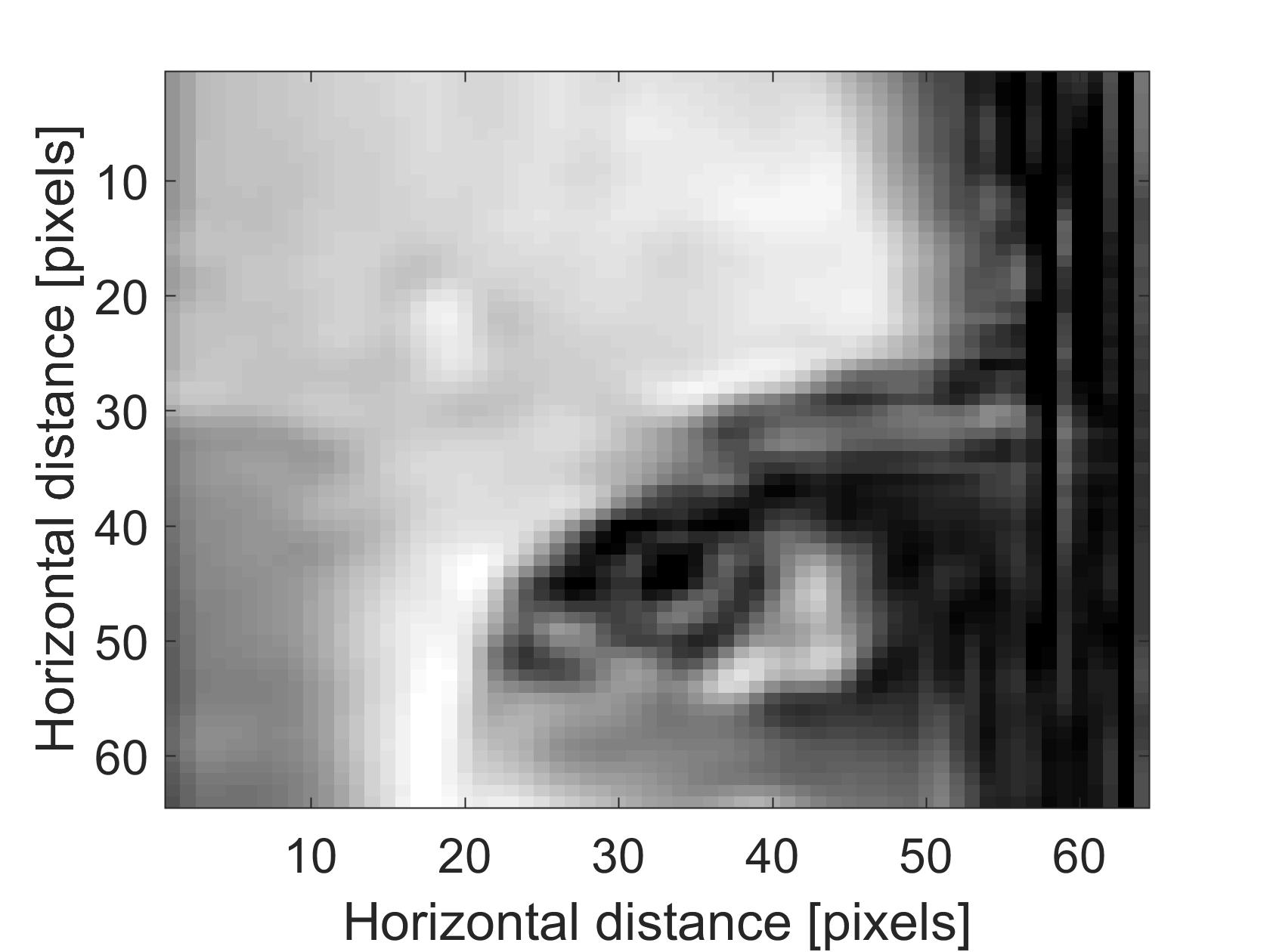}}

\begin{turn}{90}$\overbrace{~~~~~~~~~~~~~~~~~~~~~~~~~~~~~~~~~}^{\textrm{ADMM+TNRD}}$\end{turn}
\subfigure[\scriptsize Clean Lena image]{\includegraphics[width=0.21\textwidth]{Figure/simu/CleanLena.jpg}} \hfill
\subfigure[\scriptsize PSNR = 25.71 dB, SSIM = 0.72]{\includegraphics[width=0.21\textwidth]{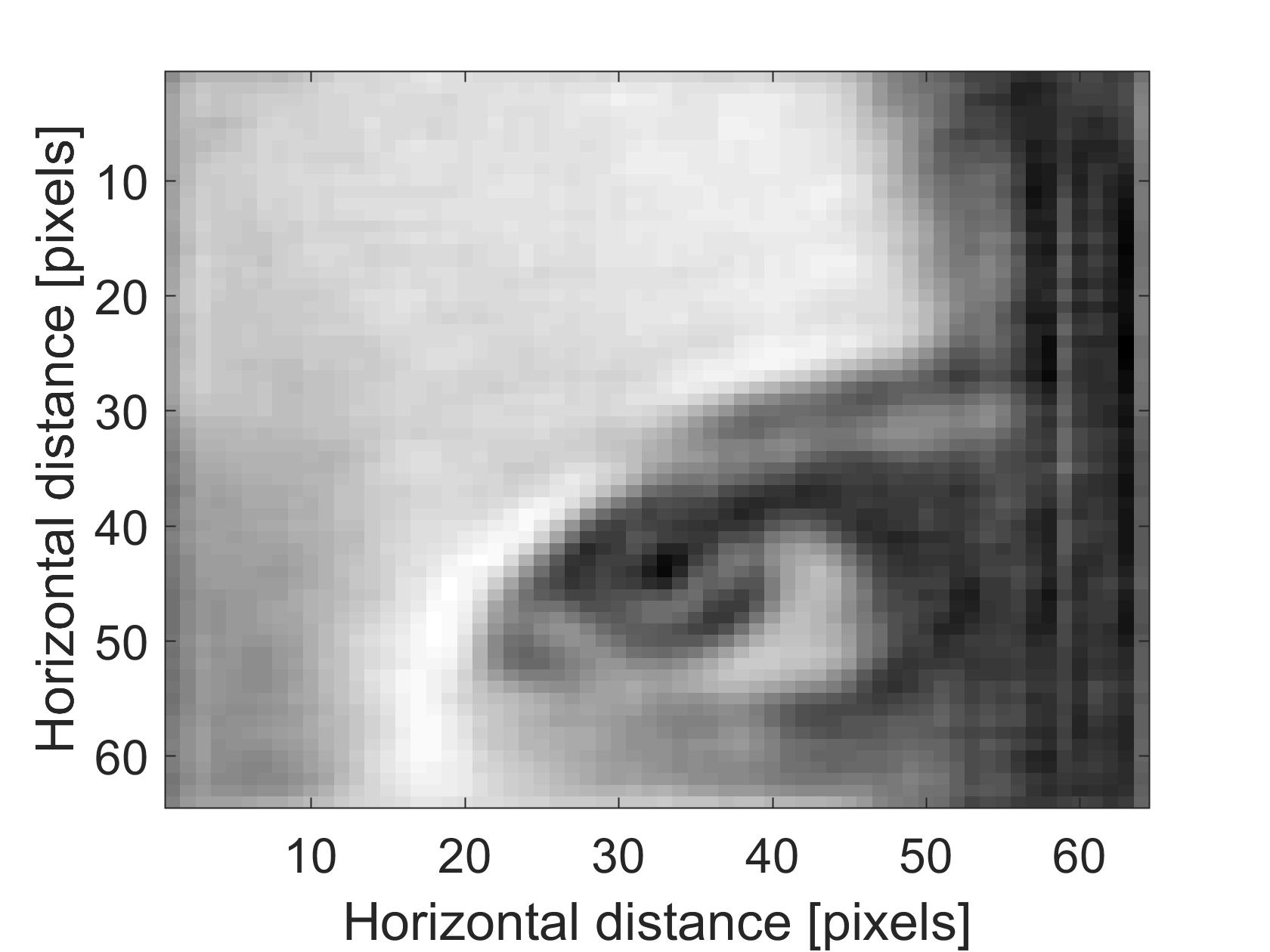}}  \hfill
\subfigure[\scriptsize PSNR = 23.38 dB, SSIM = 0.59]{\includegraphics[width=0.21\textwidth]{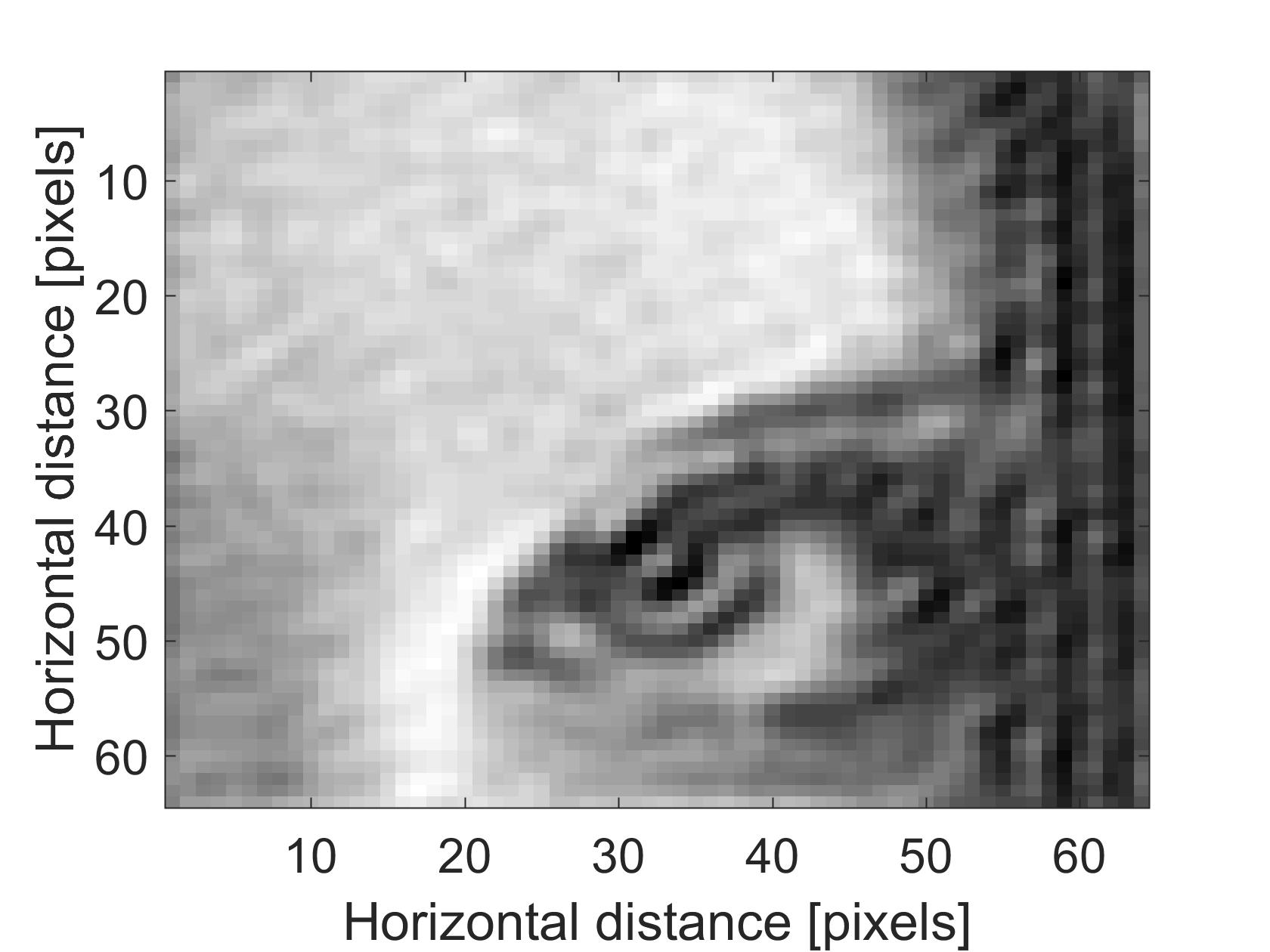}} \hfill
\subfigure[\scriptsize PSNR = 24.23 dB, SSIM = 0.67]{\includegraphics[width=0.21\textwidth]{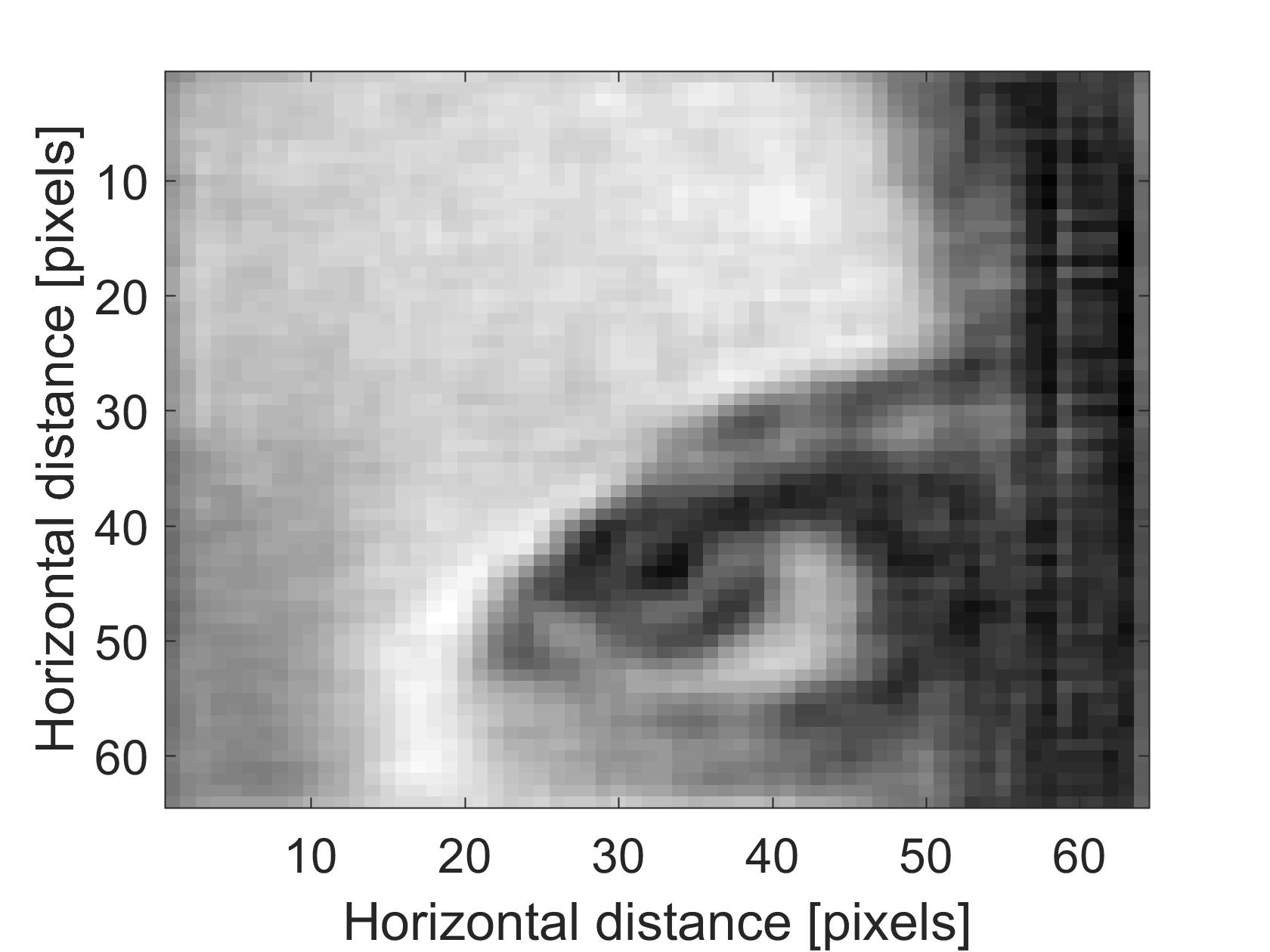}}

\begin{turn}{90}$\overbrace{~~~~~~~~~~~~~~~~~~~~~~~~~~~~~~~~~}^{\textrm{ADMM+VST+TNRD}}$\end{turn}
\subfigure[\scriptsize Clean Lena image]{\includegraphics[width=0.21\textwidth]{Figure/simu/CleanLena.jpg}} \hfill
\subfigure[\scriptsize PSNR = 25.43 dB, SSIM = 0.64]{\includegraphics[width=0.21\textwidth]{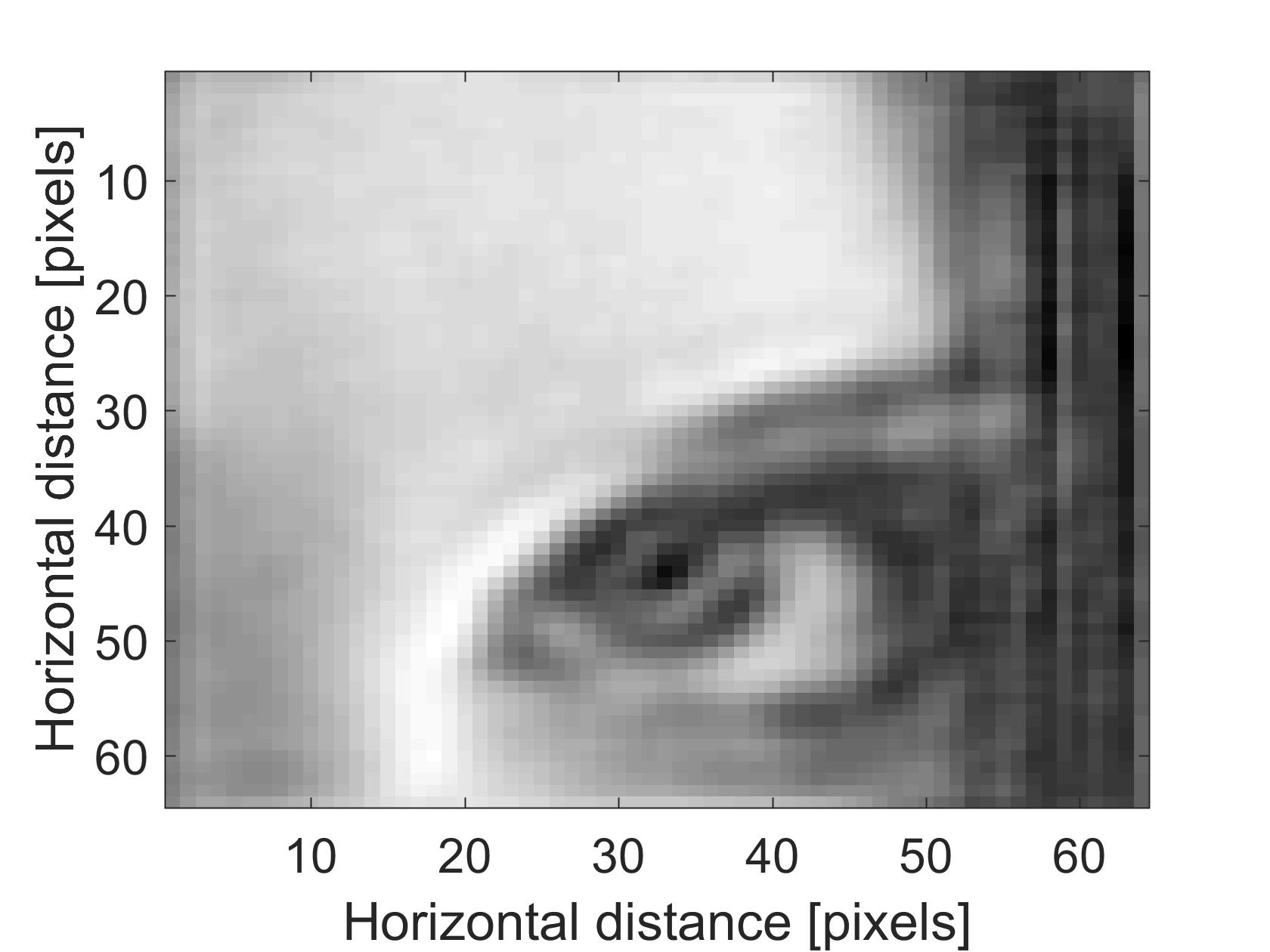}}  \hfill
\subfigure[\scriptsize PSNR = 23.66 dB, SSIM = 0.53]{\includegraphics[width=0.21\textwidth]{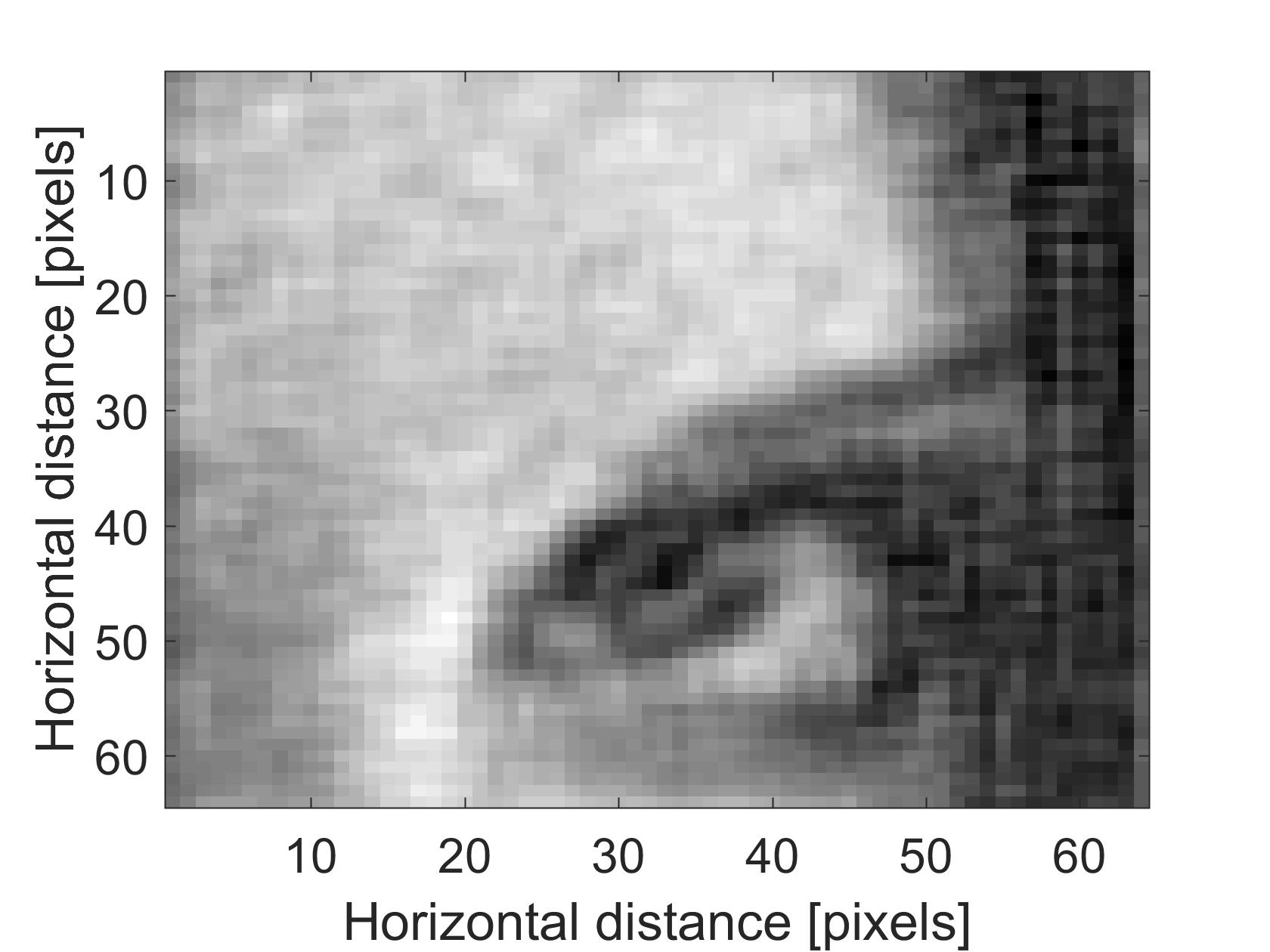}} \hfill
\subfigure[\scriptsize PSNR = 24.22 dB, SSIM = 0.61]{\includegraphics[width=0.21\textwidth]{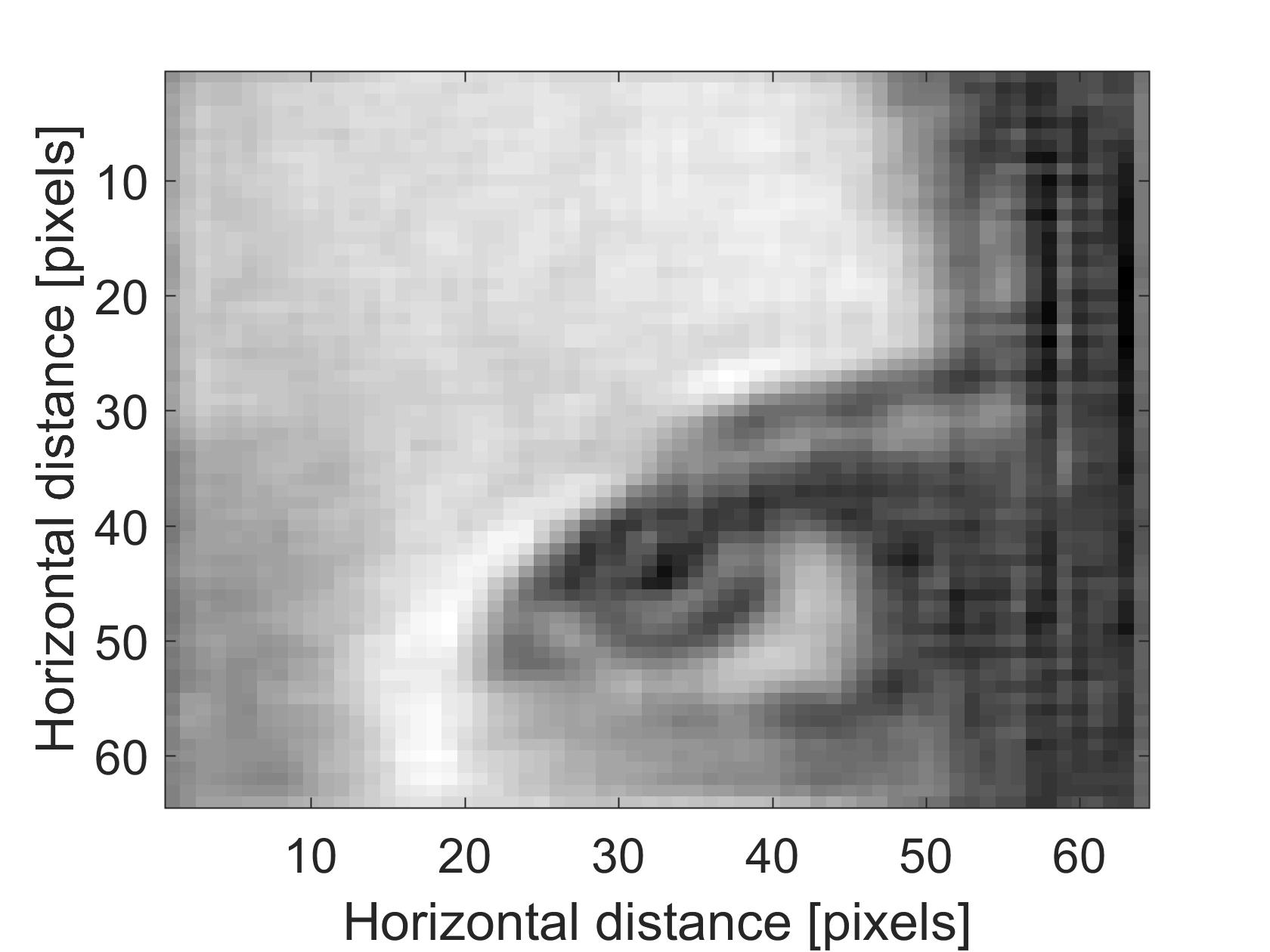}}

\begin{turn}{90}$\overbrace{~~~~~~~~~~~~~~~~~~~~~~~~~~~~~~~~~}^{\textrm{P$^4$IP}}$\end{turn}
\subfigure[\scriptsize Clean Lena image]{\includegraphics[width=0.21\textwidth]{Figure/simu/CleanLena.jpg}} \hfill
\subfigure[\scriptsize PSNR = 27.40 dB, SSIM = 0.81]{\includegraphics[width=0.21\textwidth]{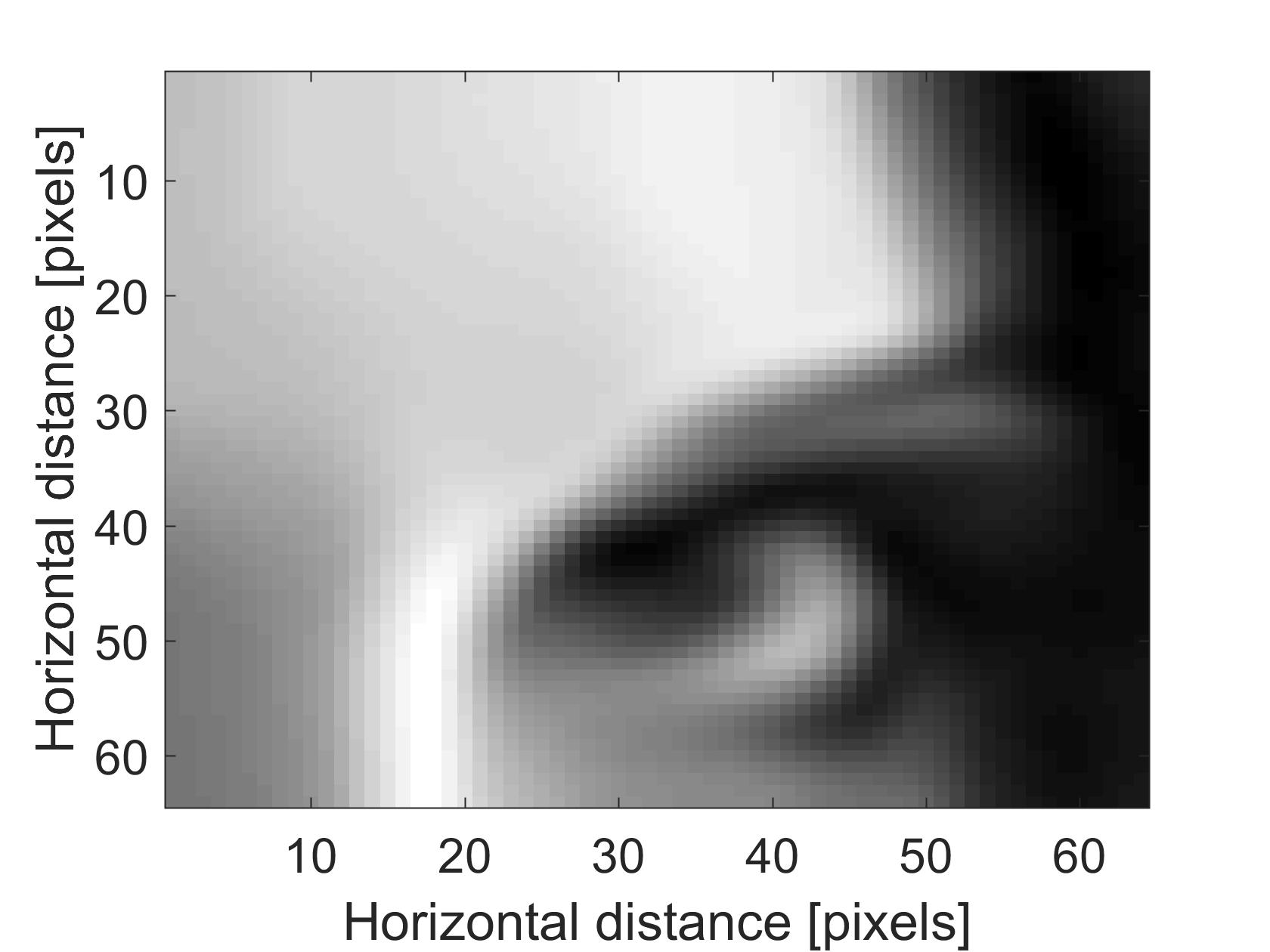}} \hfill
\subfigure[\scriptsize PSNR = 15.68 dB, SSIM = 0.56]{\includegraphics[width=0.21\textwidth]{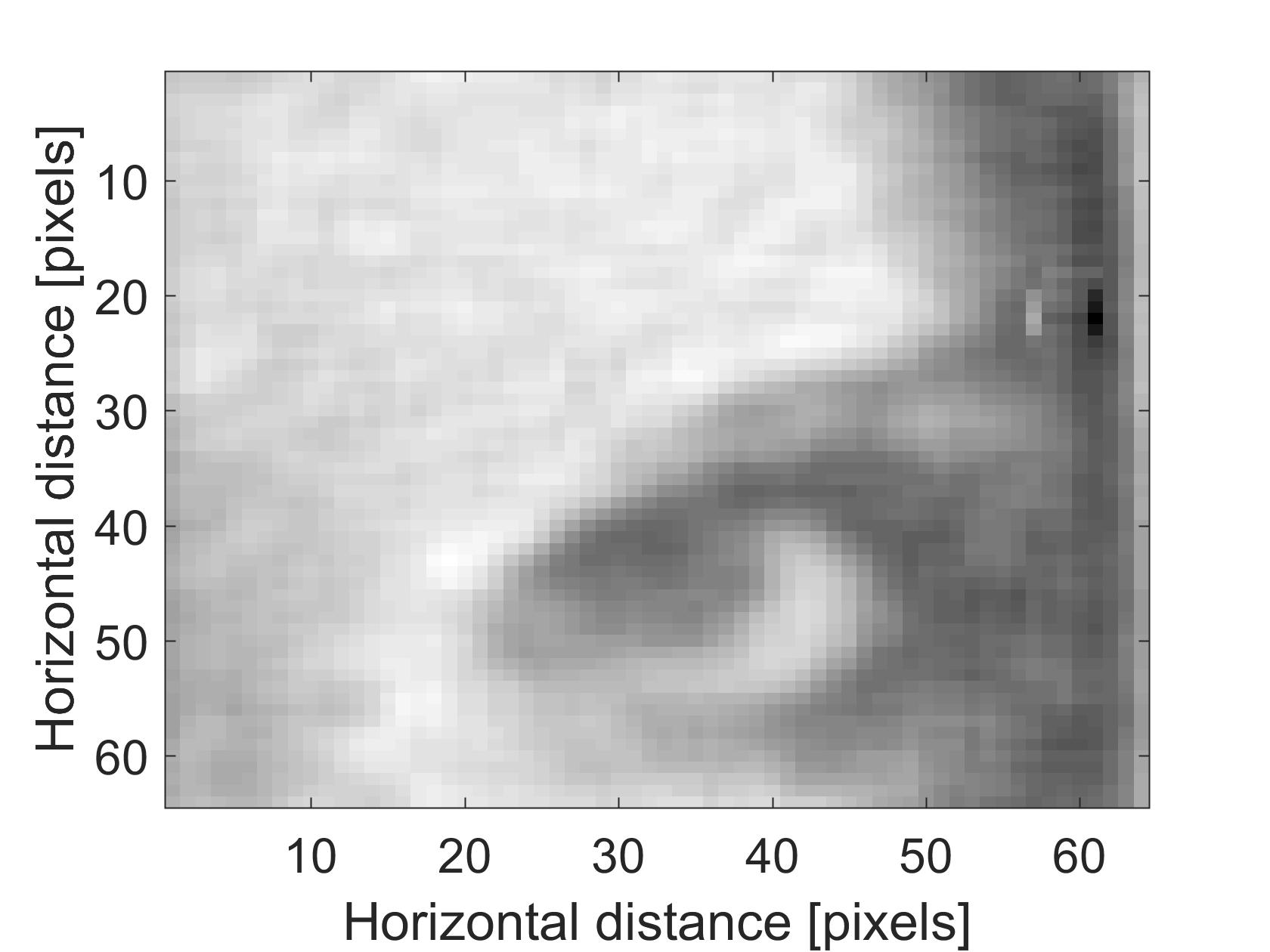}} \hfill
\subfigure[\scriptsize PSNR = 22.23 dB, SSIM = 0.68]{\includegraphics[width=0.21\textwidth]{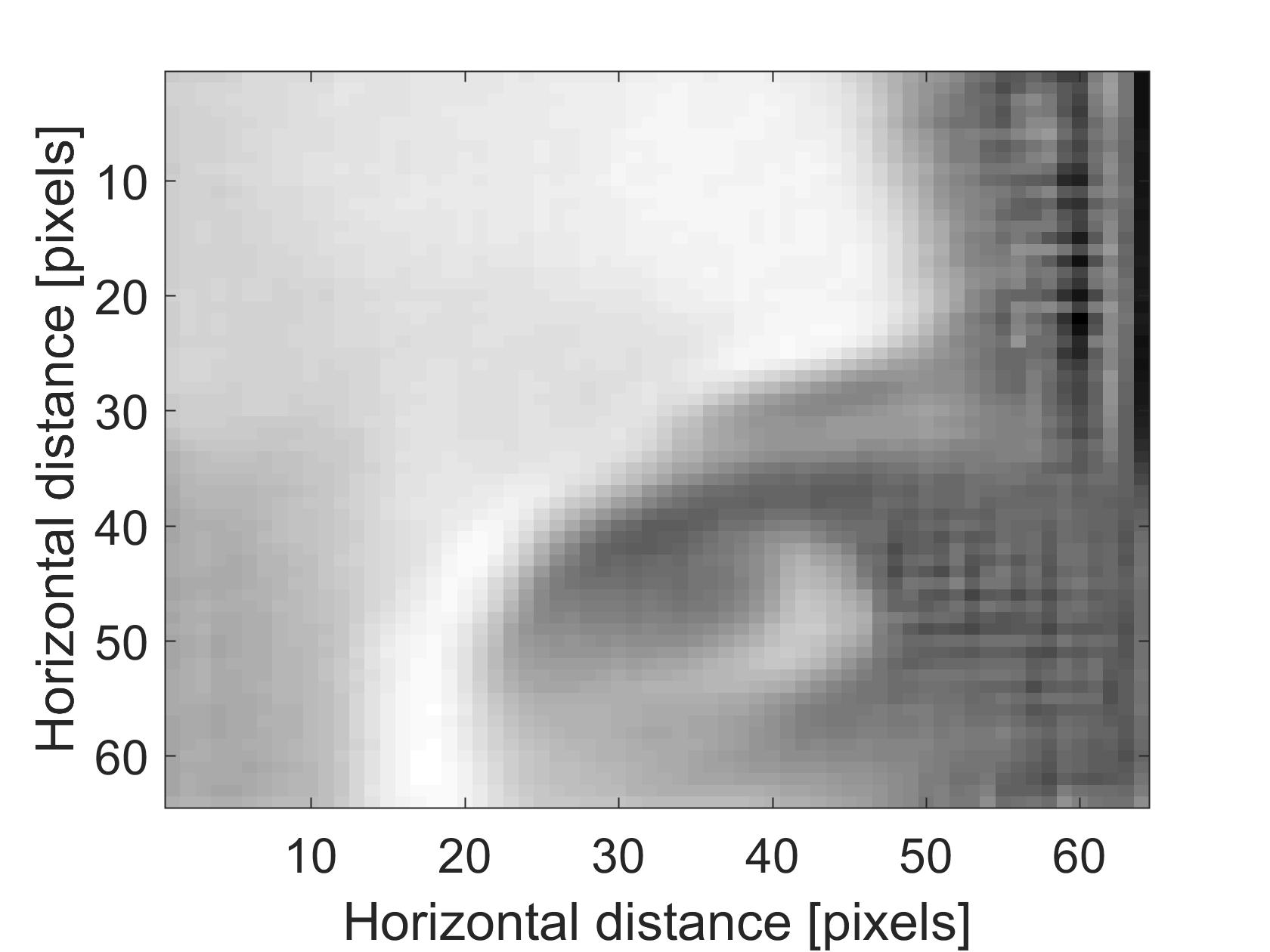}}

\begin{turn}{90}$\overbrace{~~~~~~~~~~~~~~~~~~~~~~~~~~~~~~~~~}^{\textrm{QAB-PnP}}$\end{turn}
\subfigure[\scriptsize Clean Lena image]{\includegraphics[width=0.21\textwidth]{Figure/simu/CleanLena.jpg}} \hfill
\subfigure[\scriptsize PSNR = 27.71 dB, SSIM = 0.77]{\includegraphics[width=0.21\textwidth]{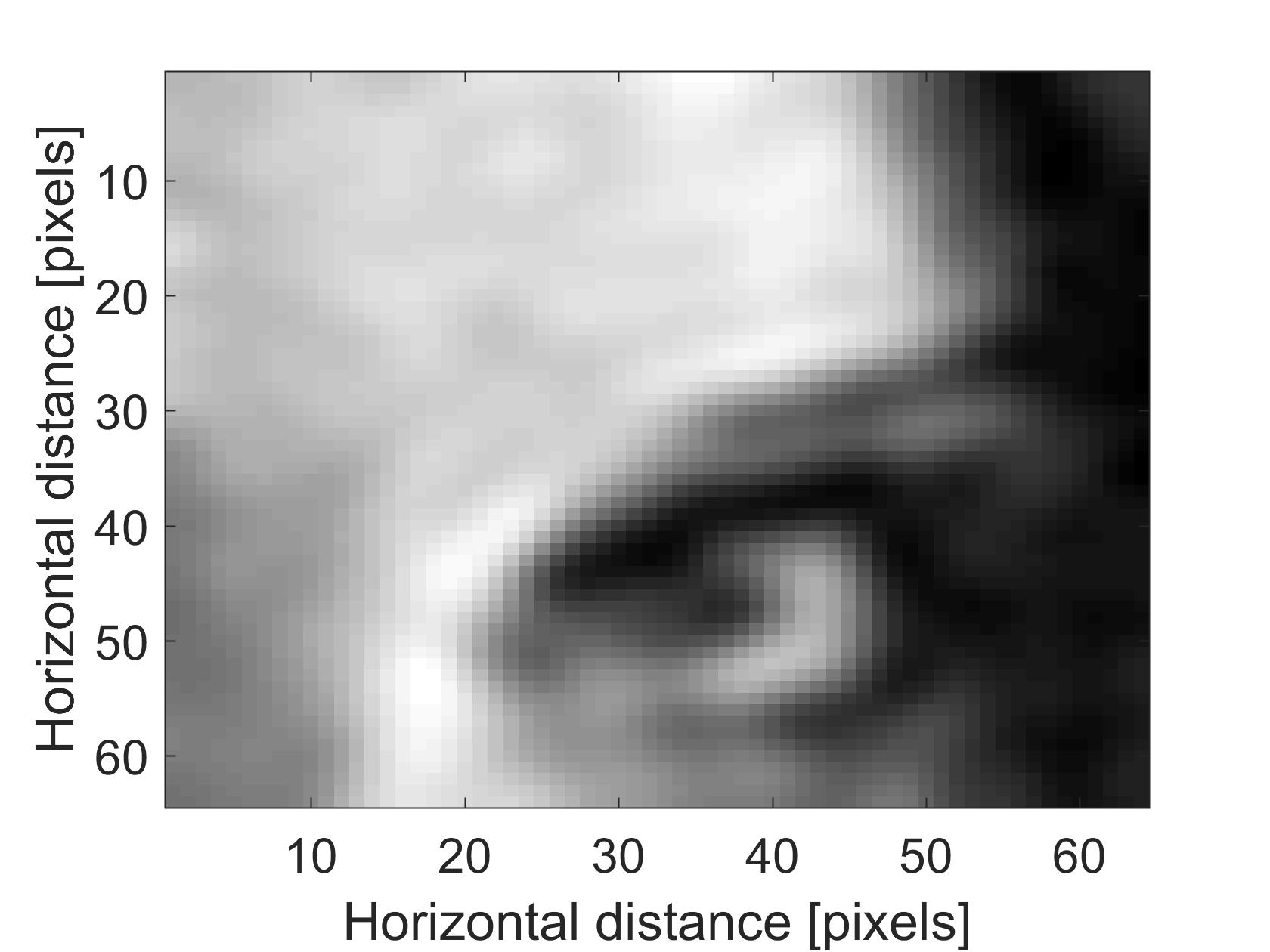}} \hfill
\subfigure[\scriptsize PSNR = 24.93 dB, SSIM = 0.74]{\includegraphics[width=0.21\textwidth]{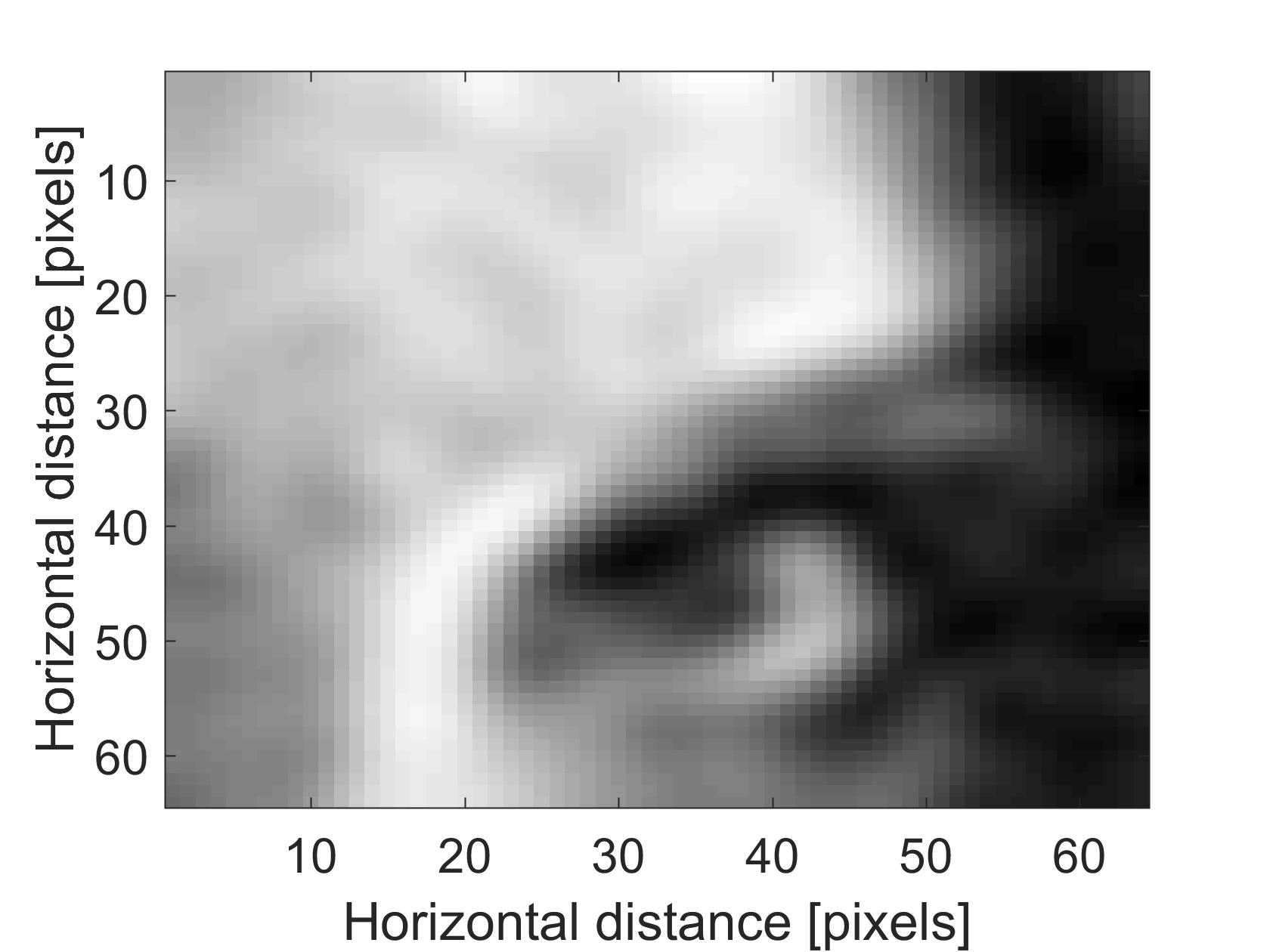}} \hfill
\subfigure[\scriptsize PSNR = 26.96 dB, SSIM = 0.76]{\includegraphics[width=0.21\textwidth]{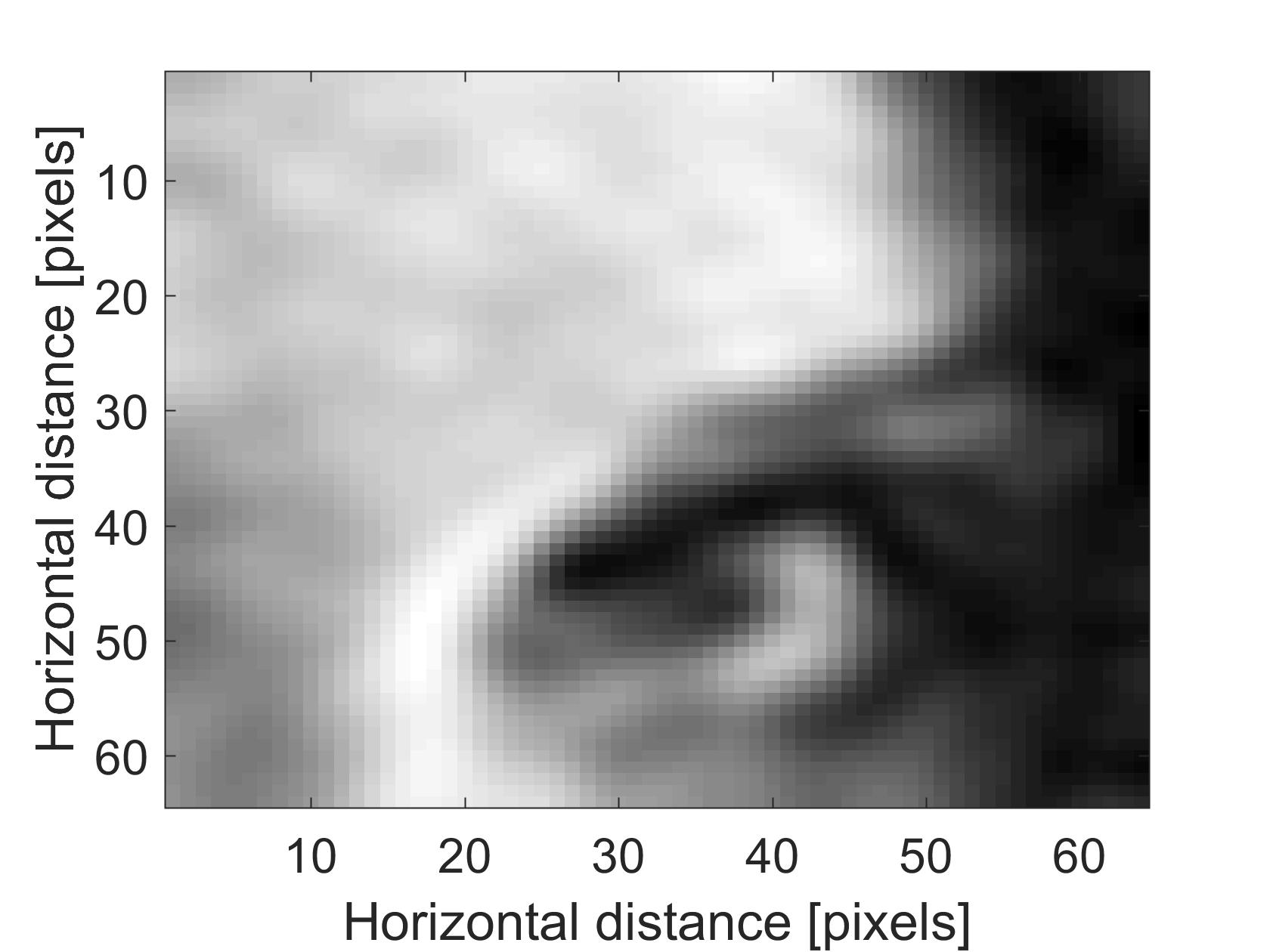}}

\caption{The best, the worst and an intermediate deconvolution results over 200 noise realizations obtained using TV-ADMM, ADMM+BM3D, ADMM+TNRD, ADMM+VST+TNRD, P$^4$IP and the proposed QAB-PnP method for Lena image degraded by a Gaussian blurring kernel $h_{\sigma = 3}^{4\times 4}$ and Poisson noise corresponding to a SNR of 15 dB.}
\label{fig:lena15}
\end{figure*}

\begin{table*}[h!]
\begin{center}
\caption{Quantitative results (average over 200 noise realizations). Best results are shown in bold.}
\label{tab:tab_pip}
\begin{tabular}{c c c c c c c c}

\multicolumn{8}{c}{Gaussian kernel ${h_{\sigma=3}}^{4\times 4}$} \\
\thickhline

\multirow{2}{*}{Sample} & \multirow{2}{*}{Method}
			 & \multicolumn{2}{c}{Poisson Noise (20 dB)} & \multicolumn{2}{c}{Poisson Noise (15 dB)} & \multicolumn{2}{c}{Poisson Noise (10 dB)}\\
			& & PSNR (dB) & SSIM & PSNR (dB) & SSIM & PSNR (dB) & SSIM \\

\thickhline

\multirow{6}{*}{Synthetic}
			
	& TV-ADMM	& 26.46$\pm$0.10 & 0.66$\pm$0.01 & 24.80$\pm$0.34 & 0.58$\pm$0.01 & 22.52$\pm$1.55 & 0.52$\pm$0.02\\
	
	& ADMM+BM3D		& 23.37$\pm$0.16 & 0.73$\pm$0.01 & 19.70$\pm$0.23 & 0.54$\pm$0.01 & 17.67$\pm$0.37 & 0.47$\pm$0.02\\

	& ADMM+TNRD		& 23.94$\pm$0.14 & 0.65$\pm$0.01 & 21.55$\pm$0.31 & 0.56$\pm$0.01 & 18.88$\pm$0.40 & 0.40$\pm$0.01 \\

	& ADMM+VST+TNRD		& 23.96$\pm$0.11 & 0.71$\pm$0.02 & 21.73$\pm$0.19 & 0.54$\pm$0.02 & 19.02$\pm$0.23 & 0.38$\pm$0.03 \\
	
	& P$^4$IP		& 23.90$\pm$1.37 & 0.74$\pm$0.06 & 20.91$\pm$2.18 & 0.59$\pm$0.11 & 18.96$\pm$3.34 & 0.48$\pm$0.18\\

\vspace{5pt}
	
	& QAB-PnP		& \textbf{29.86$\pm$0.12} & \textbf{0.92$\pm$0.00} & \textbf{27.18$\pm$0.43} & \textbf{0.86$\pm$0.01} & \textbf{24.23$\pm$1.34} & \textbf{0.74$\pm$0.03}\\


\multirow{6}{*}{Lena}

		& TV-ADMM		& 27.37$\pm$0.31 & 0.74$\pm$0.01 & 24.52$\pm$0.65 & 0.66$\pm$0.01 & 19.97$\pm$1.32 & 0.52$\pm$0.02\\
		
		& ADMM+BM3D		& 25.87$\pm$0.40 & 0.75$\pm$0.01 & 23.59$\pm$0.66 & 0.66$\pm$0.03 & 17.59$\pm$1.02 & 0.50$\pm$0.05 \\

		& ADMM+TNRD		& 25.76$\pm$0.19 & 0.71$\pm$0.01 & 24.67$\pm$0.21 & 0.69$\pm$0.01 & 19.22$\pm$0.38 & 0.50$\pm$0.02 \\

		& ADMM+VST+TNRD		& 25.85$\pm$0.23 & 0.69$\pm$0.01 & 24.73$\pm$0.39 & 0.60$\pm$0.01 & 19.11$\pm$0.80 & 0.42$\pm$0.07 \\
		
		& P$^4$IP		& 27.32$\pm$0.44 & \textbf{0.81}$\pm$0.01 & 24.87$\pm$2.76 & \textbf{0.76}$\pm$0.07 & 18.67$\pm$4.83 & 0.55$\pm$0.16\\

\vspace{5pt}
		
		& QAB-PnP		& \textbf{28.97$\pm$0.19} & \textbf{0.81$\pm$0.00} & \textbf{27.04$\pm$0.44} & 0.75$\pm$0.01 & \textbf{20.18$\pm$3.39} & \textbf{0.65$\pm$0.08}\\

\multirow{6}{*}{Fruits}
		
		& TV-ADMM		& 20.51$\pm$0.38 & 0.57$\pm$0.01 & 19.02$\pm$0.23 & 0.55$\pm$0.01 & \textbf{17.54$\pm$0.93} & 0.51$\pm$0.01\\
		
		& ADMM+BM3D		& 19.75$\pm$0.42 & 0.61$\pm$0.01 & 17.07$\pm$0.20 & 0.53$\pm$0.01 & 13.59$\pm$0.35 & 0.51$\pm$0.02 \\
	
		& ADMM+TNRD		& 19.73$\pm$1.91 & \textbf{0.64$\pm$0.02} & 17.41$\pm$0.57 & \textbf{0.59$\pm$0.01} & 16.67$\pm$0.79 & 0.51$\pm$0.06 \\

		& ADMM+VST+TNRD		& 20.65$\pm$0.39 & \textbf{0.64$\pm$0.01} & 18.40$\pm$1.19 & 0.58$\pm$0.02 & 16.51$\pm$1.36 & 0.43$\pm$0.08 \\
	
		& P$^4$IP		& 20.42$\pm$1.79 & 0.59$\pm$0.04 & 17.22$\pm$4.62 & 0.52$\pm$0.11 & 14.35$\pm$3.85 & \textbf{0.53$\pm$0.04}\\
		
		& QAB-PnP		& \textbf{21.37$\pm$0.94} & 0.62$\pm$0.01 & \textbf{19.35$\pm$0.96} & 0.57$\pm$0.02 & 17.28$\pm$3.55 & 0.51$\pm$0.12\\

\thickhline


~~\\
\multicolumn{8}{c}{Gaussian kernel ${h_{\sigma=5}}^{4\times 4}$} \\
\thickhline

\multirow{2}{*}{Sample} & \multirow{2}{*}{Method}
			 & \multicolumn{2}{c}{Poisson Noise (20 dB)} & \multicolumn{2}{c}{Poisson Noise (15 dB)} & \multicolumn{2}{c}{Poisson Noise (10 dB)}\\
			& & PSNR (dB) & SSIM & PSNR (dB) & SSIM & PSNR (dB) & SSIM \\

\thickhline

\multirow{6}{*}{Synthetic}
	
	& TV-ADMM		& 26.47$\pm$0.07 & 0.59$\pm$0.01 & 25.23$\pm$0.14 & 0.54$\pm$0.01 & 23.15$\pm$0.29 & 0.44$\pm$0.01\\
	
	& ADMM+BM3D		& 22.95$\pm$0.18 & 0.70$\pm$0.01 & 19.78$\pm$0.24 & 0.53$\pm$0.01 & 17.89$\pm$0.34 & 0.46$\pm$0.02 \\
	
	& ADMM+TNRD		& 23.81$\pm$0.18 & 0.66$\pm$0.01 & 21.72$\pm$0.22 & 0.58$\pm$0.02 & 19.03$\pm$0.44 & 0.41$\pm$0.01 \\

	& ADMM+VST+TNRD		& 23.89$\pm$0.12 & 0.69$\pm$0.01 & 21.82$\pm$0.22 & 0.52$\pm$0.02 & 18.96$\pm$0.34 & 0.37$\pm$0.04 \\
	
	& P$^4$IP		& 22.35$\pm$2.15 & 0.67$\pm$0.09 & 20.60$\pm$2.87 & 0.56$\pm$0.12 & 18.67$\pm$3.42 & 0.49$\pm$0.21\\

\vspace{5pt}

	& QAB-PnP		& \textbf{29.44$\pm$0.13} & \textbf{0.91$\pm$0.00} & \textbf{27.24$\pm$0.58} & \textbf{0.86$\pm$0.01} & \textbf{24.06$\pm$1.07} & \textbf{0.73$\pm$0.02}\\

\multirow{6}{*}{Lena}
			
	& TV-ADMM		& 27.17$\pm$0.25 & 0.74$\pm$0.01 & 25.11$\pm$0.46 & 0.61$\pm$0.01 & 19.41$\pm$0.42 & 0.44$\pm$0.01\\
	
	& ADMM+BM3D		& 25.02$\pm$0.48 & 0.73$\pm$0.01 & 23.51$\pm$0.78 & 0.65$\pm$0.02 & 17.64$\pm$1.47 & 0.48$\pm$0.06\\
	
	& ADMM+TNRD		& 25.44$\pm$0.17 & 0.71$\pm$0.01 & 24.43$\pm$0.26 & 0.68$\pm$0.02 & 19.20$\pm$0.23 & 0.51$\pm$0.02 \\

	& ADMM+VST+TNRD		& 25.46$\pm$0.29 & 0.69$\pm$0.01 & 24.53$\pm$0.32 & 0.60$\pm$0.01 & 19.41$\pm$0.49 & 0.43$\pm$0.05 \\
	
	& P$^4$IP		& 27.26$\pm$0.34 & \textbf{0.81$\pm$0.01} & 25.07$\pm$2.90 & \textbf{0.77$\pm$0.06} & 17.99$\pm$4.73 & 0.54$\pm$0.21\\

\vspace{5pt}
	
	& QAB-PnP		& \textbf{28.80$\pm$0.21} & \textbf{0.81$\pm$0.00} & \textbf{26.63$\pm$1.01} & 0.76$\pm$0.03 & \textbf{20.20$\pm$3.89} & \textbf{0.67$\pm$0.05}\\

\multirow{6}{*}{Fruits}
			
	& TV-ADMM		& 19.94$\pm$0.25 & 0.57$\pm$0.01 & 17.24$\pm$0.28 & 0.55$\pm$0.01 & 16.58$\pm$0.34 & 0.50$\pm$0.01\\
	
	& ADMM+BM3D		& 19.15$\pm$0.58 & 0.60$\pm$0.01 & 17.11$\pm$0.33 & 0.54$\pm$0.01 & 13.45$\pm$0.55 & 0.50$\pm$0.02 \\
	
	& ADMM+TNRD		& 19.68$\pm$1.10 & 0.63$\pm$0.02 & 17.95$\pm$0.96 & \textbf{0.58$\pm$0.01} & 16.13$\pm$0.74 & 0.51$\pm$0.06 \\

	& ADMM+VST+TNRD		& 20.18$\pm$0.29 & \textbf{0.65$\pm$0.01} & 18.16$\pm$0.87 & \textbf{0.58$\pm$0.01} & 16.45$\pm$1.04 & 0.45$\pm$0.03 \\
	
	& P$^4$IP		& \textbf{20.47$\pm$1.99} & 0.61$\pm$0.05 & 17.49$\pm$3.44 & 0.56$\pm$0.04 & 13.83$\pm$4.22 & 0.51$\pm$0.05\\
	
	& QAB-PnP		& 20.24$\pm$1.09 & 0.60$\pm$0.01 & \textbf{18.83$\pm$0.71} & \textbf{0.58$\pm$0.01} & \textbf{17.44$\pm$2.09} & \textbf{0.53$\pm$0.02}\\
		
\thickhline

\end{tabular}
\end{center}
\end{table*}

\begin{table}[h!]


\begin{center}
\caption{Quantitative deconvolution results when images are corrupted with high intensity noise.}
\label{tab:tab_result2}
\begin{tabular}{c c c c c}

\multicolumn{5}{c}{Gaussian kernel ${h_{\sigma=3}}^{4\times 4}$ + Poisson Noise} \\
\thickhline

\multirow{2}{*}{Sample} & \multicolumn{2}{c}{SNR $\approx$ 5 dB} & \multicolumn{2}{c}{SNR $\approx$ 0 dB}\\
		 & PSNR (dB) & SSIM & PSNR (dB) & SSIM \\
						
\thickhline

\multirow{1}{*}{Synthetic}
			
		& 18.76 & 0.41 & 16.48 & 0.35\\

\multirow{1}{*}{Lena}
		
		& 16.25 & 0.49 & 15.72 & 0.42\\

\multirow{1}{*}{Fruits}
		
		& 15.04 & 0.39 & 13.32 & 0.30 \\

\thickhline

\end{tabular}
\end{center}


\end{table}

\begin{table*}[b!]

\begin{footnotesize}
\setlength\tabcolsep{-.1pt}

\begin{center}
\caption{Quantitative results for experimental fluorescence microscopy images. Best results are shown in bold.}
\label{tab:resu_fluo}
\begin{tabular}{c c c  c c c}

\thickhline

 & \multirow{2}{*}{~~~~~~~~~~~~~~Methods~~~~~~~~~~~~~~} & \multirow{2}{*}{~~~~~~~~~~~~~~Data~~~~~~~~~~~~~~} & \multicolumn{2}{c}{Confocal microscopy} & ~~Two-photon microscopy~~\\

		& & & ~~~~Zebra Fish~~ & ~~Mouse Brain~~~~ & Mouse Brain \\
	
\thickhline


\multirow{2}{*}{ Observed Data}

		& &  PSNR (dB) &  20.20 &  27.37 &  24.07\\
\vspace{5pt}	
		& &  SSIM &  0.37 &  0.59 &  0.40\\

\multirow{14}{*}{Deblurred Results}

& \multirow{2}{*}{ TV-ADMM}

		&  PSNR (dB) &  24.27 &  30.27 &  26.57\\
\vspace{5pt}			
		& &  SSIM &  0.61 &  0.88 &  0.70\\

& \multirow{2}{*}{ ADMM+BM3D}

		 &  PSNR (dB) &  24.74 &  32.97 &  27.66\\
\vspace{5pt}			
		& &  SSIM &  0.74 &  0.90 &  0.81\\

& \multirow{2}{*}{ADMM+TNRD}

		 &  PSNR (dB) &  25.85 &  34.26 &  31.04\\
\vspace{5pt}			
		& &  SSIM &  0.79 &  0.91 &  0.89\\

& \multirow{2}{*}{ ADMM+VST+TNRD}

		 &  PSNR (dB) &  25.88 &  34.44 &  \textbf{31.23}\\
\vspace{5pt}			
		& &  SSIM &  0.79 &  0.90 &  \textbf{0.90}\\

& \multirow{2}{*}{ P$^4$IP}

		 &  PSNR (dB) &  25.18 &  33.06 &  27.09\\
\vspace{5pt}
		& &  SSIM &  0.77 &  0.92 &  0.85\\

& \multirow{2}{*}{ QAB-PnP}

		 &  PSNR (dB) &  \textbf{28.91} &  \textbf{35.68} &  30.14\\
		& &  SSIM &  \textbf{0.82} &  \textbf{0.93} &  0.79\\
		
\thickhline

\end{tabular}
\end{center}

\end{footnotesize}

\end{table*}

\subsection{Poisson deconvolution results}
\label{sec:application}

Poisson deconvolution is a well discussed domain in the literature where PnP algorithms implanting a Gaussian denoiser with or without a VST transformation have exhibited promising outcomes \cite{azzari2017variance, rond2016poisson}. The proposed method is intrinsically adaptive, which makes it well-adapted to different noise statistics for the problem addressed and does not require using any additional transformation in the denoising step.

This subsection regroups image deconvolution results obtained with the proposed method and five approaches from the literature. The experiments consisted in recovering the images in Fig.~\ref{fig:sample} from degraded versions by Gaussian blurring kernels with different variances and Poisson noise at different SNRs. The first comparative method is a standard Poisson deconvolution method that consists in estimating the image that minimizes a cost function formed by the data fidelity term in \eqref{eq:datafidelity} and the classical total variation regularization \cite{de2011alternating}. This method will be denoted by TV-ADMM hereafter. The second method denoted by ADMM+BM3D is an integration of the BM3D denoiser in the PnP-ADMM algorithm.
Similarly, a deep learning denoiser trained on natural images was integrated into the PnP-ADMM scheme and used as comparison method. In particular, the CNN-based flexible learning method, known as the trainable nonlinear reaction diffusion (TNRD) \cite{chen2017trainable}, was used given its efficiency within regularization by denoising approaches \cite{romano2017little}. Finally, a PnP-ADMM algorithm coupled with an Anscombe transformation (VST) and a BM3D denoiser, denoted by P$^4$IP in \cite{rond2016poisson} was used for comparison. Note that TNRD has been also used with and without VST. The resulting algorithms are denoted by ADMM+TNRD and ADMM+VST+TNRD. It is important to mention that the methods used for comparisons such as TV-ADMM, P$^4$IP and ADMM+VST+TNRD are particularly designed for handling data degraded by Poisson noise, and are therefore appropriate choices as comparative methods to the proposed Poisson deconvolution algorithm.

As explained previsouly, the proposed method does not require such a VST-like transformation due to the adaptive nature of the embedded denoiser \cite{dutta2021quantum}. Therefore, the proposed algorithm is expected to present better generic convergence properties compared to P$^4$IP.  In the example in Fig.~\ref{fig:p4ipvsprop}, where P$^4$IP had fast convergence, the rate of convergence of QAB-PnP is similar to P$^4$IP and faster than TV-ADMM, ADMM+BM3D, ADMM+TNRD and ADMM+VST+TNRD. To evaluate the computational complexity of the proposed algorithm in comparison with other standard techniques, the average computational time and required number of iterations before convergence are given in Table~\ref{tab:iterVStime} with respect to different images. The results confirm the faster convergence of the proposed method, albeit, at the cost of higher computational time per iteration.

The deconvolution results obtained with the six methods can be visually appreciated in Figs.~\ref{fig:lena10},~\ref{fig:sinimag15} and~\ref{fig:fruimag20}. The PSNR and the structure similarity (SSIM) \cite{wang2004image} were used to evaluate the deconvolution accuracy. The resulting numerical results, for two different blurring kernels and three different SNRs, are regroupped in Table~\ref{tab:tab_pip}. In particular, average and standard deviation values are reported for 200 noise realizations. For further investigation, the quantitative results obtained with the proposed method in presence of very high-intensity noise, in particular, with SNRs close to 5 dB and 0 dB,  are provided in Table~\ref{tab:tab_result2}.

One may observe that the proposed scheme is capable to adapt both to low and high level of noise and outperforms the five other methods in almost all the simulations. It is important to note that QAB-PnP not only provides the best average values, but also the lowest standard deviations, in particular compared to P$^4$IP. This observation is confirmed by the results in Fig.~\ref{fig:lena15}, that displays, for a given simulation, the best, the worst and an intermediate result over 200 noise realizations. While the difference between these three results is barely observable for the proposed method, this is not the case for P$^4$IP. Finally, one may observe the big accuracy difference between the proposed method and the five others for the synthetic image. 

\subsection{Application to fluorescence microscopy imaging}
\label{sec:applifluo}

This section highlights the applicability of the proposed deconvolution method to real-life imaging applications, in particular to fluorescence microscopy imaging using, e.g., confocal \cite{pawley2006handbook} or two-photon \cite{denk1990two} microscopes. Fluorescence microscopy images are intrinsically noisy, contaminated by Poisson-Gaussian noise. Poisson noise is the dominating source of noise \cite{de2011alternating, zhang2019poisson, nam2016holistic}, due to a limited number ($\sim 10^2$ per pixel) of quantized photons captured by a microscopic detector compared to normal photography ($\sim 10^5$ per pixel). Therefore, enhancing such contaminated fluorescence images is of interest for many modern biological studies.

\begin{figure*}[h!]
\centering

Zebra Fish (Confocal microscopy imaging)\\
\subfigure[Ground truth]{\includegraphics[width=0.17\textwidth]{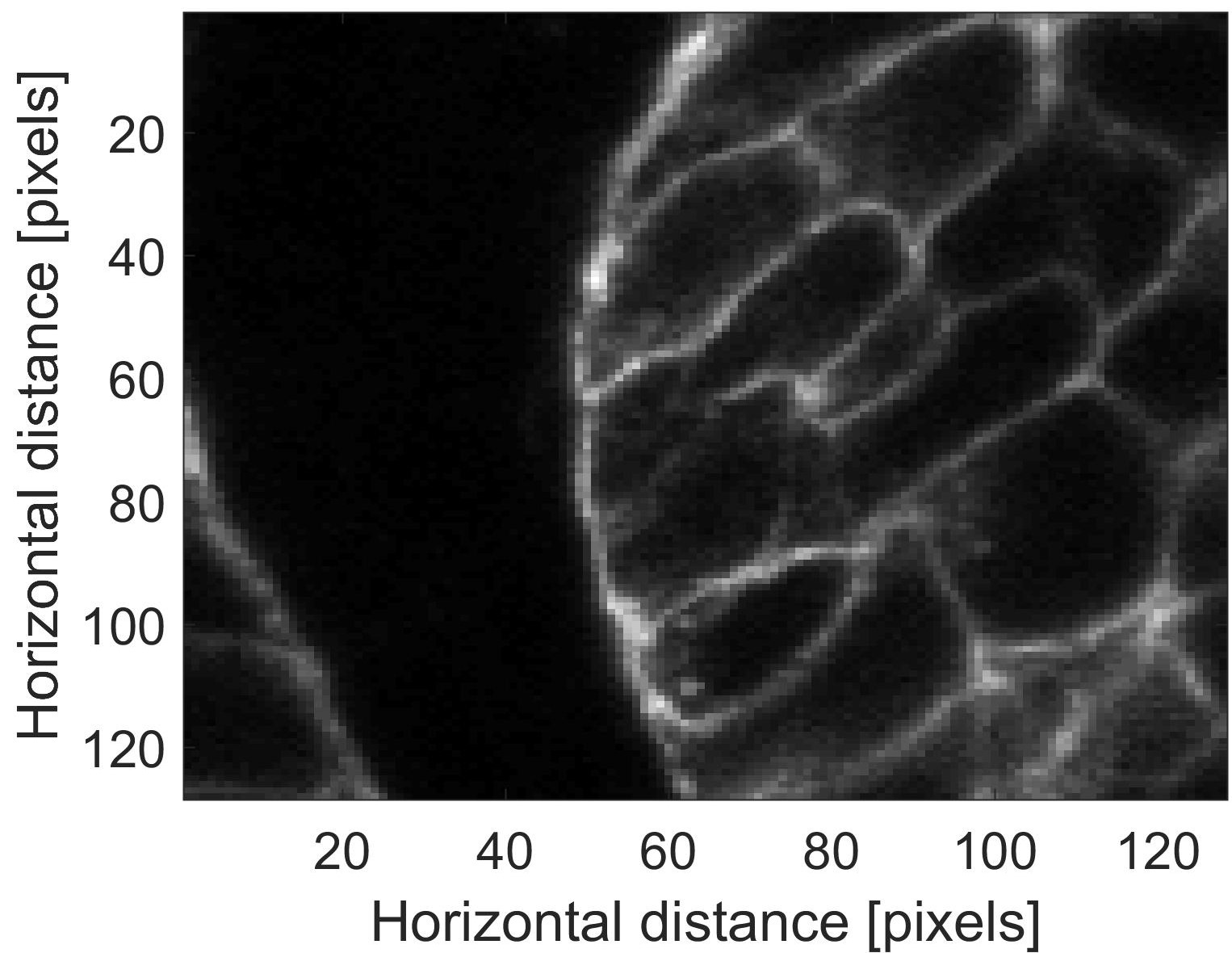}}
\subfigure[Observed image]{\includegraphics[width=0.17\textwidth]{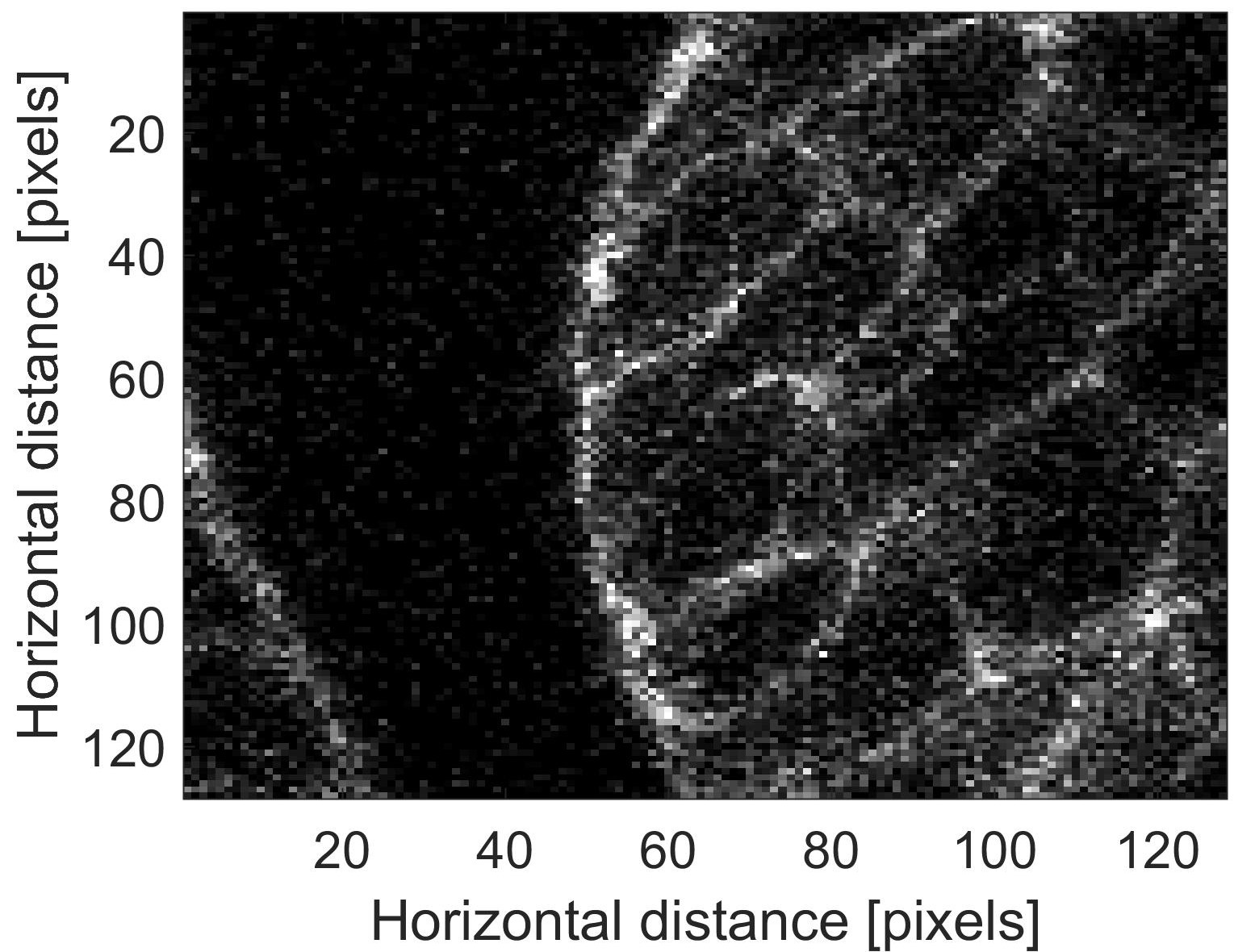}}
\subfigure[TV-ADMM]{\includegraphics[width=0.17\textwidth]{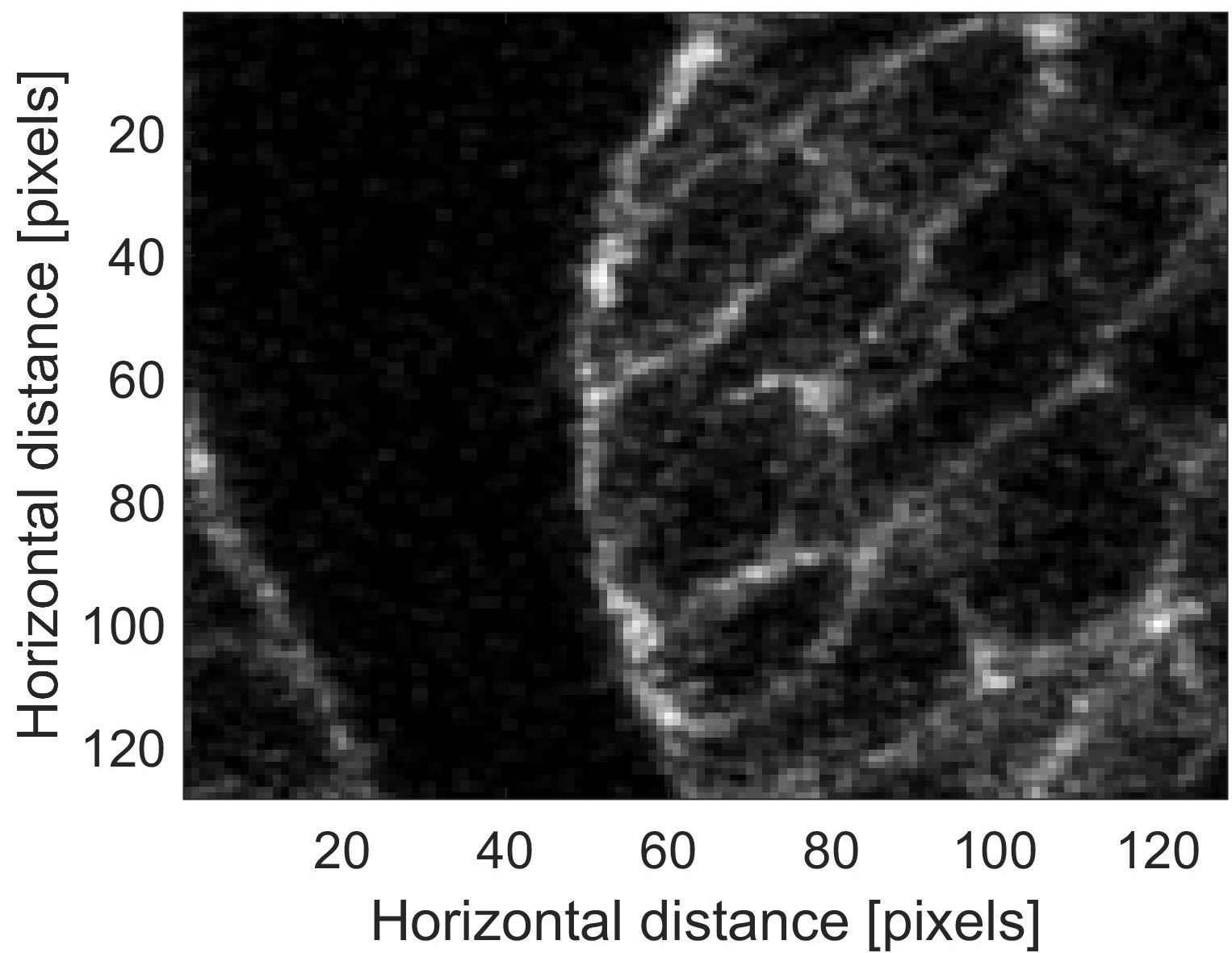}}
\subfigure[ADMM+BM3D]{\includegraphics[width=0.17\textwidth]{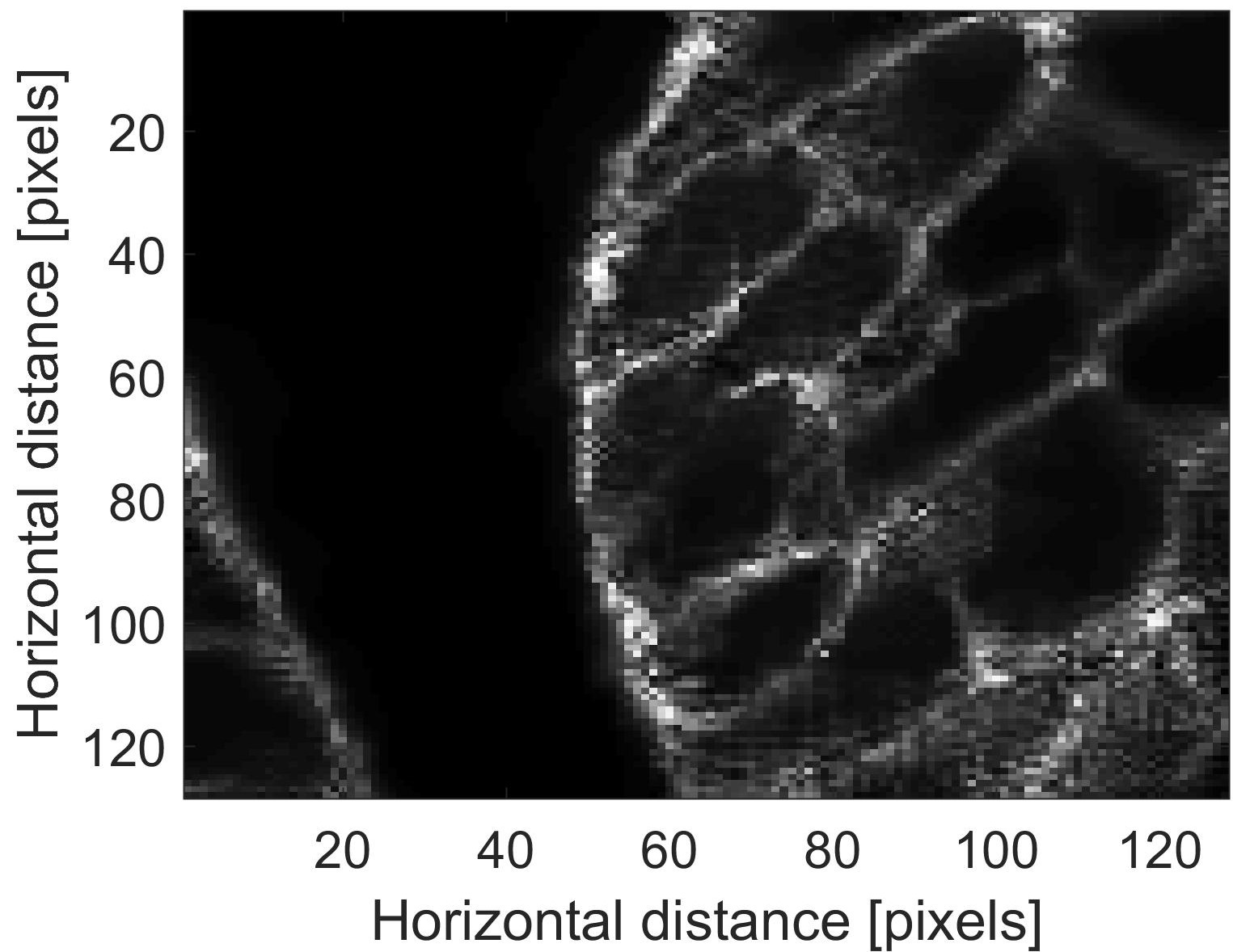}}
\subfigure[ADMM+TNRD]{\includegraphics[width=0.17\textwidth]{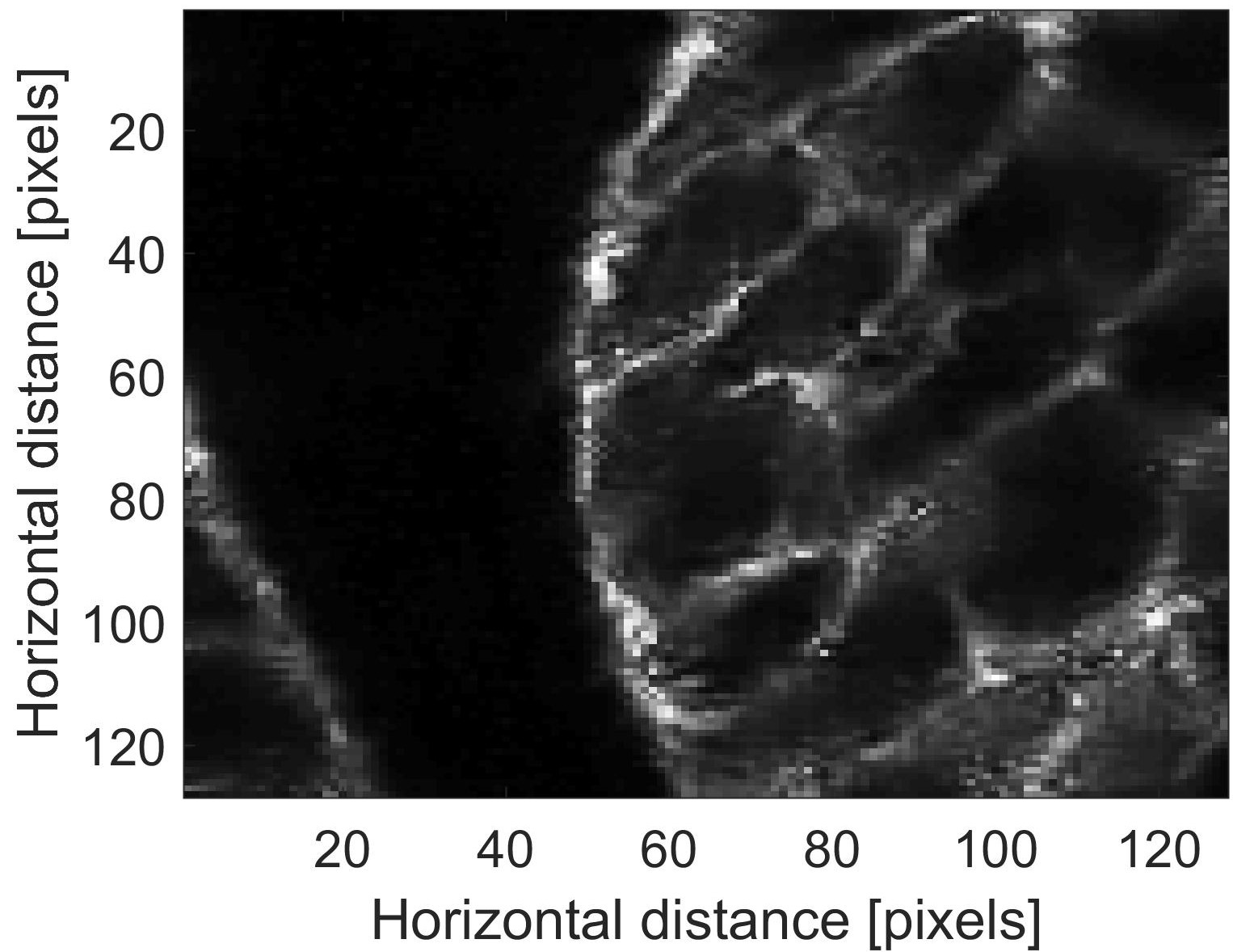}}
\subfigure[ADMM+VST+TNRD]{\includegraphics[width=0.17\textwidth]{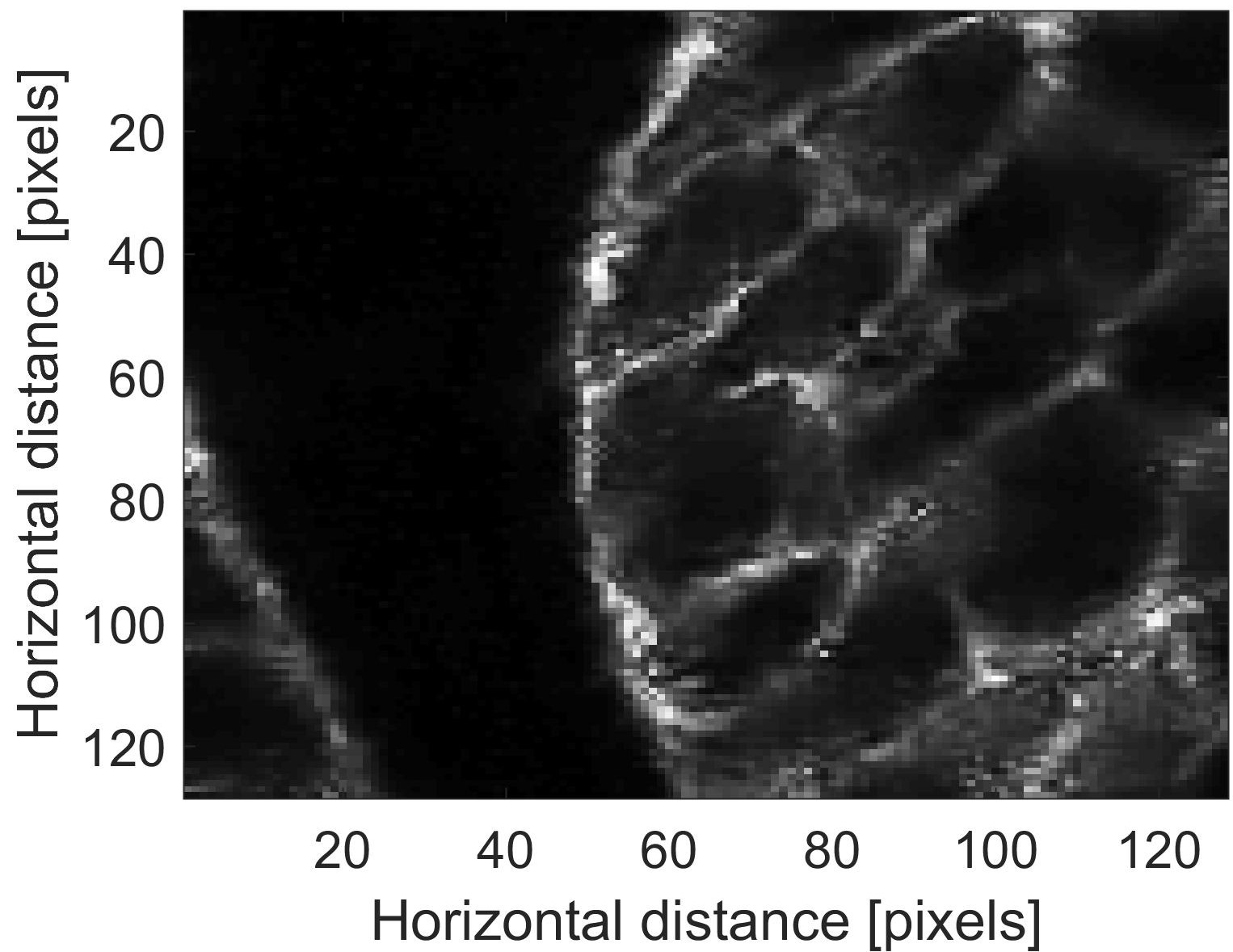}}
\subfigure[P$^4$IP]{\includegraphics[width=0.17\textwidth]{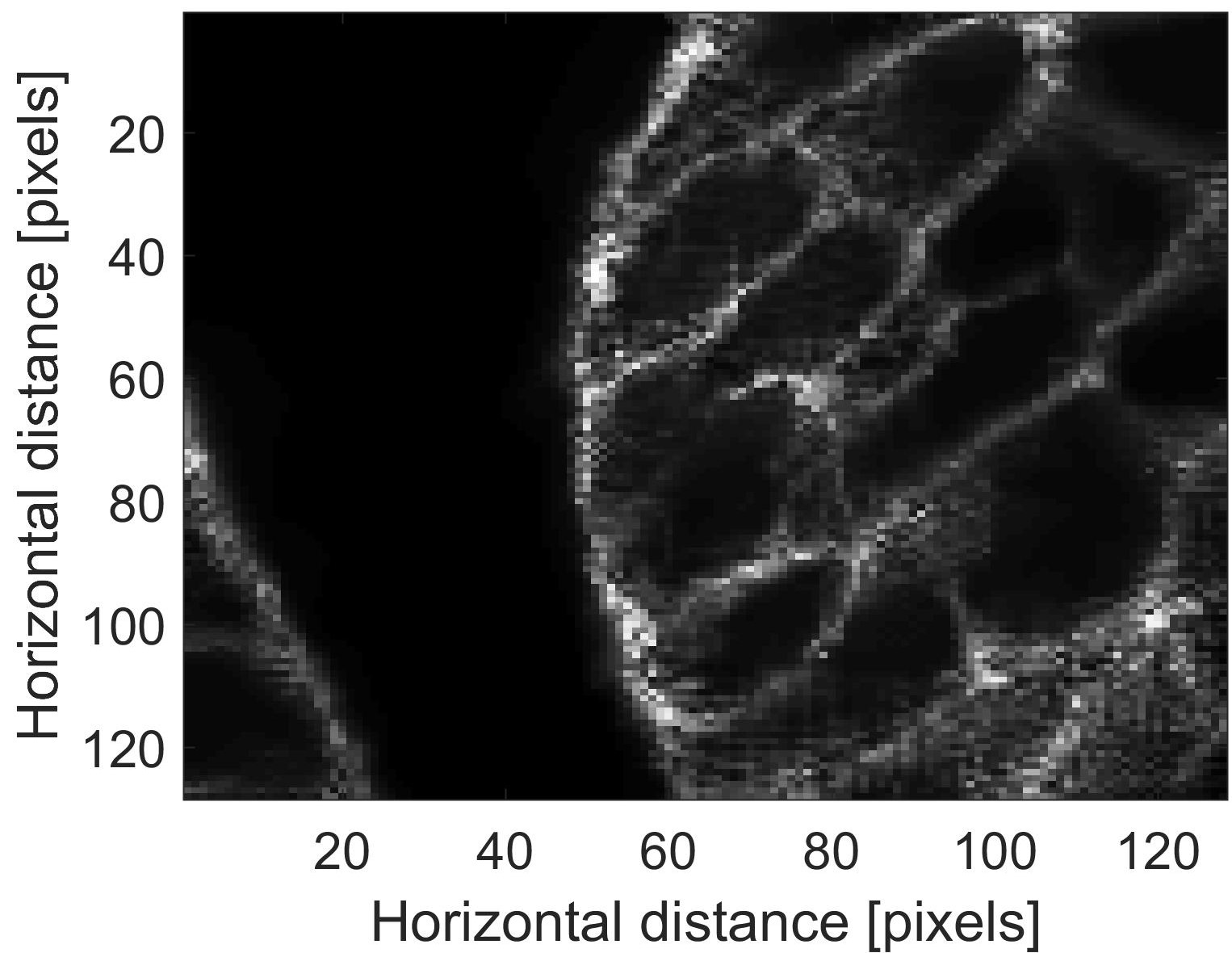}}
\subfigure[QAB-PnP]{\includegraphics[width=0.17\textwidth]{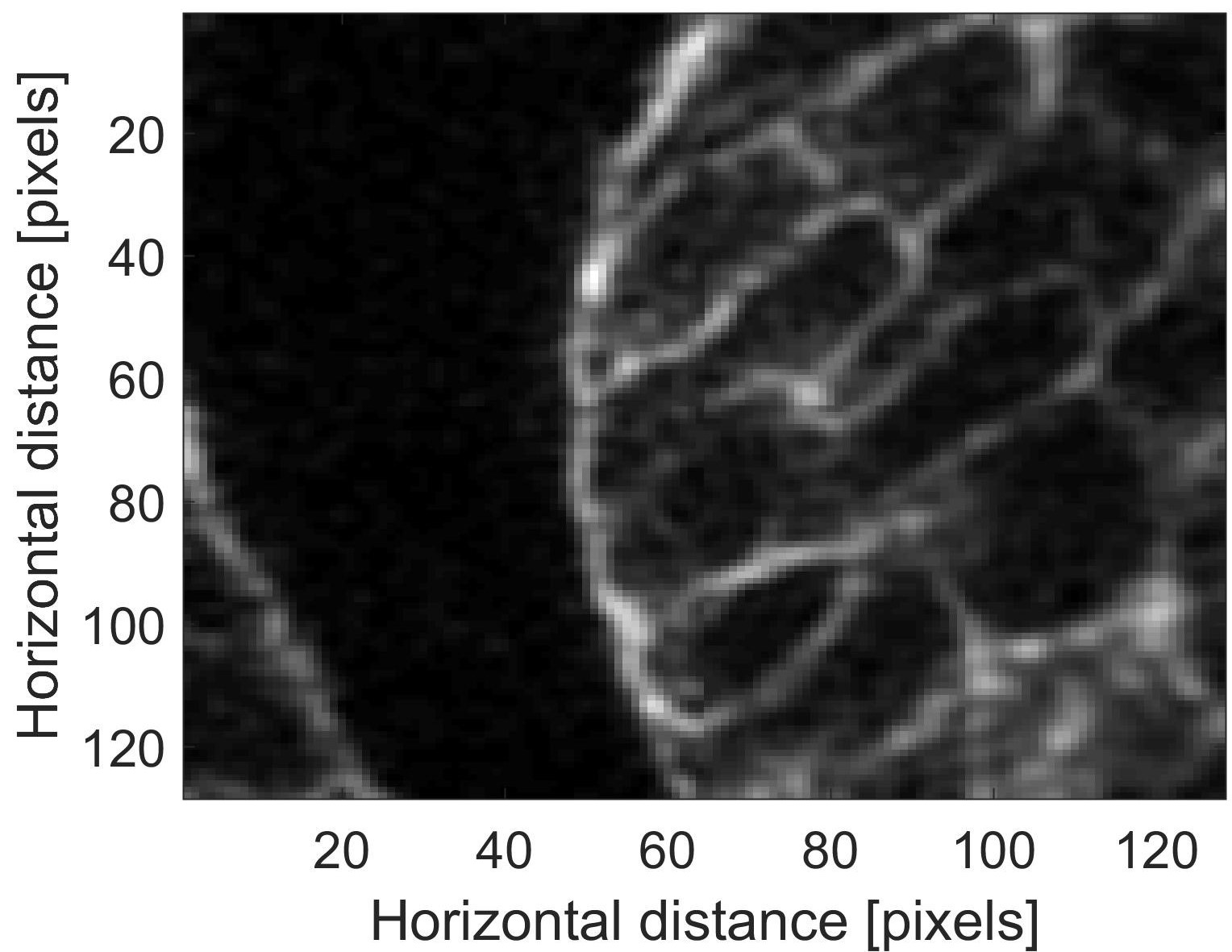}}

\vspace{5pt}
  Mouse Brain (Confocal microscopy imaging)\\
\subfigure[Ground truth]{\includegraphics[width=0.17\textwidth]{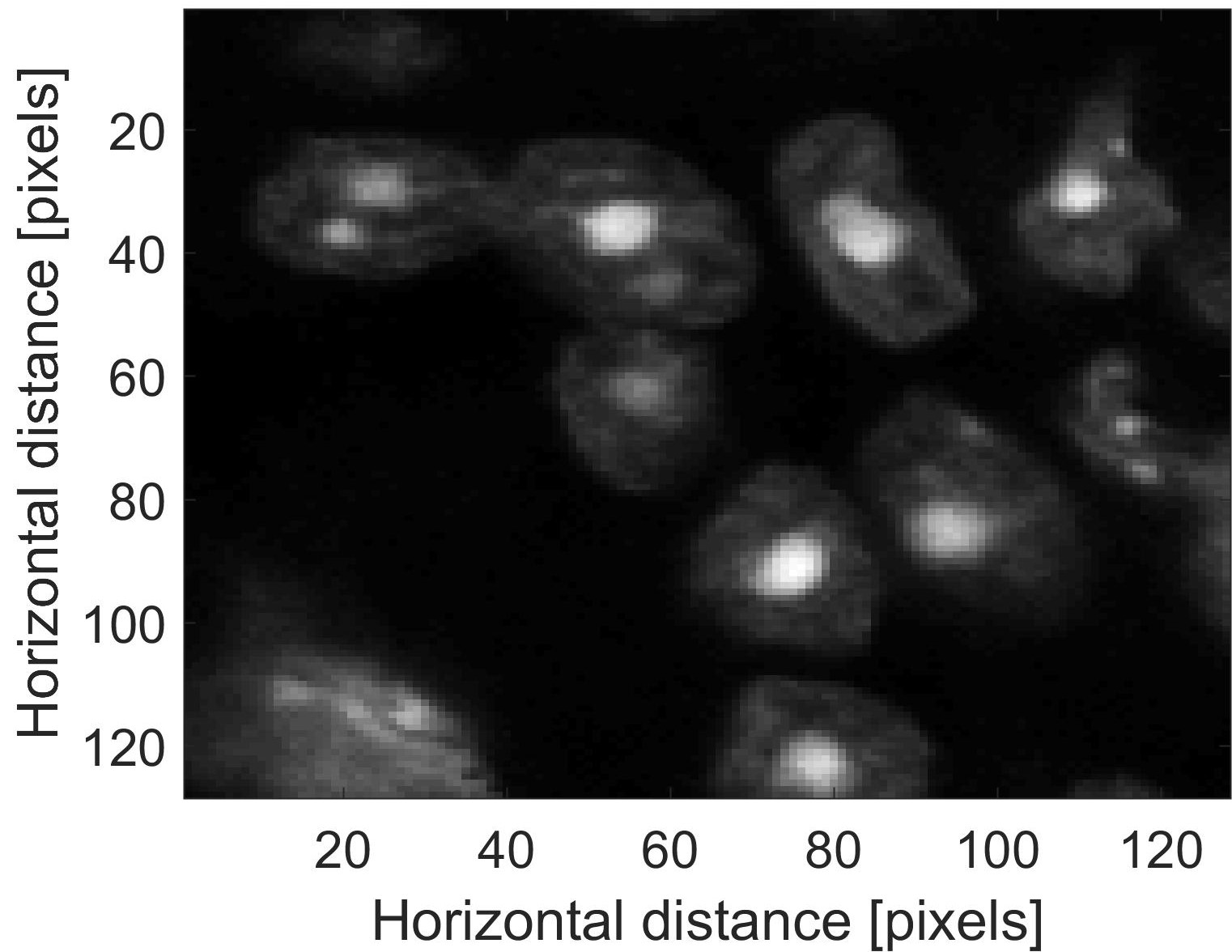}}
\subfigure[Observed image]{\includegraphics[width=0.17\textwidth]{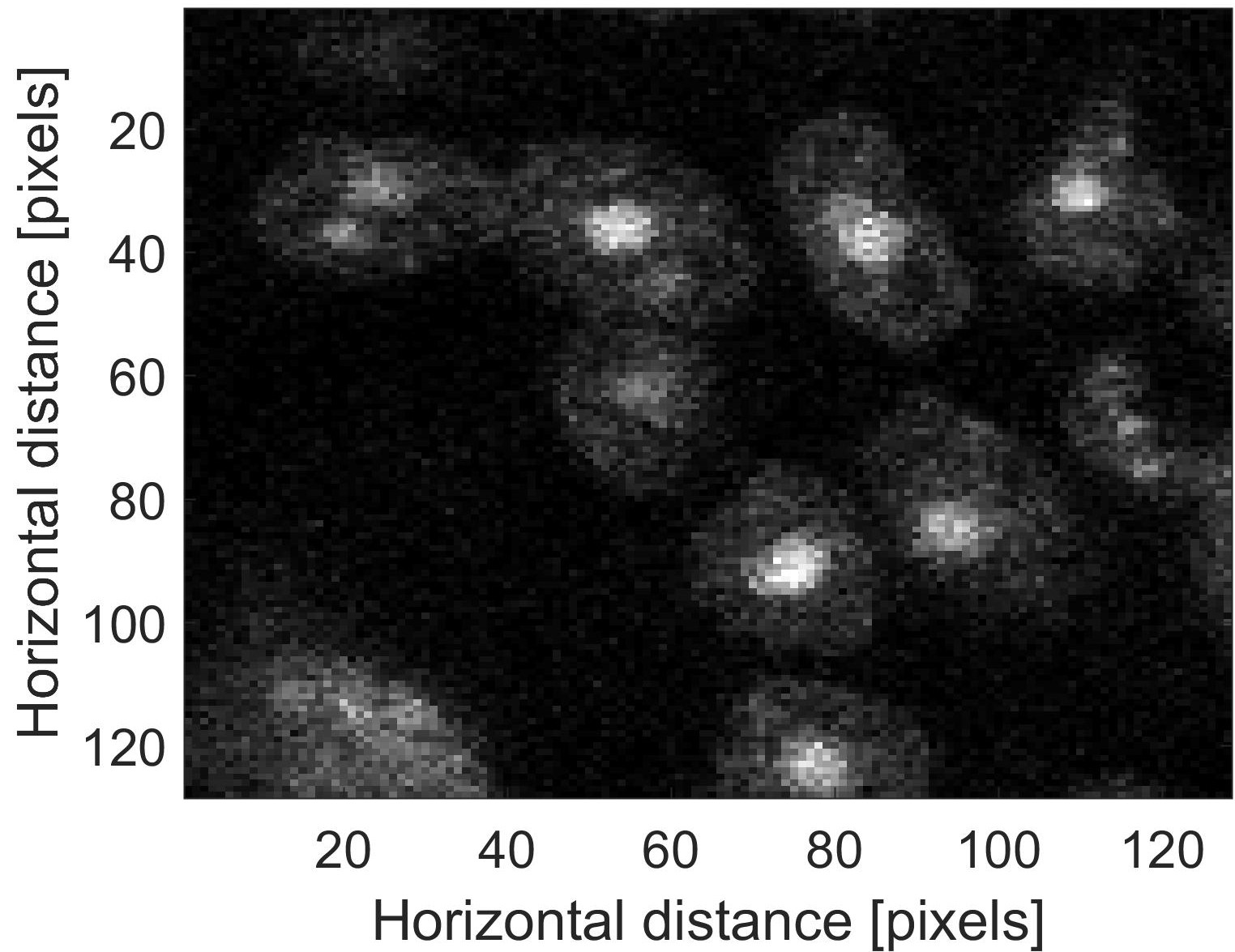}}
\subfigure[TV-ADMM]{\includegraphics[width=0.17\textwidth]{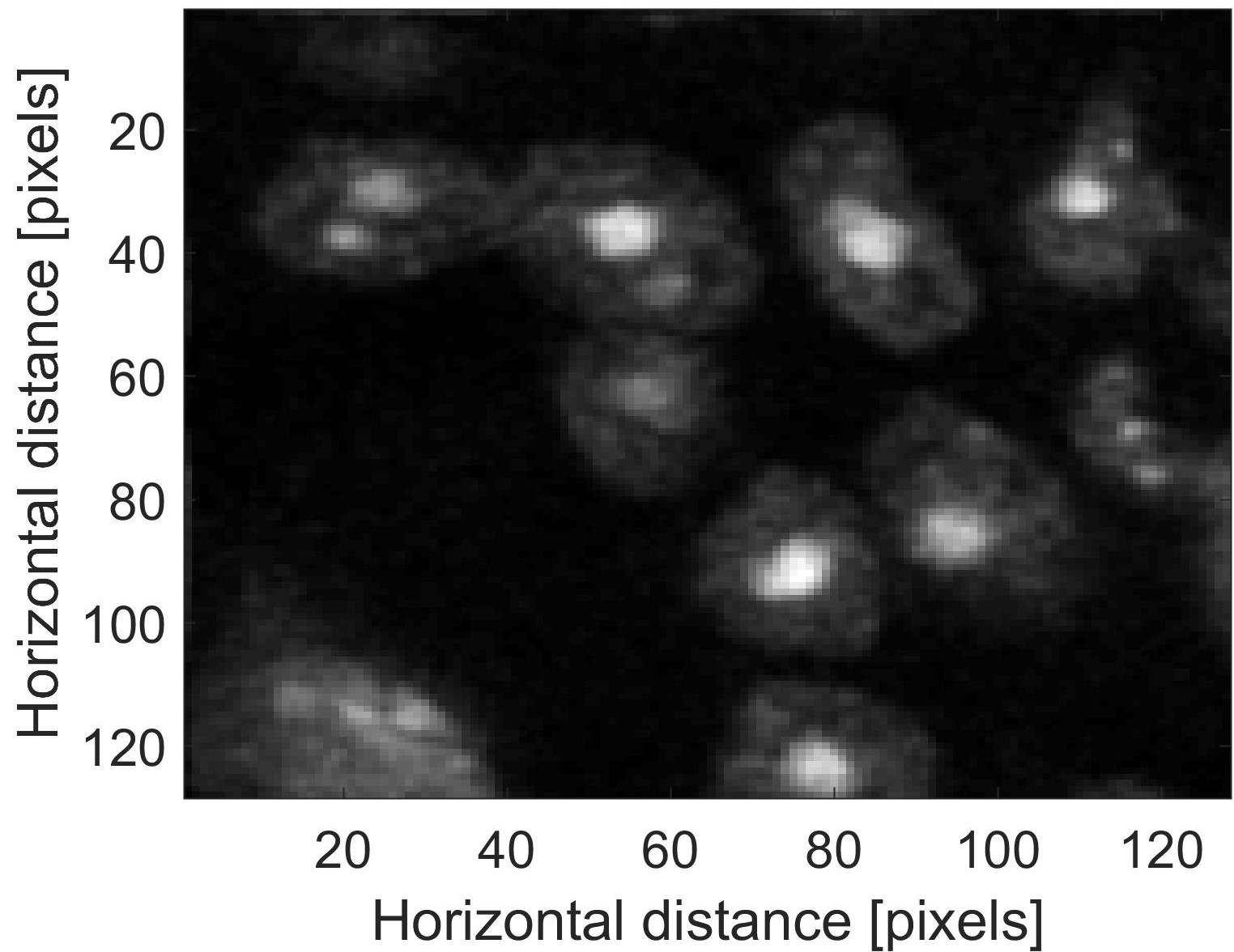}}
\subfigure[ADMM+BM3D]{\includegraphics[width=0.17\textwidth]{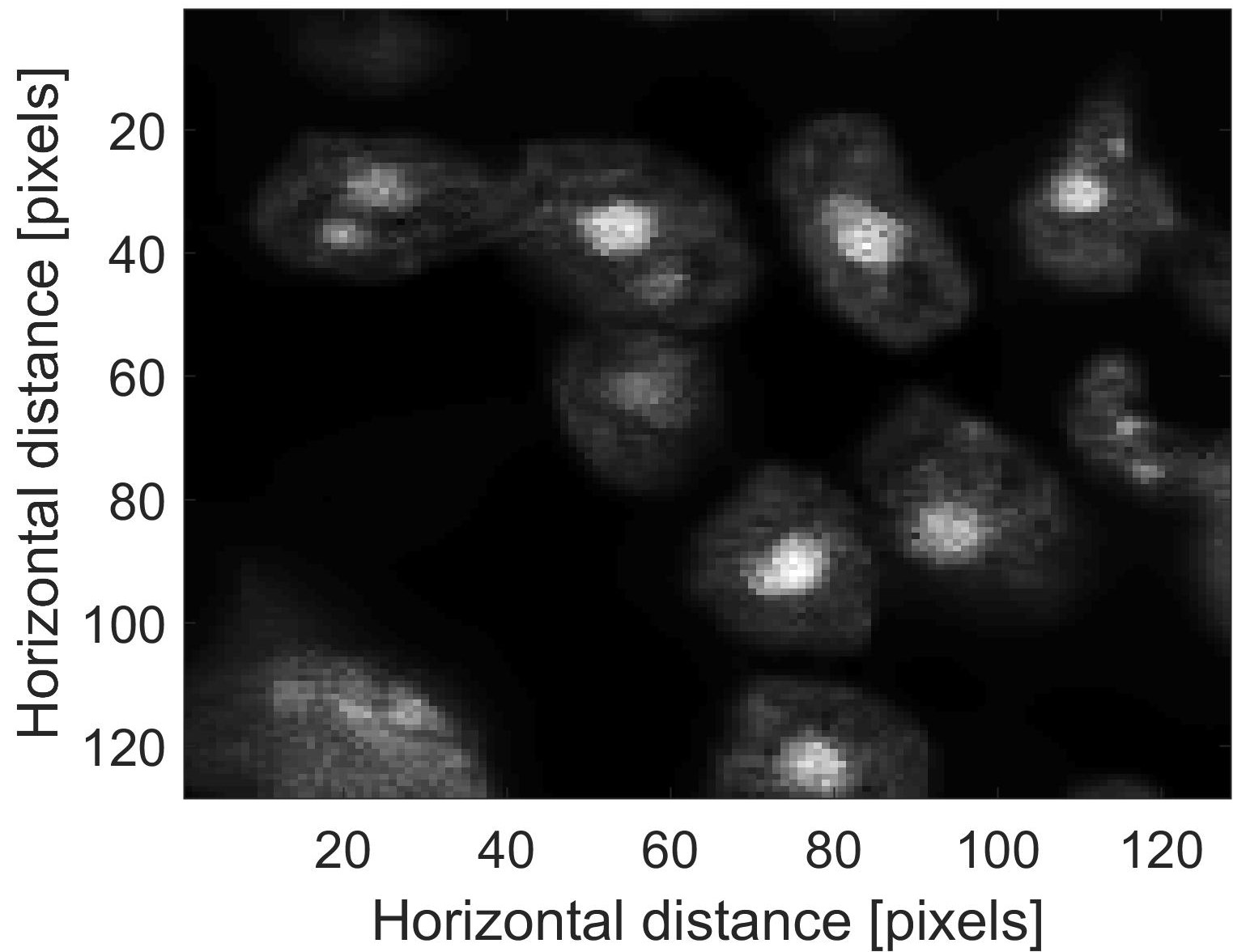}}
\subfigure[ADMM+TNRD]{\includegraphics[width=0.17\textwidth]{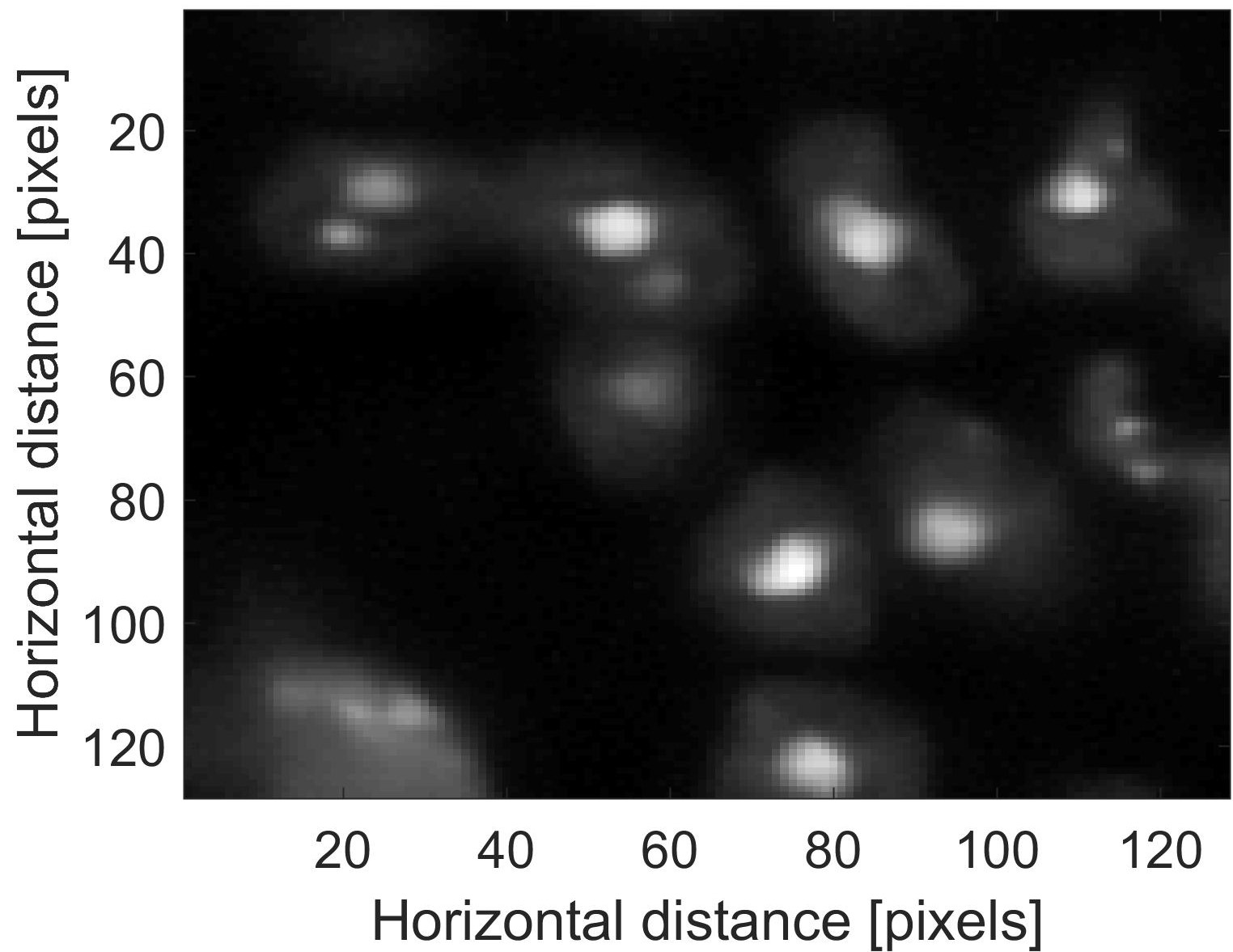}}
\subfigure[ADMM+VST+TNRD]{\includegraphics[width=0.17\textwidth]{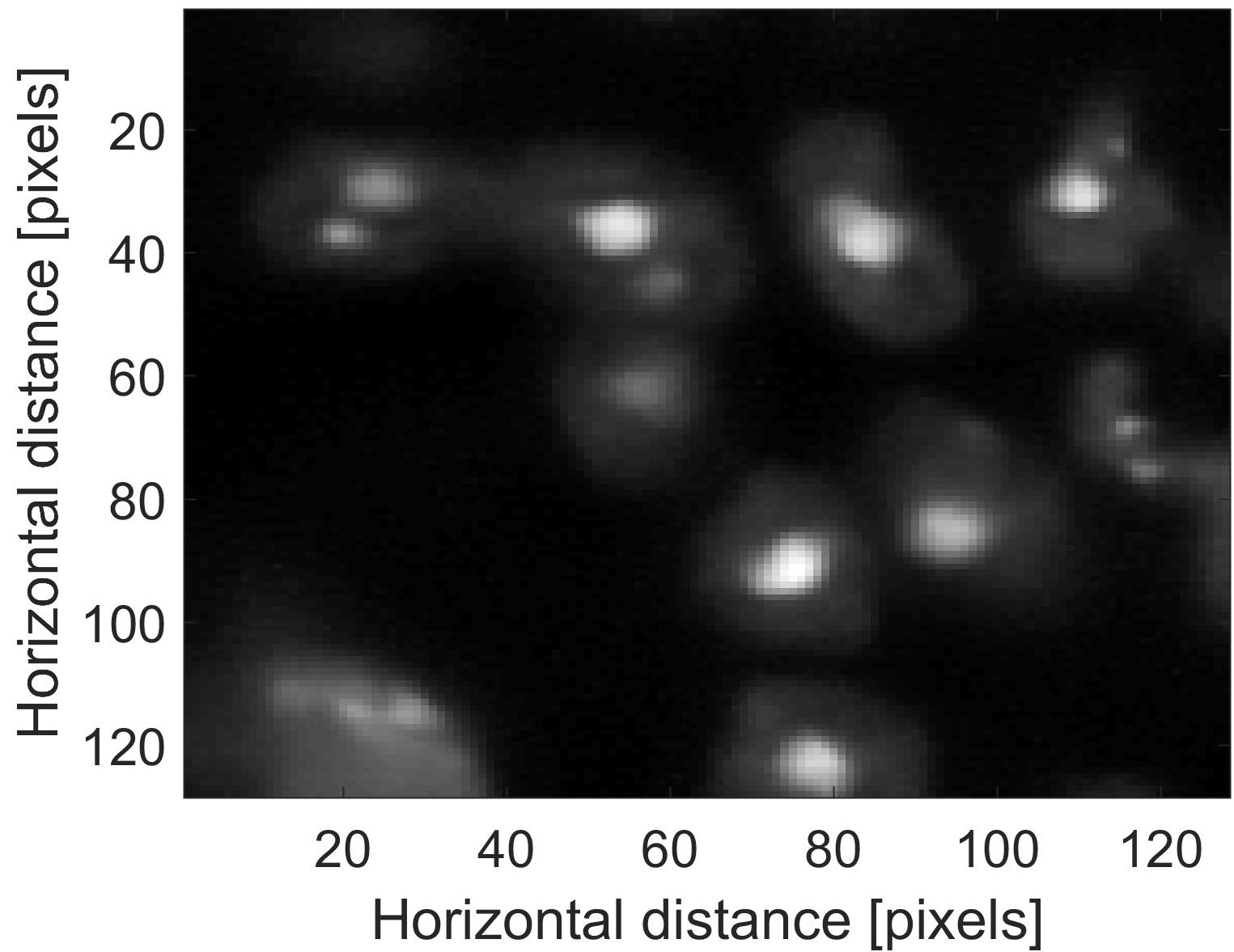}}
\subfigure[P$^4$IP]{\includegraphics[width=0.17\textwidth]{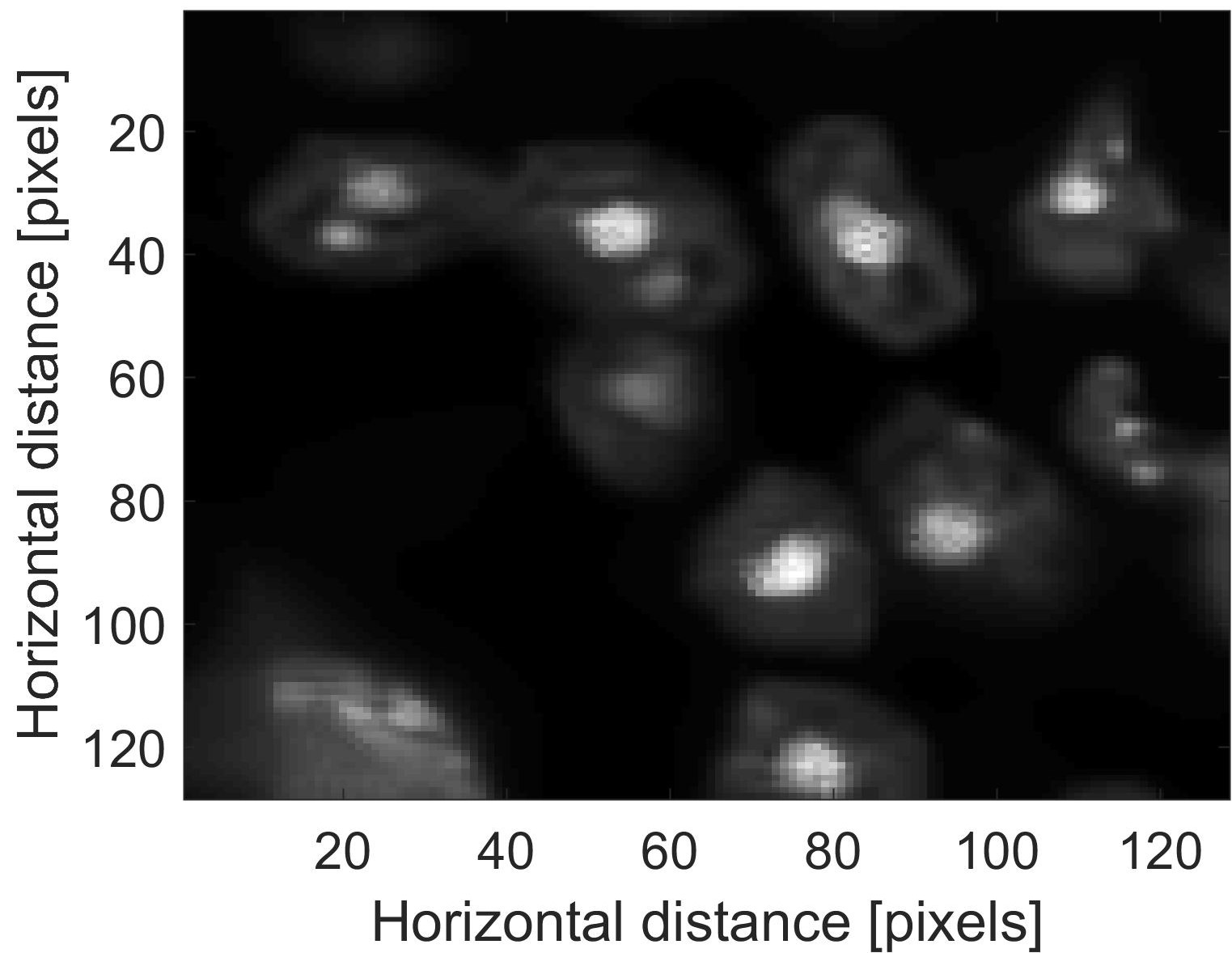}}
\subfigure[QAB-PnP]{\includegraphics[width=0.17\textwidth]{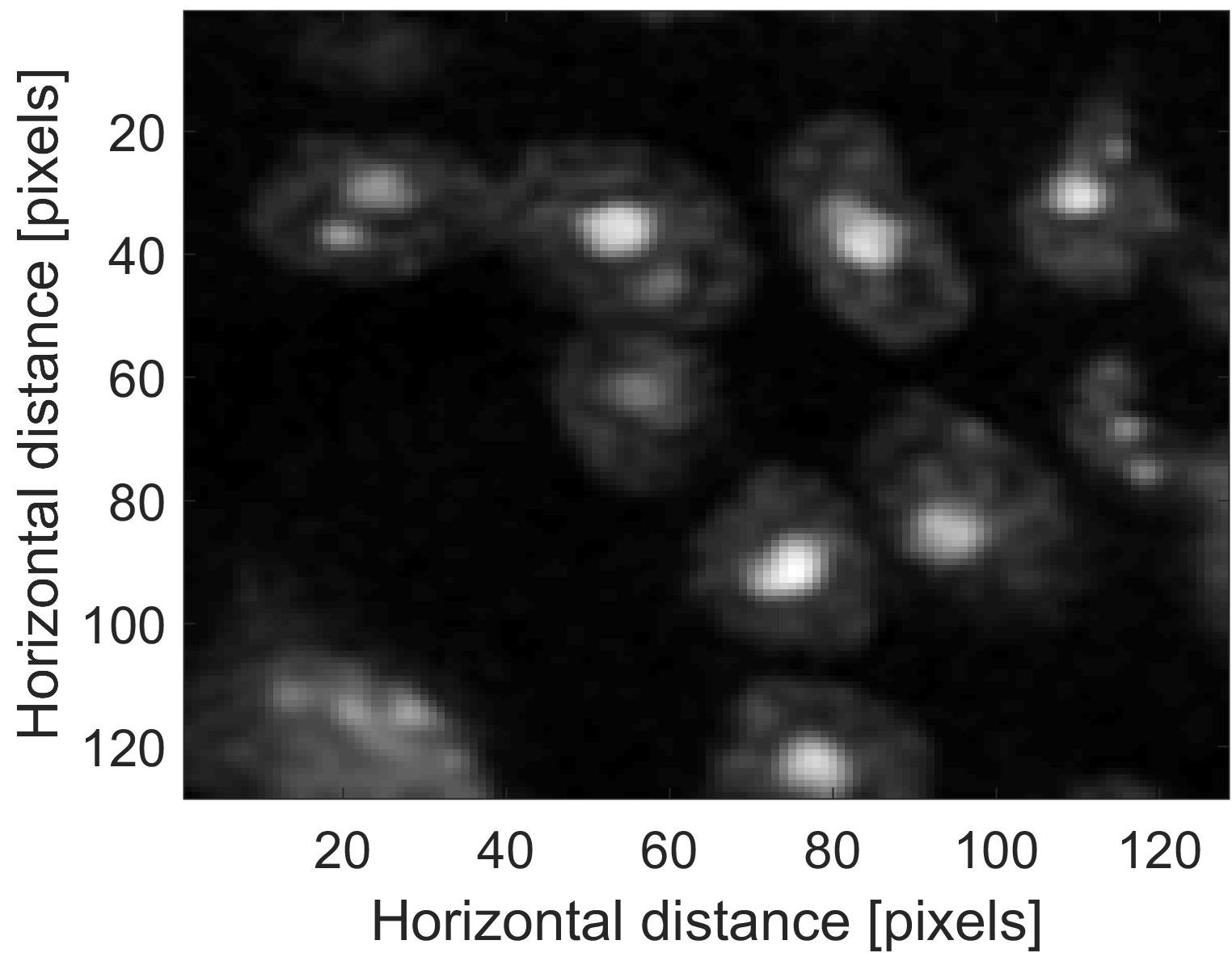}}

\vspace{5pt}
  Mouse Brain (Two-photon microscopy imaging)\\
\subfigure[Ground truth]{\includegraphics[width=0.17\textwidth]{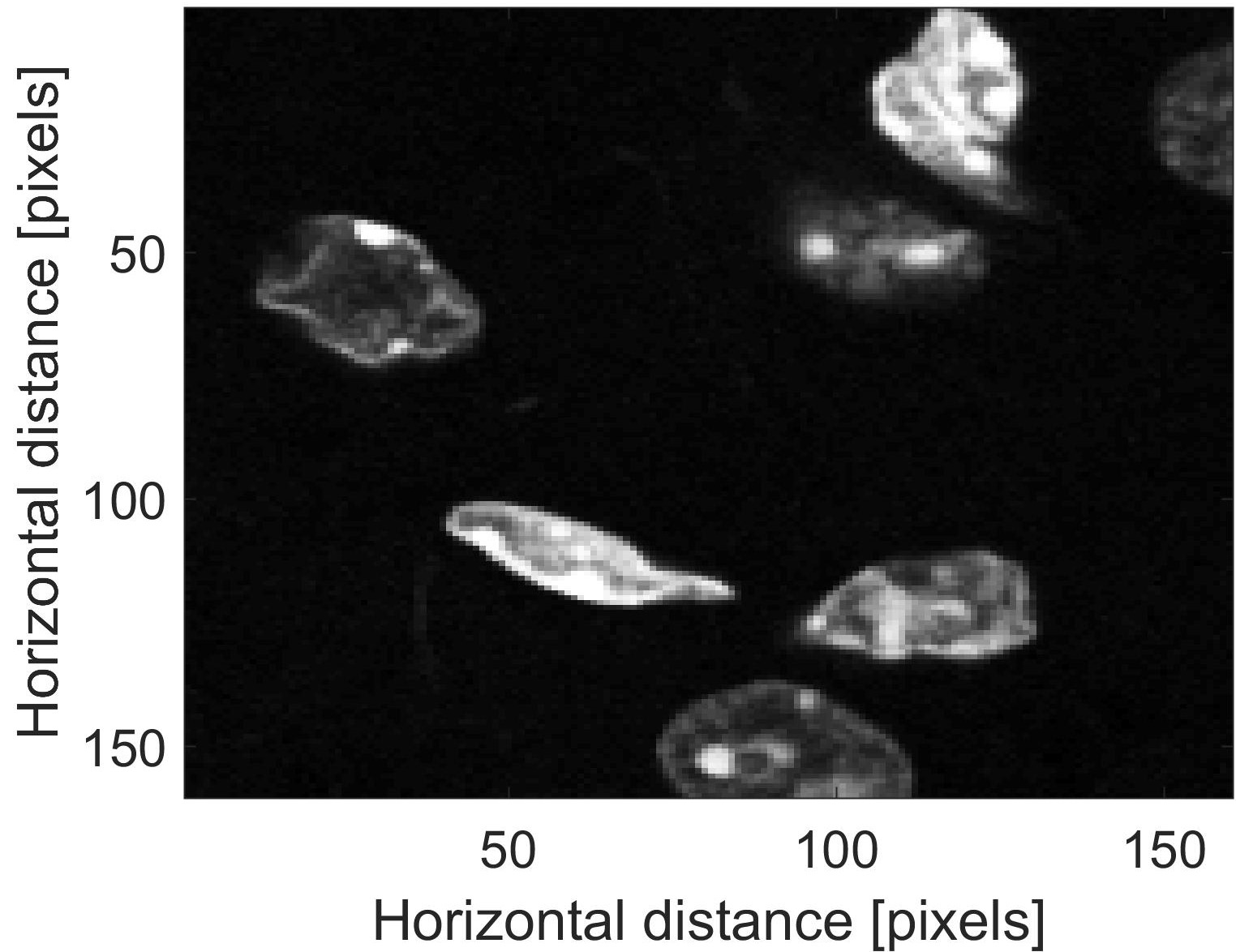}}
\subfigure[Observed image]{\includegraphics[width=0.17\textwidth]{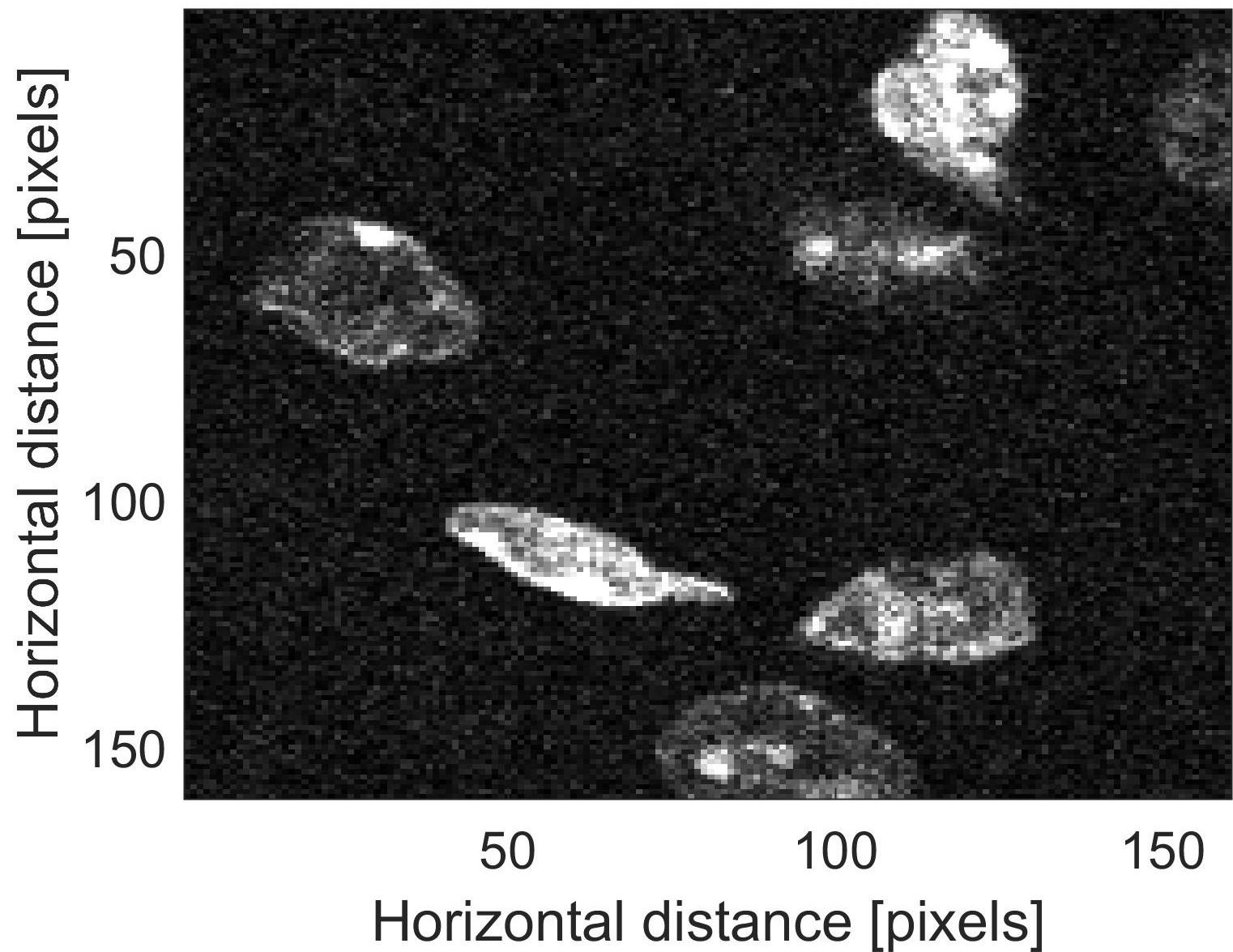}}
\subfigure[TV-ADMM]{\includegraphics[width=0.17\textwidth]{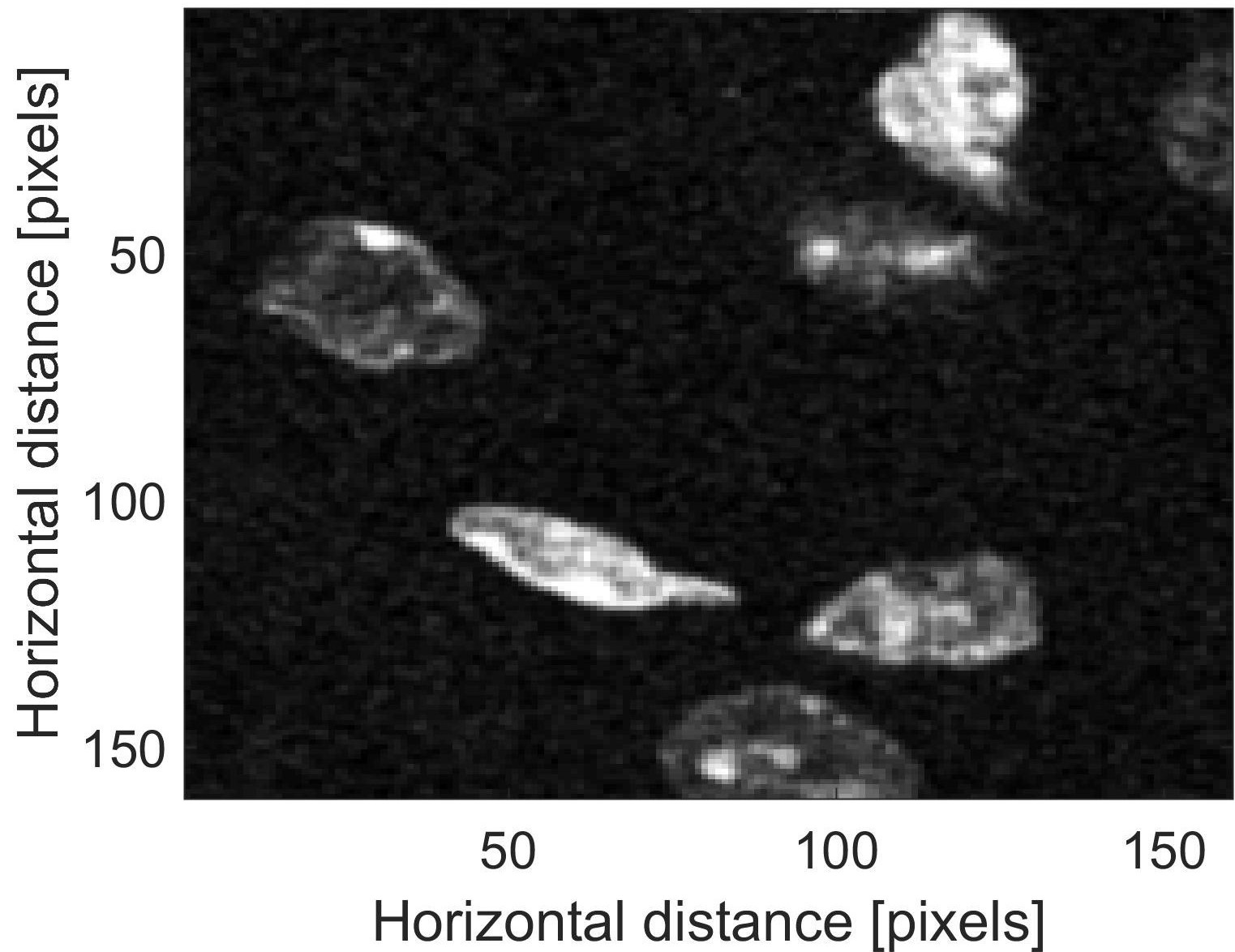}}
\subfigure[ADMM+BM3D]{\includegraphics[width=0.17\textwidth]{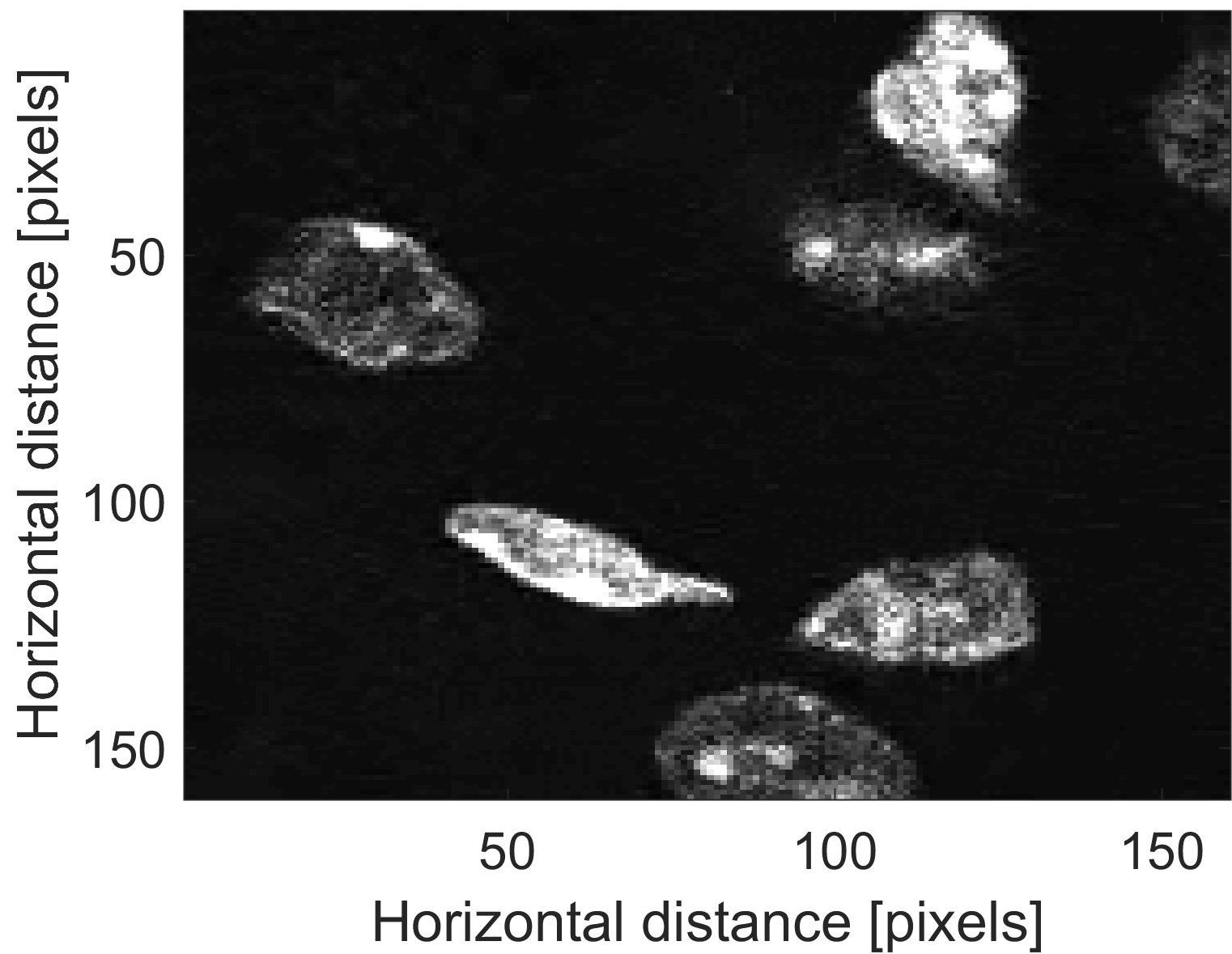}}
\subfigure[ADMM+TNRD]{\includegraphics[width=0.17\textwidth]{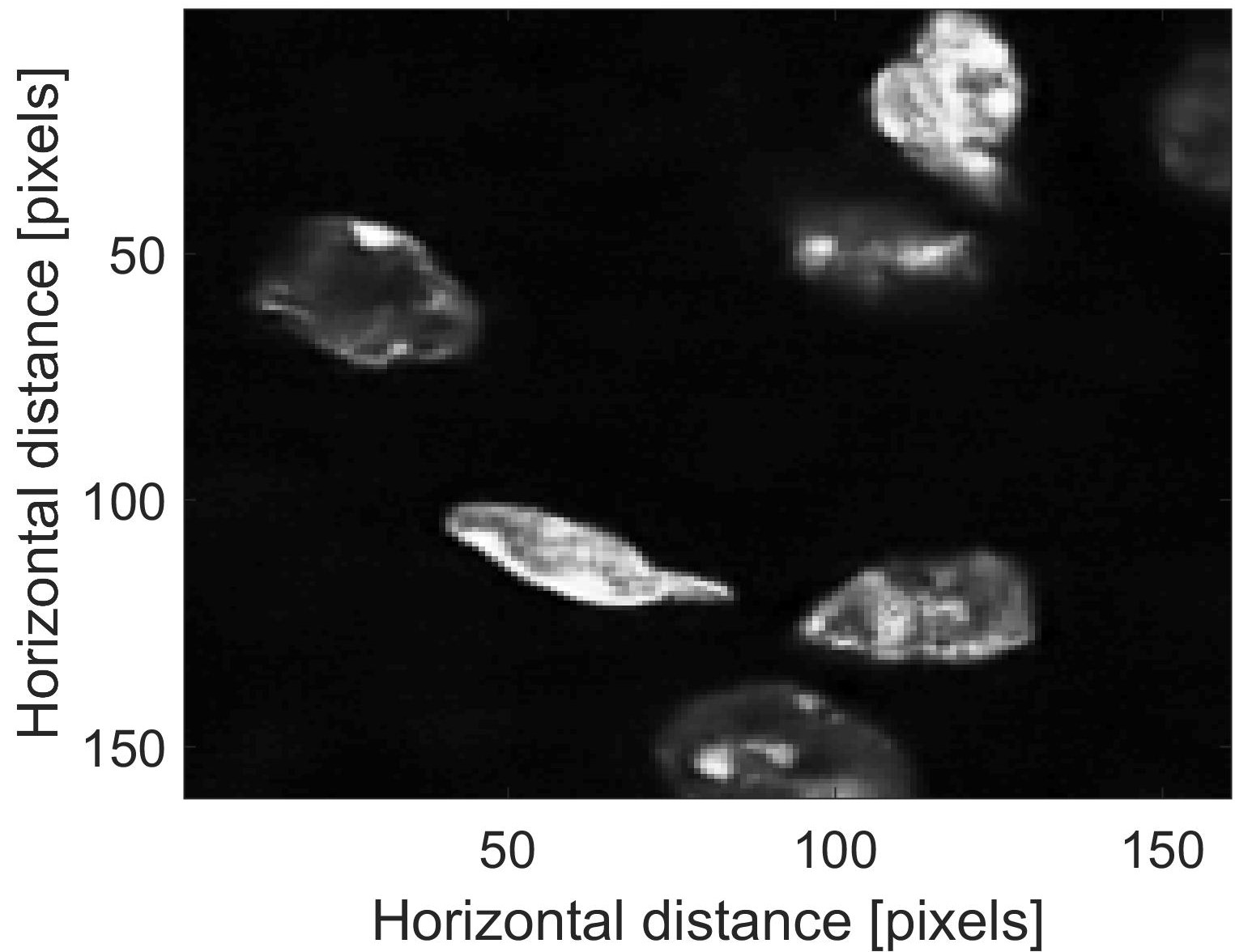}}
\subfigure[ADMM+VST+TNRD]{\includegraphics[width=0.17\textwidth]{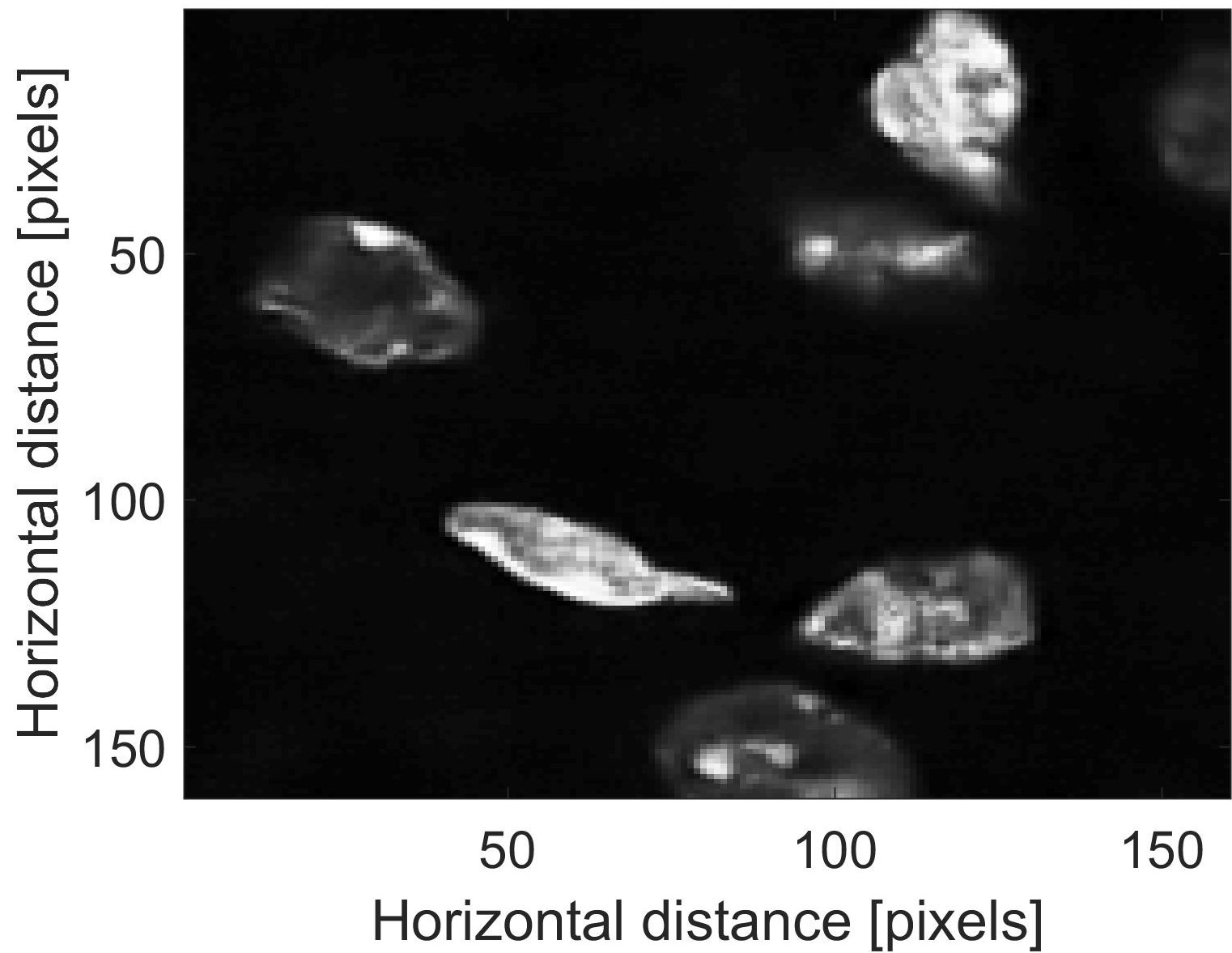}}
\subfigure[P$^4$IP]{\includegraphics[width=0.17\textwidth]{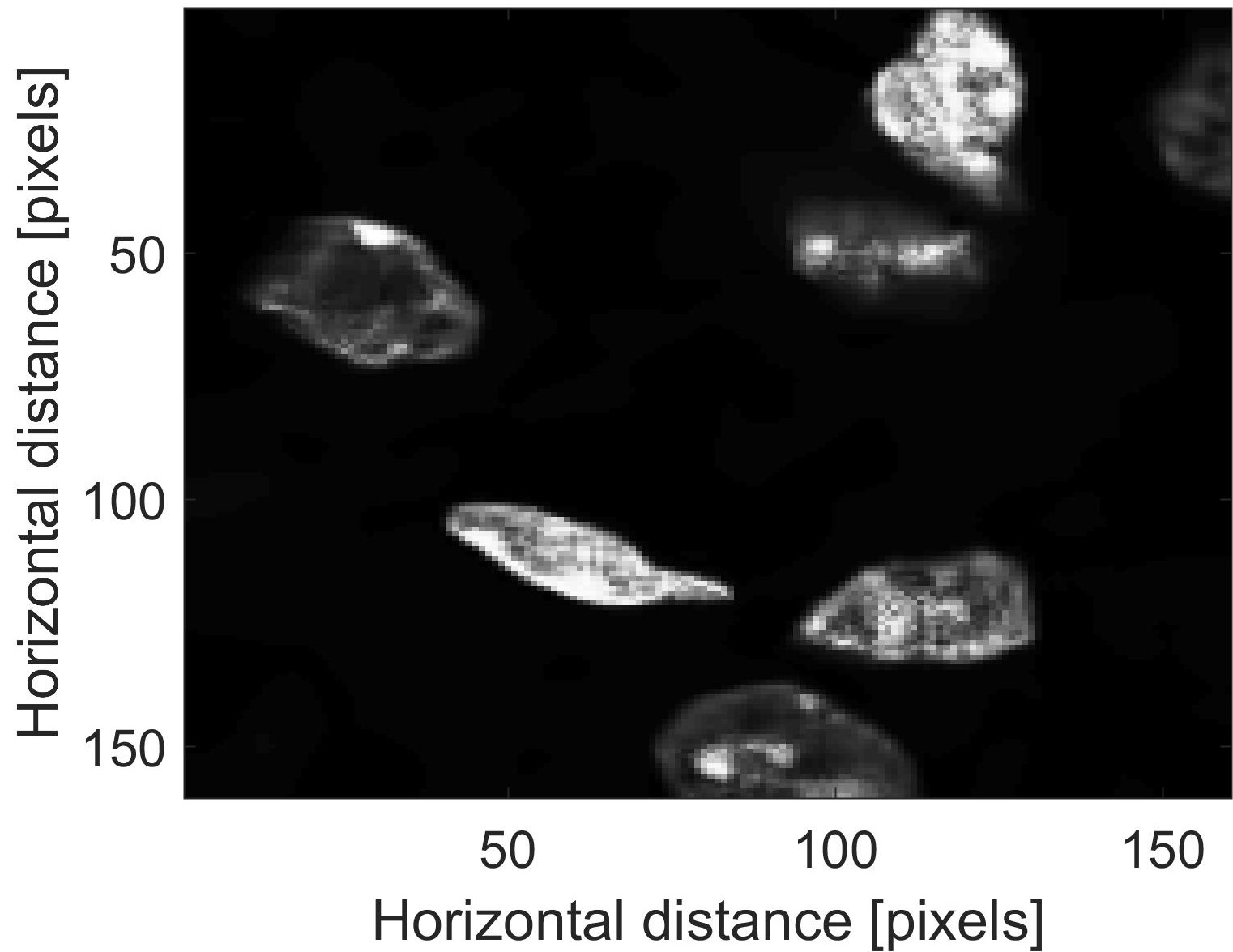}}
\subfigure[QAB-PnP]{\includegraphics[width=0.17\textwidth]{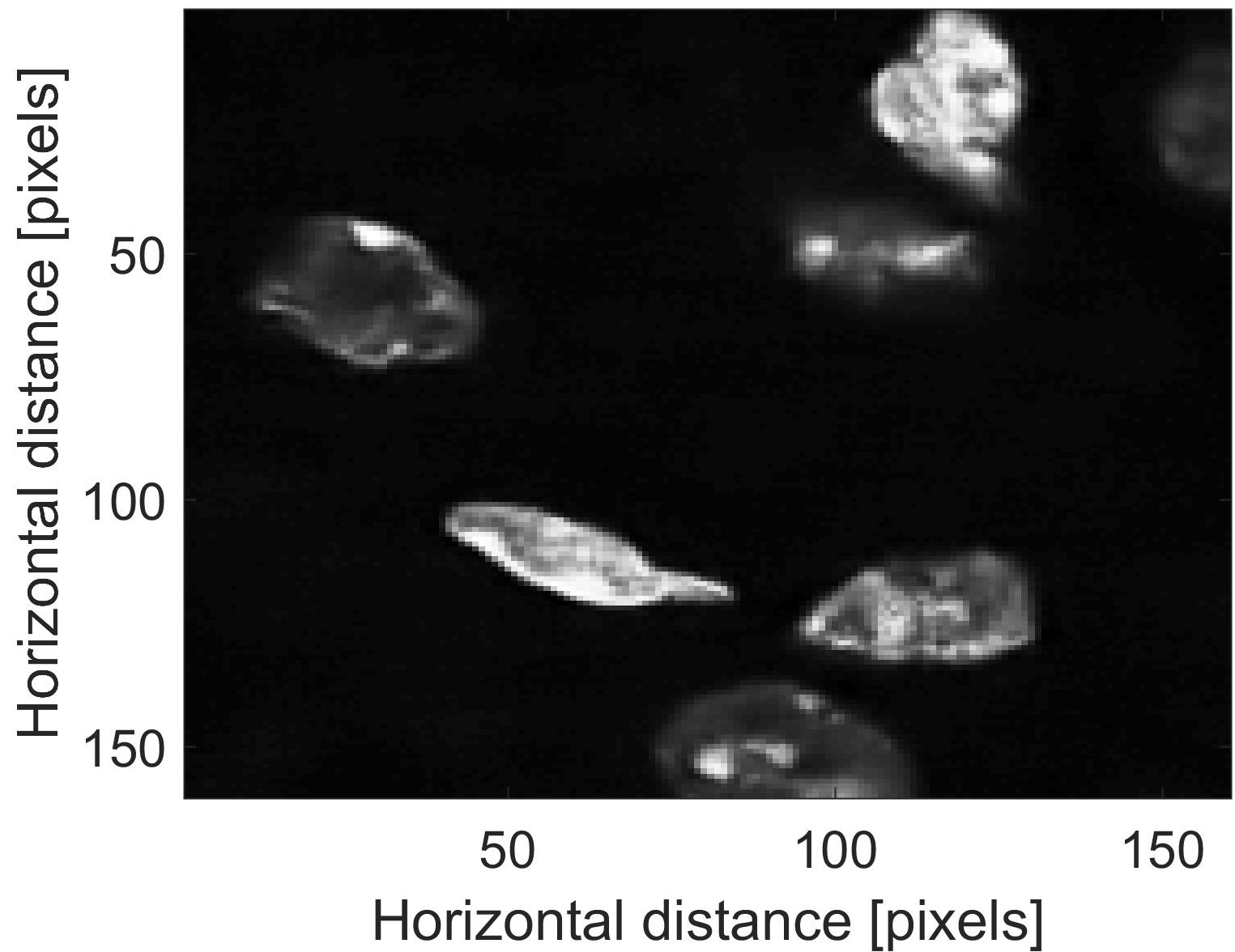}}

\caption{Deconvolution results for experimental fluorescence microscopy images using TV-ADMM, ADMM+BM3D, ADMM+TNRD, ADMM+VST+TNRD, P$^4$IP and the proposed QAB-PnP method. The proposed QAB-PnP algorithm used $\mathcal{E} = 4.1$, $\lambda_0 = 1.3$, $\hbar ^2/2m = 4$ and $\gamma = 1.01$, $\sigma_{\mathcal{QAB}} = 7$.}
\label{fig:flu_results}
\end{figure*}

Herein, we used three microscopy images from the online available data-set\footnote{\url{http://tinyurl.com/y6mwqcjs}} to illustrate the potential of the proposed method. Fig.~\ref{fig:flu_results} regroups the observed distorted images, their corresponding ground truth, and the deblurred images estimated by the six methods. PSNR and SSIM values comparing the observed and the deblurred images to the clean ones are given in Table~\ref{tab:resu_fluo}. These results clearly show the efficiency of the proposed algorithm in real fluorescence microscopy image enhancement.

\section{Conclusions}
\label{sec:conclusion}

This paper proposed a new PnP-ADMM scheme to handle Poisson deconvolution problems. Although Gaussian denoiser-based PnP-ADMM algorithms have achieved enormous success in this domain of image restoration, they are still facing a theoretical limitation related to the Anscombe transformation used to approximately transform the Poisson noise into additive Gaussian noise. Under this transformation, the convolution operation is not invariant. To overcome this drawback, we proposed in this work the QAB denoiser derived from principles of quantum mechanics, whose architecture makes it well adapted to different noise statistics, explaining its good behavious as denoiser embedded in a PnP-ADMM algorithm.

The simulation results allowed to provide an in-depth analysis of the impact of the hyperparameters on the accuracy and computation efficiency of the proposed method. They also allowed to show its interest compared to five existing methods. An issue of our proposal is the computational burden. The use of the OMP algorithm already dramatically decreases this time compared to earlier implementations \cite{dutta2021quantum}, but other improvements are certainly possible. As shown in our previous work \cite{dutta2021quantum} in the proposed quantum adaptive basis is equally efficient for Gaussian, Poisson and speckle noise removal problems without considering any prior information about the noise statistics. Therefore, the proposed deconvolution method could be suitable for other noise degradation than Poisson, and its evaluation in such conditions represents an interesting perspective. As another future perspective of this work one may think of implementing a more advanced inversion algorithm for a Poissonian model (\textit{e.g.,} SPIRAL-TAP \cite{harmany2012this}) instead of using a gradient descent method. Moreover, blind deconvolution is also an interesting perspective for future study, by coupling the proposed deconvolution algorithm with a PSF estimation method \cite{zhaohui2021convolutional, yu2012ablind}. Finally, such a PnP scheme can be further extended to other reconstruction problems, such as compressed sensing or super-resolution, using more efficient quantum mechanics based algorithms or by absorbing the patch-based procedure to the proposed framework, using for example the many-body quantum theory.

\section{Acknowledgments}
\label{sec:Acknowledge}

We would like to thank Dr Yoann Altmann, Assistant Professor at Heriot-Watt University, Edinburgh, Scotland, for his valuable inputs in this work. We also thank CNRS for funding through the 80 prime program.

\bibliographystyle{IEEEbib}
\bibliography{FINAL_Article_Dutta}

\begin{IEEEbiography}[{\includegraphics[width=1in,height=1.25in,clip,keepaspectratio]{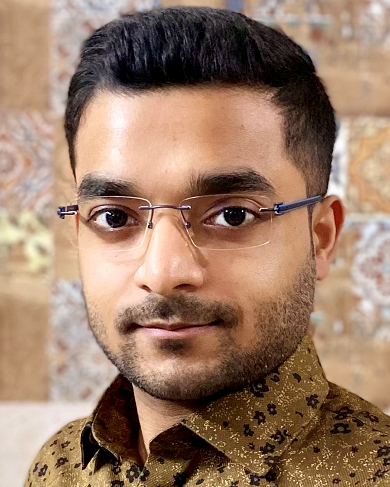}}]
{\textbf{Sayantan Dutta}} (Student Member, IEEE) received the B.Sc. degree in mathematics from the University of Burdwan, Bardhaman, India, in 2016, the M.Sc. degree in applied mathematics from the Visva-Bharati University, Santiniketan, India, in 2018, and the M.S. degree in fundamental physics from the University of Tours, Tours, France, in 2019. He is currently working toward the Ph.D. degree with the Institut de Recherche en Informatique de Toulouse and Laboratoire de Physique Th\'eorique de Toulouse laboratories, University Paul Sabatier Toulouse 3, Toulouse, France. His research interests include quantum computing, quantum image processing and inverse problems, particularly denoising, deblurring, and compressed sensing.

\end{IEEEbiography}

\begin{IEEEbiography}[{\includegraphics[width=1in,height=1.25in,clip,keepaspectratio]{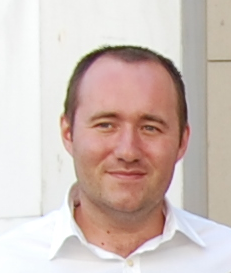}}]
{\textbf{Adrian Basarab}} (Senior Member, IEEE) received the M.S. and Ph.D. degrees in signal and image processing from the National Institute for Applied Sciences, Lyon, France, in 2005 and 2008, respectively. Between 2009 (respectively 2016) and 2021 he was assistant (respectively associate) professor at the University Paul Sabatier Toulouse 3 and a member of IRIT laboratory (UMR CNRS 5505). Since 2021, he is full professor at the University of Lyon and a member of CREATIS laboratory (UMR CNRS 5220). His research interests include computational medical imaging and more particularly inverse problems (deconvolution, super-resolution, compressive sampling, beamforming, image registration and fusion) applied to ultrasound image formation, ultrasound elastography, cardiac ultrasound, quantitative acoustic microscopy, computed tomography and magnetic resonance imaging. He is currently an Associate Editor for Digital Signal Processing and was a member of the French National Council of Universities Section 61 – Computer sciences, Automatic Control and Signal Processing from 2010 to 2015. In 2017, he was a Guest Co-Editor for the IEEE TUFFC special issue on "Sparsity driven methods in medical ultrasound." Since 2018, he has been the head of "Computational Imaging and Vision" Group, IRIT Laboratory. Since 2019, he has been a member of the EURASIP Technical Area Committee Biomedical Image \& Signal Analytics. Since 2020, he has been a member of the IEEE Ultrasonics Symposium TPC.
\end{IEEEbiography}

\begin{IEEEbiography}[{\includegraphics[width=1in,height=1.25in,clip,keepaspectratio]{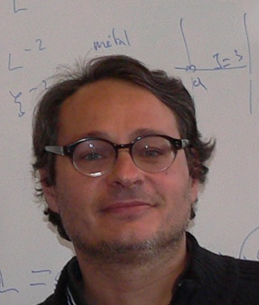}}]
{\textbf{Bertrand Georgeot}} is Directeur de Recherche (group leader) at CNRS in Laboratoire de Physique Th\'eorique, Université Paul Sabatier, Toulouse. After studies at the Ecole Polytechnique, he obtained his PhD in Orsay in 1993 then was postdoctoral associate at the University of Maryland (USA) and Nils Bohr Institute (Denmark). He is CNRS researcher in Toulouse since 1996. Since 2016 he is head of the laboratoire de physique th\'eorique.

He has authored or coauthored more than 90 publications in peer-reviewed journals, mostly in quantum physics (quantum chaos, Anderson localization, cold atom physics, multifractal quantum states) but also in interdisciplinary research in dynamical systems, astrophysics, classical and quantum computer sciences, network theory.

\end{IEEEbiography}

\begin{IEEEbiography}[{\includegraphics[width=1in,height=1.25in,clip,keepaspectratio]{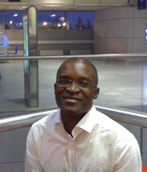}}]
{\textbf{Denis Kouam\'e}} (Senior Member, IEEE) is a professor in medical imaging and signal processing at Paul Sabatier University of Toulouse, France, since 2008. From 1998 to 2008 he was assistant then associate professor at University of Tours. From 1996 to 1998 he was senoir engineer at GIP Tours, France. He received the M.Sc, the Ph.D. and the habilitation to supervise research works(HDR) in signal processing and medical ultrasound imaging from the University of Tours in 1993, 1996 and 2004 respectively.

He was head of signal and image processing group, and then head of Ultrasound imaging group at Ultrasound and Signal Lab at University of Tours respectively from 2000 to 2006 and from 2006 to 2008. From 2009-2015 He was head of Health and Information Technology (HIT) strategic field at the Institut de Recherche en Informatique IRIT Laboratory, Toulouse. He currently leads the Signals and Image department at IRIT. His research interests cover the following areas: Medical imaging, ultrasound imaging, high resolution imaging, Doppler signal processing, Multidimensional biomedical Signal and image analysis including parametric modeling, spectral analysis and application to flow estimation, Sparse representation, Inverse problems.

He was invited for talks or in charge of different invited special sessions or tutorials at several IEEE conferences : ICASSP, ISBI, ISSPIT, ICIP. He has served on several international conferences technical program committees in signal, image processing or medical imaging, and also chaired various sessions at different international conferences. He was/is/ invited for talks in different universities inside and outside France. He was/is involved, as principal investigator or as member, in different European or French research projets (ANR,FUI, INSERM,…).

He is an Associate Editor for the IEEE Transactions on Ultrasonics Ferroelectrics and Frequency Control and for for the IEEE Transactions on Image Processing.

\end{IEEEbiography}

\EOD

\end{document}